\documentstyle[epsf,aps,psfig]{revtex}
\renewcommand{\thefootnote}{\fnsymbol{footnote}}
\begin{document}
\begin{flushright}
RIKEN-BNL preprint
\end{flushright}
\vspace*{1cm} 
\setcounter{footnote}{1}
\begin{center}
{\Large\bf Forming Disoriented Chiral Condensates through Fluctuations}
\\[1cm]
Dirk H.\ Rischke
\\ ~~ \\
{\it RIKEN-BNL Research Center, Physics Department} \\
{\it Brookhaven National Laboratory, Upton, New York 11973, U.S.A.}
\\ ~~ \\ ~~ \\
\end{center}
\begin{abstract} 
Using the influence functional formalism,
classical equations of motion for the $O(N)$ model are derived in the 
presence of a heat bath, in both the symmetric phase as well as the
phase of spontaneously broken symmetry. The heat bath leads to 
dissipation and fluctuation terms in the classical equations of motion,
which are explicitly computed to lowest order in perturbation theory.
In the broken phase these terms are found to be large for 
the $\sigma$ field, even at zero temperature,
due to the decay process $\sigma \rightarrow \pi\pi$,
while they are small for the $\pi$ fields at temperatures
below $T_c \simeq 160$ MeV. It is shown that in large volumes
the presence of dissipation and fluctuations suppresses the formation of 
disoriented chiral condensates (DCC's). In small volumes, however,
fluctuations become sufficiently large to induce the 
formation of DCC's even if chiral symmetry has not been restored in the 
initial stage of the system's evolution.
\\ \\ PACS number(s): 25.75.-q, 11.30.Qc, 11.30.Rd, 12.39.Fe, 12.38.Mh
\end{abstract}
\renewcommand{\thefootnote}{\arabic{footnote}}
\setcounter{footnote}{0}

\section{Introduction and Conclusions}

At vanishing net-baryon number density and temperatures above
$T_c \simeq 160 $ MeV, lattice calculations of quantum 
chromodynamics (QCD) predict the existence of a phase of nuclear matter 
where quarks and gluons are deconfined and chiral symmetry is 
restored \cite{lattice}.
One of the primary goals of relativistic heavy-ion physics is to
create and study this phase in nuclear collisions \cite{harris}. 

The formation of a so-called disoriented chiral condensate (DCC) has been 
proposed as a possible signature for the restoration of chiral 
symmetry \cite{DCC}. The idea is the following: in the phase where chiral 
symmetry is restored the quark condensate vanishes, 
$\langle \bar{q} q \rangle \simeq 0$. If at all, in a heavy-ion collision
this state can only be transiently created.
Once the system cools below $T_c$, chiral symmetry is spontaneously broken 
and the system has to evolve back into the true ground state where $\langle
\bar{q} q \rangle \neq 0$. If in the course of this evolution
the pseudoscalar condensate $\langle \bar{q} \mbox{\boldmath $\tau$} 
\gamma_5 q \rangle$ assumes non-vanishing values (instead of remaining zero, 
as in the ground state), one speaks of a disoriented chiral condensate.

This mechanism becomes physically most transparent in the framework
of the $O(4)$ model.
One identifies $\phi_1 \sim \langle \bar{q} q \rangle,\, \phi_i \sim
\langle \bar{q} \tau_i \gamma_5 q \rangle,\,i=2,3,4$, and spontaneously broken
symmetry is realized by a potential 
$U(\underline{\phi})$ which looks like a (tilted) ``mexican hat'', with
minimum at $\underline{\phi} = (f_\pi,{\bf 0})$ for $T=0$. For
increasing $T$, the ``hat'' becomes shallower and the minimum
moves towards the origin, such that $\underline{\phi} \rightarrow 0$
for $T \rightarrow T_c$, and chiral symmetry is restored.
The creation of DCC's is most likely in the so-called ``quench scenario''
\cite{quench}. Here, it is assumed that, after restoration of chiral
symmetry, the system cools instantaneously 
to $T=0$. If the fields $\underline{\phi}$ are assumed to
follow classical equations of motion in the potential $U(\underline{\phi})$,
the evolution of the system can be visualized as ``rolling down'' from the
initial state with restored chiral symmetry, $\underline{\phi} =0$, into the 
true ground state $\underline{\phi} = (f_\pi, {\bf 0})$. If that
happens on a ``path'' where $\phi_i \neq 0, \,i=2,3,4$, the
chiral condensate becomes ``disoriented''.

The ratio $R$ of neutral pions to the sum of neutral and charged pions was
suggested as experimental observable \cite{DCC}. 
If a single domain of DCC is formed,
the probability $P(R) \sim 1/\sqrt{R}$, which is
drastically different from the case where pions are emitted in
a statistically independent manner,
$P(R) \sim \delta (R-1/3)$ (for large, isospin-symmetric systems).

The formation and decay of DCC's has been studied in a variety of approaches
over the last couple of years \cite{approaches}, and the
original idea has undergone several refinements.
One obvious effect that has an influence on DCC formation in heavy-ion
collisions is the presence of a background of a multitude of other particles.
Most of these are pions with typical transverse momenta on the order of a 
couple of hundred MeV \cite{QM97}. In \cite{cgtb} it was assumed that these
pions constitute a background (``heat bath'') of unobserved, thermalized 
degrees of freedom. Their presence leads to temperature-dependent
dissipation and fluctuation terms
in the classical equations of motion which correspond to four--particle
interactions between classical fields and particles in the heat bath
and can be rigorously derived \cite{cgbm} (cf.\ also \cite{cgsl}) via the
influence functional formalism \cite{Feynman}. For a typical (average)
temperature evolution in a heavy-ion collision it was then studied 
in \cite{cgtb}, how DCC's form in an expanding system which is in
contact with this (steadily cooling) heat bath of particles. 
The main result was that, on the average, dissipation and fluctuation tend to
suppress the formation of DCC's. 
However, fluctuations grow $\sim 1/\sqrt{V}$ for $V \rightarrow 0$. Thus,
in small volumes and in a single event, the fluctuations can be large enough 
to destabilize the system and actually enhance the likelihood to form a
DCC.

The authors of \cite{cgtb} made two approximations.
The first was to compute the dissipation terms
in the chirally {\em symmetric\/} phase and then to use them
for the evolution of the system in the phase where
chiral symmetry is spontaneously {\em broken}. 
This leads to considerable simplifications, since in the symmetric phase 
these terms are straightforward generalizations of results obtained 
previously in $\phi^4$ theory \cite{cgbm}. For dissipation arising from
four--particle interactions which are present both in the symmetric
and the broken phase, this approximation is probably 
justified in the initial stage of the evolution, where temperatures 
are high and close to $T_c$. It becomes questionable at smaller
temperatures due to the fact that not all particles have the same mass 
in the broken phase (the $\sigma$ is heavy, while the $\pi$'s are light). 

There is, however, another reason to reconsider this approximation
at temperatures below $T_c$. In the broken phase the 
structure of the underlying Lagrangian is fundamentally different:
there are additional three--particle interactions. As will be shown in detail
in the following, this has the
consequence that, while the dissipation is $\sim \lambda^2$ in the
symmetric phase and of equal magnitude for all fields 
$\phi_a,\, a=1,\ldots,4$, dissipative corrections arise already to 
{\em first\/} order in $\lambda$ in the broken phase, and are of
sizable magnitude for the $\sigma$ degree of freedom and rather small for 
pions. The former correspond physically
to the decay of a $\sigma$ into two $\pi$'s (cf.\ also \cite{boya}), 
and are non-vanishing even at $T=0$. Therefore, the dissipation 
coefficients below $T_c$ are different from an extrapolation
of the results obtained in the symmetric phase.

The second approximation made in \cite{cgtb} was to infer the variance of
the fluctuation terms $\xi_a,\, a=1,\ldots, 4,$ from the dissipation 
coefficient $\eta$ via 
\begin{equation} \label{variance1}
\langle \xi_a(t)\, \xi_b(t') \rangle = \frac{2\, T\, \eta}{V} \, \delta(t-t')\,
\delta_{ab}\,\, .
\end{equation} 
Apart from the fact that this equation does not account for
different dissipation coefficients for $\sigma$ and $\pi$'s, one has to note
that the factor $2T$ stems from the high-temperature (i.e.\ classical) 
limit of a more general expression (cf.\ \cite{cgbm} and below). 
For temperatures $T$ smaller than the typical mass scale of the theory,
i.e., for $T < m_\pi \simeq T_c$, one therefore expects sizable deviations
from (\ref{variance1}).
In particular, whenever $\eta$ happens to be finite at $T=0$ (which, as
mentioned above and shown below, is indeed the case for the $\sigma$ 
field in the broken phase), eq.\ (\ref{variance1}) predicts that 
fluctuations vanish even in the presence of dissipation, in 
contradiction to the dissipation--fluctuation theorem.

The aim of this paper is to make a first step towards a consistent
treatment of dissipation and fluctuation in the framework of the
$O(4)$ model in the phase where chiral symmetry
is broken. The outline of the paper, as well as the main results and 
conclusions are as follows. In section II, the derivation of the 
influence functional is presented for a system of $N$ real-valued scalar
fields $\phi_1, \ldots, \phi_N$.
Short wavelength modes, i.e., those with ``hard'' momenta $|{\bf k}| >
k_c$, where $k_c$ is an arbitrary momentum scale, are separated
from long wavelength modes, i.e., those with ``soft'' momenta 
$|{\bf k}|\leq k_c$ \cite{boede}, and the reduced density matrix for 
the soft modes is obtained from the full density matrix by tracing 
over the hard degrees of freedom, which thus are thought to
constitute the unobserved background (``heat bath'') mentioned above. 
The influence functional enters
as a phase factor in the reduced density matrix. This section is a
straightforward generalization of the treatment in \cite{cgbm}. It
serves merely to introduce the notation, and can be skipped by readers 
familiar with the subject. 

Section III contains the derivation of the classical
equations of motion for the soft fields by expanding the reduced
density matrix around its diagonal elements. The main difference as compared
to previous treatments of the subject \cite{cgbm,son,gleiser} 
is that functional derivatives of the influence functional are expressed 
as averages over functional derivatives of 
the action characterizing the interaction between soft, 
classical fields and the unobserved hard degrees of freedom. This has the 
advantage that, in a perturbative computation of the correction terms to the
classical equations of motion up to some given order $n$ in the
coupling constant, expectation values of the hard degrees 
of freedom have to computed only to order $n-1$. It is furthermore
shown that for systems with more than one field degree of freedom, 
there can be cross correlations between the noise terms for different fields.

In Section IV, the general framework derived in the previous sections
is applied to the $O(N)$ model in the symmetric case. The corrections
to the classical equation of motion are first computed to order $\lambda$,
where they only change the mass term in the
classical equation of motion, and then also to order $\lambda^2$,
where they lead to dissipation and fluctuations. The treatment
is fairly cursory, since this case is rather similar to $\phi^4$ theory
(or, in other words, the $O(1)$ case) discussed in detail in \cite{cgbm}. 
The main focus is to demonstrate the applicability of the method 
developed in Section III. It is shown that the damping coefficient 
agrees with previous results in the cases $N=1$ \cite{cgbm} and 
$N=4$ \cite{cgtb}.

In Section V the $O(N)$ model is discussed for the case of spontaneously
broken symmetry. In this case, only corrections up to first order in
the coupling constant are considered, but due to the presence of
two interaction vertices in the Lagrangian (one proportional to $\lambda$ and
the other proportional to $\lambda\, f_\pi \sim \lambda^{1/2}$), there is 
dissipation and fluctuation already to this order in $\lambda$. 
In particular, the dissipation
coefficient for the $\sigma$ field is shown to be large even at $T=0$, 
$\eta_\sigma \simeq m_\sigma \simeq 600$ MeV. The physical process 
responsible for this is the decay of a $\sigma$ into two $\pi$'s. 
This has important consequences which are discussed in Section VI. 
On the other hand,
the dissipation coefficient for the $\pi$ fields is small for the
temperature range of interest, the reason being that
scattering of a $\pi$ or $\sigma$ from the heat bath
off a classical $\pi$ field is strongly suppressed by phase space.
For $T \rightarrow 0$, as well as in the chiral limit $m_\pi \rightarrow 0$
for arbitrary $T$, one even has $\eta_\pi \rightarrow 0$.
The classical equations of motion for $\sigma$'s and $\pi$'s are derived 
and studied in detail for the ${\bf k} =0$ modes of the classical fields. 
It is shown that the static solution for the $\sigma$ field corresponds to 
the well-known shift of the vacuum ground state at finite temperature.
Moreover, the validity of Goldstone's theorem at the classical level is 
checked.

In Section VI, arguments are presented that
the formation of DCC's is most likely in a quench scenario,
i.e., at $T=0$. Numerical solutions of the equations of motion
for the homogeneous modes of the fields at zero temperature are then
presented, which show that, in large volumes, 
the large dissipation coefficient for the $\sigma$ field leads to a rapid 
damping of oscillations of all classical fields, including the pions.
The formation of DCC's seems thus not very likely in large systems. 
In small volumes, on the other hand, the
fluctuations associated with the dissipation are large enough 
to disorient the pion fields and possibly lead to the formation of DCC's.
These results are in agreement with those found in \cite{cgtb}, except
that here they apply even at $T=0$.
This may have the experimentally interesting implication that
DCC's are perhaps formed more readily in collisions of lighter ions, or
even in $pp$--collisions (which constitutes a possible explanation for
the CENTAURO events \cite{centauro}), while they are presumably less likely 
to be formed in collisions of heavy ions. A definite conclusion, however,
can only be drawn after performing dynamical simulations including 
modes with finite ${\bf k}$ and taking the overall 
expansion of the system into account \cite{krdhr}.

There is, however, another possible consequence of the results found here.
As long as the volume is small, $\sim 10\, {\rm fm}^3$, the
disorientation of the $\pi$ fields (and possibly DCC formation)
becomes likely even if the system's evolution starts near the true 
ground state (even small perturbations in the $\pi$ fields
suffice). As a consequence, in small systems 
restoration of chiral symmetry does not seem to be a necessary prerequisite
to observe disorientation of classical pion fields. It is, however,
rather likely that this fluctuation-induced phenomenon is related to 
ordinary fluctuations in finite volumes.

The results of the present work have to be viewed in the light of
the following two comments: (a) the classical approximation works well in the
limit of large occupation numbers. For instance, in thermodynamical 
equilibrium this is achieved for modes with energy $\omega \ll T$. 
However, the lowest-energy mode for (non-interacting) 
$\sigma$ particles has $\omega = m_\sigma
\simeq 600$ MeV which is much larger than the temperature in the broken 
phase ($T \leq T_c \simeq 160$ MeV). Therefore, at least in thermodynamical
equilibrium, the $\sigma$ field should not seriously be considered classically.
In a sense, the expected large quantum corrections become manifest
in the fluctuations induced by the decay $\sigma \rightarrow
\pi \pi$ as calculated in the present work.
(b) The dissipation and fluctuation terms are here computed to first
order in perturbation theory. However, for realistic 
parameters of the $O(N)$ model the coupling constant $\lambda \simeq
20$, which renders the perturbative expansion uncontrollable. Future
studies will have to improve on this point.

Possible other extensions of the present work are (a) the inclusion of
collisional interactions between $\sigma$'s and $\pi$'s which are
of order $\sim \lambda^2$ \cite{dhr}, (b) the study
of long wavelength modes with finite momenta ${\bf k}$ instead of
the homogeneous modes only, as well as (c) the study of DCC formation 
including these effects with a realistic temperature evolution \cite{krdhr}.

Units are $\hbar=c=k_B=1$, and the metric tensor is 
$g^{\mu\nu} = {\rm diag} (+,-,-,-)$.

\section{Derivation of the influence functional}

Let us consider a quantum system of $N$ real-valued scalar
fields $\underline{\phi} = (\phi_1, \ldots,\phi_N)$,
characterised by a Lagrangian density ${\cal L}(\underline{\phi})$.
The Hamilton operator $\hat{H} \equiv \int {\rm d}^3 {\bf x}\,\, {\cal H}
(\hat{\underline{\phi}}(x))\,, \,\, {\cal H} \equiv \underline{\pi}
\cdot \partial_t \underline{\phi} - {\cal L},\, \underline{\pi} \equiv
\partial {\cal L}/\partial (\partial_t \underline{\phi})$, 
is assumed to have no explicit time dependence. 
The time evolution of the density matrix reads:
\begin{equation}
i\, \partial_t \, \hat{\rho} = [ \hat{H}, \hat{\rho}]\,\, ,
\end{equation}
with initial condition $\hat{\rho}(t_i) \equiv \hat{\rho}_i$. 
The formal solution at time $t_f$ is:
\begin{equation} \label{solution}
\hat{\rho} (t_f) = \hat{U}(t_f,t_i) \, \hat{\rho}_i \, \hat{U}(t_i,t_f)\,\, ,
\end{equation}
where
\begin{equation}
\hat{U}(t_f,t_i) \equiv \exp \left\{ -i\, \hat{H}\, (t_f-t_i)\right\}
\equiv \hat{U}^\dagger (t_i,t_f) \equiv \hat{U}^{-1} (t_i,t_f)
\end{equation}
is the time evolution operator. 
Let us choose a basis $\left\{ | \underline{\phi}_f \rangle \right\} $ 
of eigenfunctions of the Schr\"odinger field operator 
$\hat{\underline{\phi}}_f({\bf x})$. In this basis, one finds 
for the density matrix element
\begin{equation} \label{densmatrix}
\rho(\underline{\phi}_f,\underline{\phi}_f';t_f)  
\equiv \langle \underline{\phi}_f | \, \hat{\rho}(t_f) \, 
| \underline{\phi}_f' \rangle
= \int {\rm D}\underline{\phi}_i \, {\rm D} \underline{\phi}_i' \, 
\rho(\underline{\phi}_i,\underline{\phi}_i';t_i)
\int_{\underline{\phi}_i}^{\underline{\phi}_f} {\cal D}\underline{\phi} 
\int_{\underline{\phi}_i'}^{\underline{\phi}_f'} {\cal D} \underline{\phi}' \,
\exp \left\{ i \left( S[\underline{\phi}] - S[\underline{\phi}'] \right) 
\right\} \,\, ,
\end{equation}
where one has employed eq.\ (\ref{solution}), the completeness relation
\begin{equation}
1 = \int {\rm D} \underline{\phi}_i \, | \underline{\phi}_i 
\rangle \, \langle \underline{\phi}_i |
\end{equation}
(here ${\rm D} \underline{\phi} \equiv \prod_{a=1}^N \prod_{\bf x} {\rm d}
\phi_a({\bf x})$),
and the path integral representation
\begin{equation}
\langle \underline{\phi}_f |\, \hat{U}(t_f,t_i) \, | \underline{\phi}_i 
\rangle \equiv \int_{\underline{\phi}_i}^{\underline{\phi}_f} {\cal D} 
\underline{\phi} \, \exp \left\{ i \, S[\underline{\phi}] \right\}\,\, ,
\end{equation}
where ${\cal D} \underline{\phi} \equiv \prod_{a=1}^N 
\prod_{t,{\bf x}} {\rm d} \phi_a(t,{\bf x})$, and
$S[\underline{\phi}] = \int_{t_i}^{t_f} {\rm d}t \int {\rm d}^3{\bf x} \, 
{\cal L}(\underline{\phi}(t,{\bf x})) \equiv \int_{t_i}^{t_f} 
{\rm d}^4x\, {\cal L} (\underline{\phi}(x))$.
Note that it is customary \cite{cgbm,son,gleiser} to employ the 
closed-time-path formalism
\cite{ctp} to simplify the right-hand side of eq.\ (\ref{densmatrix}).
Although this is an elegant bookkeeping device, for the sake of clarity
we shall continue to work in terms of the fields $\underline{\phi}$ and
$\underline{\phi}'$.

At each time $t$, a particular field component with 3--momentum ${\bf k}$ has
the Fourier representation
\begin{equation} \label{FR}
\underline{\phi} (t,{\bf k}) \equiv \int {\rm d}^3 {\bf x} \, 
e^{-i\, {\bf k} \cdot {\bf x}} \, \underline{\phi} (t,{\bf x})\,\, .
\end{equation}

Let us now separate ``hard'' from ``soft'' degrees of freedom
\cite{cgbm,boede}. More precisely, let us define soft fields 
\begin{equation} \label{soft}
\underline{\varphi}(t,{\bf x}) \equiv \int 
\frac{{\rm d}^3{\bf k}}{(2 \pi)^3}\,
e^{i \, {\bf k}\cdot {\bf x}} \, \underline{\phi}(t,{\bf k}) \, 
\Theta(k_c - |{\bf k}|) \,\, ,
\end{equation}
and hard fields
\begin{equation} \label{hard}
\underline{\Phi}(t,{\bf x}) \equiv \int \frac{{\rm d}^3{\bf k}}{(2 \pi)^3}\,
e^{i \, {\bf k}\cdot {\bf x}} \, \underline{\phi}(t,{\bf k}) \, 
\Theta(|{\bf k}|-k_c) \,\,,
\end{equation}
where $k_c$ is an arbitrary momentum scale separating hard from soft
momentum modes. Obviously, the space spanned by $\{ | \underline{\phi} 
\rangle \}$ is the product space spanned by $\{ | \underline{\varphi} 
\rangle \}$  and $\{ | \underline{\Phi} \rangle \}$, 
$\{ | \underline{\phi} \rangle \}= \{ | \underline{\varphi} \rangle \}
\otimes \{ | \underline{\Phi} \rangle \}$. 
(This is most easily seen in the space of functions 
$\underline{\phi}(t, {\bf k})$.) Also, $\underline{\phi}(x) = 
\underline{\varphi}(x) + \underline{\Phi}(x)$,
and $S[\underline{\phi}] = S[\underline{\varphi}] + S[\underline{\Phi}] + 
S_I[\underline{\varphi}, \underline{\Phi}]$, where 
$S_I[\underline{\varphi},\underline{\Phi}]$ 
is the action characterizing interactions between soft and hard fields.

The density matrix (\ref{densmatrix}) then assumes the form
\begin{eqnarray}
\lefteqn{ \rho (\underline{\varphi}_f, \underline{\Phi}_f; 
\underline{\varphi}_f', \underline{\Phi}_f';t_f)  = 
\int {\rm D} \underline{\varphi}_i \, {\rm D} \underline{\Phi}_i \, 
{\rm D}\underline{\varphi}_i' \, {\rm D} \underline{\Phi}_i' \,\, 
\rho (\underline{\varphi}_i,\underline{\Phi}_i; \underline{\varphi}_i', 
\underline{\Phi}_i';t_i) } \nonumber \\ 
& \times & \int_{\underline{\varphi}_i}^{\underline{\varphi}_f} 
{\cal D}\underline{\varphi}  \int_{\underline{\Phi}_i}^{\underline{\Phi}_f} 
{\cal D}\underline{\Phi} \int_{\underline{\varphi}_i'}^{\underline{\varphi}_f'}
{\cal D}\underline{\varphi}' \int_{\underline{\Phi}_i'}^{\underline{\Phi}_f'}
{\cal D}\underline{\Phi}' \,\, \exp \left\{ i \left( S[\underline{\varphi}] 
+ S[\underline{\Phi}] +S_I [\underline{\varphi},\underline{\Phi}]
- S[\underline{\varphi}'] - S[\underline{\Phi}'] - S_I [\underline{\varphi}',
\underline{\Phi}'] \right) \right\}
 \,\, . \label{DM}
\end{eqnarray}
Let us now assume that the interactions between soft and hard fields
vanish at the initial time $t_i$. Then, the
initial density matrix $\hat{\rho}_i$ is block-diagonal, $\hat{\rho}_i
= \hat{\rho}_i^{(\varphi)} \otimes \hat{\rho}_i^{(\Phi)}$, and
\begin{equation}
\rho(\underline{\varphi}_i,\underline{\Phi}_i; \underline{\varphi}_i', 
\underline{\Phi}_i';t_i) \equiv \rho^{(\varphi)}(\underline{\varphi}_i, 
\underline{\varphi}_i';t_i) \, \rho^{(\Phi)}(\underline{\Phi}_i, 
\underline{\Phi}_i';t_i)\,\,.
\end{equation} 
The object of interest in the following is the {\em reduced\/}
density matrix $\rho^{(\varphi)} (\underline{\varphi}_f, 
\underline{\varphi}_f' ;t_f)$ for the soft fields, 
which is obtained by tracing (\ref{DM})
over the degrees of freedom of the hard fields. 
This reduced density matrix has the form
\begin{equation} \label{reducedDM}
\rho^{(\varphi)}(\underline{\varphi}_f, \underline{\varphi}_f';t_f)  =  
\int {\rm D}\underline{\varphi}_i \, {\rm D}\underline{\varphi}_i'
\, \rho^{(\varphi)}(\underline{\varphi}_i, \underline{\varphi}_i';t_i)
\int_{\underline{\varphi}_i}^{\underline{\varphi}_f} 
{\cal D}\underline{\varphi} 
\int_{\underline{\varphi}_i'}^{\underline{\varphi}_f'} 
{\cal D}\underline{\varphi}' 
\, \exp \left\{ i \left( S[\underline{\varphi}] - S[\underline{\varphi}'] 
+ S_{IF} [\underline{\varphi},\underline{\varphi}'] \right) \right\} \,\, ,
\end{equation}
where the {\em influence functional\/} $S_{IF}[\underline{\varphi}, 
\underline{\varphi}']$ is defined by \cite{Feynman}
\begin{eqnarray} 
\lefteqn{\exp \left\{ i \, S_{IF}[\underline{\varphi}, \underline{\varphi}'] 
\right\}  }\nonumber \\
& \equiv  & \int {\rm D} \underline{\Phi}_f \, {\rm D} \underline{\Phi}_i \, 
{\rm D} \underline{\Phi}_i' \, \rho^{(\Phi)}(\underline{\Phi}_i, 
\underline{\Phi}_i';t_i) \int_{\underline{\Phi}_i}^{\underline{\Phi}_f} 
{\cal D}\underline{\Phi} \int_{\underline{\Phi}_i'}^{\underline{\Phi}_f} 
{\cal D}\underline{\Phi}' \, \exp \left\{ i \left( S[\underline{\Phi}] 
+S_I [\underline{\varphi},\underline{\Phi}]
 - S[\underline{\Phi}'] - S_I [\underline{\varphi}',\underline{\Phi}'] 
\right) \right\} \,\, .\label{infl}
\end{eqnarray}
An obvious property of the influence functional (for real-valued
scalar fields) is 
\begin{equation} \label{propa}
S_{IF}[\underline{\varphi}, \underline{\varphi}'] = 
- S^*_{IF}[\underline{\varphi}', \underline{\varphi}]\,\, .
\end{equation}
For the proof, note that for any element of the (hermitean) density matrix 
one has
$\rho^*(\phi, \phi';t) \equiv \langle \phi |\, \hat{\rho}(t)\, 
|\phi' \rangle^* = \langle \phi' |\, \hat{\rho}^\dagger(t)\, |\phi \rangle 
\equiv \langle \phi' |\, \hat{\rho}(t)\, |\phi \rangle
\equiv \rho(\phi',\phi;t)$.

Another property is 
\begin{equation} \label{propb}
S_{IF}[\underline{\varphi},\underline{\varphi}] =0\,\, .
\end{equation} 
For the proof, note that in $S_{IF}$, $\underline{\varphi}$ 
plays the role of an external, and thus fixed, field. Defining the action 
$S_{\varphi} [\underline{\Phi}] \equiv S[\underline{\Phi}] 
+ S_I[\underline{\varphi}, \underline{\Phi}]$ for hard 
fields in the presence of this external field $\underline{\varphi}$, 
and a corresponding time evolution operator $\hat{U}_{\varphi} (t_f,t_i)$
with matrix elements
\begin{equation} \label{Uphi}
\langle \underline{\Phi}_f | \hat{U}_{\varphi} (t_f,t_i) 
| \underline{\Phi}_i \rangle   \equiv 
\int_{\underline{\Phi}_i}^{\underline{\Phi}_f}
{\cal D} \underline{\Phi} \, \exp \left\{ i\, S_{\varphi} 
[\underline{\Phi}] \right\} \,\, ,
\end{equation}
one then reverts in eq.\ (\ref{infl}) the steps which led to 
eq.\ (\ref{densmatrix}). 
Then, $\exp \left\{ i \, S_{IF} [\underline{\varphi},\underline{\varphi}]
\right\} = \int {\rm D} \underline{\Phi}_f \, 
\rho^{(\Phi)}_{\varphi} (\underline{\Phi}_f, 
\underline{\Phi}_f;t_f) \equiv {\rm Tr}\, 
\hat{\rho}^{(\Phi)}_{\varphi}(t_f) \equiv 1$, 
consequently, $S_{IF} [\underline{\varphi},\underline{\varphi}] \equiv 0$.
Here, $\hat{\rho}^{(\Phi)}_{\varphi} (t_f)$ is the density matrix
for the hard degrees of freedom as it evolved from its
initial value $\hat{\rho}^{(\Phi)}_i$ to the final time $t_f$ subject
to the time evolution operator $\hat{U}_{\varphi} (t_f,t_i)$. 
This evolution in general differs from the one when the external field
$\underline{\varphi}$ is absent.

\section{Derivation of classical equations of motion for the soft fields}

The classical equations of motion for the soft fields $\underline{\varphi}$ 
are determined by expanding the phase
$S[\underline{\varphi}] - S[\underline{\varphi}'] + 
S_{IF} [\underline{\varphi}, \underline{\varphi}']$ in eq.\ (\ref{reducedDM}) 
around field configurations $\underline{\varphi} \equiv \underline{\varphi}'$,
i.e., where this phase vanishes according to (\ref{propb}). Let us introduce
\begin{equation}
\underline{\bar{\varphi}} \equiv \frac{1}{2} \left( \underline{\varphi} + 
\underline{\varphi}'\right) \,\,\,\,\, ,
\,\,\,\,\,\,\, \Delta \underline{\varphi} \equiv \underline{\varphi} 
- \underline{\varphi}' \,\,\, ,
\end{equation}
and expand $S[\underline{\varphi}] - S[\underline{\varphi}'] + 
S_{IF} [\underline{\varphi}, \underline{\varphi}']$ to
quadratic order in $\Delta \underline{\varphi}$:
\begin{eqnarray}
\lefteqn{S[\underline{\varphi}] - S[\underline{\varphi}'] + 
S_{IF} [\underline{\varphi}, \underline{\varphi}'] } \nonumber \\
& \simeq & \int_{t_i}^{t_f} {\rm d}^4 x \, 
\left[ \frac{\delta S[\underline{\bar{\varphi}}]}{\delta 
\underline{\bar{\varphi}} (x)}  
+ \frac{1}{2} \left( \frac{ \delta S_{IF} [\underline{\varphi},
\underline{\varphi}'] }{\delta \underline{\varphi} (x)} 
- \frac{ \delta S_{IF}[\underline{\varphi},\underline{\varphi}'] }{
\delta \underline{\varphi}' (x)} \right)_{\underline{\varphi} = 
\underline{\varphi}' = \underline{\bar{\varphi}}} \right] \cdot 
\Delta \underline{\varphi}(x) + \frac{1}{8} \int_{t_i}^{t_f} 
{\rm d}^4x\, {\rm d}^4y    \label{exp2ndorder} \\
& \times & \sum_{a,b=1}^N  \Delta \varphi_a(x) \, \left( 
\frac{ \delta^2 S_{IF}[\underline{\varphi},\underline{\varphi}'] }{
\delta \varphi_a (x) \, \delta \varphi_b (y)} -
\frac{ \delta^2 S_{IF}[\underline{\varphi},\underline{\varphi}'] }{
\delta \varphi_a (x) \, \delta \varphi_b' (y)} -
\frac{ \delta^2 S_{IF}[\underline{\varphi},\underline{\varphi}'] }{
\delta \varphi_a' (x) \, \delta \varphi_b (y)} +
\frac{ \delta^2 S_{IF}[\underline{\varphi},\underline{\varphi}'] }{
\delta \varphi_a' (x) \, \delta \varphi_b' (y)} \right)_{\underline{\varphi} 
= \underline{\varphi}' = \underline{\bar{\varphi}}} \Delta \varphi_b (y)
\,\, .  \nonumber
\end{eqnarray}
Previous derivations of classical equations of motion in 
a background of hard fields usually compute the functional
derivatives of the influence functional in eq.\ (\ref{exp2ndorder})
directly for the specific system under consideration.
However, these functional derivatives can be further evaluated also
in the general case. To this end, let us define an ``$n$--point function'':
\begin{equation} \label{Npoint}
A(\underline{\Phi}, \underline{\Phi}') 
\equiv \Phi_{a_1}^{n_1} (x_1)\, \Phi_{a_2}^{n_2}(x_2) \cdots 
\Phi_{a_k}^{n_k}(x_k)\, {\Phi'}_{b_1}^{m_1}(x_1') \, {\Phi'}_{b_2}^{m_2}(x_2')
\cdots {\Phi'}_{b_l}^{m_l}(x_l')\,\,\, , 
\,\,\,\, n= \sum_{i=1}^k n_i + \sum_{j=1}^l m_j\,\,,
\end{equation}
of the fields $\underline{\Phi}, \, \underline{\Phi}'$. Here,
$a_i$ and $b_j$ label components of these $N$--dimensional fields.
Without loss of generality one may assume that $t_1 \geq t_2 \geq \cdots$ and
$t_1'\geq t_2' \geq \cdots$. Let us then define
the average of $A\left(\underline{\Phi}, \underline{\Phi}'\right)$ 
in the presence of the ``background field '' $\underline{\bar{\varphi}}$ as:
\begin{eqnarray}
\langle A (\underline{\Phi}, \underline{\Phi}') 
\rangle_{\underline{\bar{\varphi}}} 
& \equiv & \int {\rm D} \underline{\Phi}_f \, {\rm D} \underline{\Phi}_i \, 
{\rm D} \underline{\Phi}_i' \, \rho^{(\Phi)}(\underline{\Phi}_i, 
\underline{\Phi}_i';t_i) \nonumber \\
& \times & \int_{\underline{\Phi}_i}^{\underline{\Phi}_f} 
{\cal D}\underline{\Phi} \int_{\underline{\Phi}_i'}^{\underline{\Phi}_f} 
{\cal D}\underline{\Phi}' \, \exp \left\{ i \left( S[\underline{\Phi}] 
+S_I [\underline{\bar{\varphi}},\underline{\Phi}]
 - S[\underline{\Phi}'] - S_I [\underline{\bar{\varphi}},\underline{\Phi}'] 
\right) \right\} \, A( \underline{\Phi}, \underline{\Phi}' )
\,\, . \label{average}
\end{eqnarray}
Using the time evolution operator (\ref{Uphi}) (with
$\underline{\varphi} \equiv \underline{\bar{\varphi}})$, 
one notices that this is equivalent to
\begin{equation} \label{Npointaverage}
\langle A (\underline{\Phi},\underline{\Phi}') 
\rangle_{\underline{\bar{\varphi}}} \equiv
{\rm Tr} \, \left\{ \hat{\rho}_{\bar{\varphi}}^{(\Phi)}(t_f) 
A ( \hat{\underline{\Phi}}, \hat{\underline{\Phi}}' ) \right\}\,\, ,
\end{equation}
i.e., the usual expectation value of the $n$--point function $A$ in
the (in general, non-equilibrium) ensemble characterized by the
density matrix $\hat{\rho}_{\bar{\varphi}}^{(\Phi)}(t_f)$. Note
that in this expectation value all $\underline{\Phi}$ fields are
{\em time-ordered}, while all $\underline{\Phi}'$ fields are
{\em anti-time-ordered\/}. 

Writing $S_{IF} = -i\,  \ln \, \exp \{ i\, S_{IF} \}$, one now observes 
with the definition (\ref{infl}) and the property (\ref{propb}) of the 
influence functional, and the definition of the average (\ref{average}) that
\begin{mathletters} \label{22}
\begin{eqnarray}
\left. \frac{\delta S_{IF}[\underline{\varphi},\underline{\varphi}']}{
\delta \underline{\varphi}(x)} \right|_{\underline{\varphi} = 
\underline{\varphi}' = \underline{\bar{\varphi}}} & = & \left\langle
\frac{\delta S_I[\underline{\bar{\varphi}},\underline{\Phi}]}{
\delta \underline{\bar{\varphi}}(x)} \right\rangle_{\underline{\bar{\varphi}}}
\,\,, \\
\left. \frac{\delta S_{IF}[\underline{\varphi},\underline{\varphi}']}{
\delta \underline{\varphi}'(x)} \right|_{\underline{\varphi} = 
\underline{\varphi}' = \underline{\bar{\varphi}}} & = &  - \left\langle
\frac{\delta S_I[\underline{\bar{\varphi}},\underline{\Phi}']}{
\delta \underline{\bar{\varphi}}(x)} \right\rangle_{\underline{\bar{\varphi}}}
\,\,,  \\
\left. \frac{\delta^2 S_{IF}[\underline{\varphi},\underline{\varphi}']}{
\delta \varphi_a(x) \delta \varphi_b(y)} \right|_{\underline{\varphi} = 
\underline{\varphi}' = \underline{\bar{\varphi}}} & = &  
\left\langle \frac{\delta^2 S_I[\underline{\bar{\varphi}},\underline{\Phi}]}{
\delta \bar{\varphi}_a(x) \delta \bar{\varphi}_b(y)} 
\right\rangle_{\underline{\bar{\varphi}}} \nonumber \\ 
& +  & i 
\left\langle \frac{\delta S_I[\underline{\bar{\varphi}},\underline{\Phi}]}{
\delta \bar{\varphi}_a(x)}\, 
\frac{\delta S_I[\underline{\bar{\varphi}},\underline{\Phi}]}{
\delta \bar{\varphi}_b(y)} \right\rangle_{\underline{\bar{\varphi}}} - i 
\left\langle \frac{\delta S_I[\underline{\bar{\varphi}},\underline{\Phi}]}{
\delta \bar{\varphi}_a(x)}\right\rangle_{\underline{\bar{\varphi}}} 
\left\langle \frac{\delta S_I[\underline{\bar{\varphi}},\underline{\Phi}]}{
\delta \bar{\varphi}_b(y)} \right\rangle_{\underline{\bar{\varphi}}}\,\,, 
\\
\left. \frac{\delta^2 S_{IF}[\underline{\varphi},\underline{\varphi}']}{
\delta \varphi_a(x) \delta \varphi_b'(y)} \right|_{\underline{\varphi} = 
\underline{\varphi}' = \underline{\bar{\varphi}}} & = &  - i 
\left\langle \frac{\delta S_I[\underline{\bar{\varphi}},\underline{\Phi}]}{
\delta \bar{\varphi}_a(x)}\, 
\frac{\delta S_I[\underline{\bar{\varphi}},\underline{\Phi}']}{\delta 
\bar{\varphi}_b(y)} \right\rangle_{\underline{\bar{\varphi}}} + i 
\left\langle \frac{\delta S_I[\underline{\bar{\varphi}},\underline{\Phi}]}{
\delta \bar{\varphi}_a (x)} \right\rangle_{\underline{\bar{\varphi}}} 
\left\langle \frac{\delta S_I[\underline{\bar{\varphi}},\underline{\Phi}']}{
\delta \bar{\varphi}_b(y)} \right\rangle_{\underline{\bar{\varphi}}}\,\,, 
\\
\left. \frac{\delta^2 S_{IF}[\underline{\varphi},\underline{\varphi}']}{
\delta \varphi_a'(x) \delta \varphi_b'(y)} \right|_{\underline{\varphi} = 
\underline{\varphi}' = \underline{\bar{\varphi}}} & = &  - 
\left\langle \frac{\delta^2 S_I[\underline{\bar{\varphi}},\underline{\Phi}']}{
\delta \bar{\varphi}_a(x) \delta \bar{\varphi}_b(y)} 
\right\rangle_{\underline{\bar{\varphi}}} 
\nonumber \\ 
& + &  i 
\left\langle \frac{\delta S_I[\underline{\bar{\varphi}},\underline{\Phi}']}{
\delta \bar{\varphi}_a(x)}\, 
\frac{\delta S_I[\underline{\bar{\varphi}},\underline{\Phi}']}{
\delta \bar{\varphi}_b(y)} \right\rangle_{\underline{\bar{\varphi}}} - i 
\left\langle \frac{\delta S_I[\underline{\bar{\varphi}},\underline{\Phi}']}{
\delta \bar{\varphi}_a(x)} \right\rangle_{\underline{\bar{\varphi}}} 
\left\langle \frac{\delta S_I[\underline{\bar{\varphi}},\underline{\Phi}']}{
\delta \bar{\varphi}_b(y)} \right\rangle_{\underline{\bar{\varphi}}}\,\,.
\end{eqnarray}
\end{mathletters}
Since the fields $\underline{\bar{\varphi}},\, \underline{\Phi}, \, 
\underline{\Phi}'$ are assumed to be real-valued, eq.\ (\ref{exp2ndorder})
can be further simplified using symmetry properties of the average 
(\ref{average}) with respect to exchange of $\underline{\Phi} 
\leftrightarrow \underline{\Phi}'$, for instance:
\begin{equation}
\left\langle \frac{\delta S_I[\underline{\bar{\varphi}},\underline{\Phi}']}{
\delta \underline{\bar{\varphi}}(x)} \right\rangle_{\underline{\bar{\varphi}}} 
\equiv \left\langle \frac{\delta S_I[\underline{\bar{\varphi}},
\underline{\Phi}]}{\delta \underline{\bar{\varphi}}(x)} 
\right\rangle_{\underline{\bar{\varphi}}}^*\,\, .
\end{equation}
Then, eq.\ (\ref{exp2ndorder}) becomes:
\begin{equation}
S[\underline{\varphi}] - S[\underline{\varphi}'] + 
S_{IF} [\underline{\varphi}, \underline{\varphi}'] \simeq 
\int_{t_i}^{t_f} {\rm d}^4 x \, \underline{\cal E} (x) \cdot 
\Delta \underline{\varphi} (x) 
 +   \frac{i}{2} \int_{t_i}^{t_f} 
{\rm d}^4x\, {\rm d}^4y \sum_{a,b=1}^N \Delta \varphi_a(x) \, 
{\cal I}_{ab} (x,y) \, \Delta \varphi_b (y)\,\, ,
\end{equation}
where
\begin{mathletters}
\begin{eqnarray}
\underline{\cal E} (x) & = & 
\frac{\delta S[\underline{\bar{\varphi}}]}{ \delta \underline{\bar{\varphi}} 
(x)} + \underline{\cal R} (x) \,\,, \\
\underline{\cal R} (x) & \equiv &   
{\rm Re} \, \left\langle \frac{ \delta S_{I}
[\underline{\bar{\varphi}},\underline{\Phi}] }{\delta 
\underline{\bar{\varphi}} (x)} \right\rangle_{\underline{\bar{\varphi}}}
\,\, , \label{R} \\
{\cal I}_{ab} (x,y) & \equiv & 
\frac{1}{2} \,  {\rm Im}\, \left\langle 
\frac{ \delta^2 S_{I}[\underline{\bar{\varphi}},\underline{\Phi}] }{\delta 
\bar{\varphi}_a (x) \, \delta \bar{\varphi}_b (y)} 
\right\rangle_{\underline{\bar{\varphi}}} - {\rm Re}\, \left\langle 
\frac{ \delta S_{I} [\underline{\bar{\varphi}},\underline{\Phi}] }{
\delta \bar{\varphi}_a (x)} \right\rangle_{\underline{\bar{\varphi}}}
\, {\rm Re} \, \left\langle 
\frac{ \delta S_{I} [\underline{\bar{\varphi}},\underline{\Phi}] }{
\delta \bar{\varphi}_b (y)} \right\rangle_{\underline{\bar{\varphi}}}
\nonumber \\
& +  & \frac{1}{4} \left\langle \left(
\frac{ \delta S_{I}[\underline{\bar{\varphi}},\underline{\Phi}] }{
\delta \bar{\varphi}_a (x)} + 
\frac{ \delta S_{I}[\underline{\bar{\varphi}},\underline{\Phi}'] }{
\delta \bar{\varphi}_a (x)} \right) \left(
\frac{ \delta S_{I}[\underline{\bar{\varphi}},\underline{\Phi}] }{
\delta \bar{\varphi}_b (y) } + 
\frac{ \delta S_{I}[\underline{\bar{\varphi}},\underline{\Phi}'] }{
\delta \bar{\varphi}_b (y)} \right) \right\rangle_{\underline{\bar{\varphi}}}
\label{Iofphi}
\end{eqnarray}
\end{mathletters}
are real-valued functions. Inserting this into expression
(\ref{reducedDM}) and changing the integration variables
$\underline{\varphi},\, \underline{\varphi}'$ to $\underline{\bar{\varphi}},\,
\Delta \underline{\varphi}$, one obtains
\begin{eqnarray}
\lefteqn{\rho^{(\varphi)}(\underline{\bar{\varphi}}_f, 
\Delta \underline{\varphi}_f;t_f) \simeq 
\int {\rm D} \underline{\bar{\varphi}}_i \, {\rm D} \Delta 
\underline{\varphi}_i\, \rho^{(\varphi)}(\underline{\bar{\varphi}}_i,
\Delta \underline{\varphi}_i;t_i) \int_{\underline{\bar{\varphi}}_i}^{
\underline{\bar{\varphi}}_f} {\cal D} \underline{\bar{\varphi}}
\int_{\Delta \underline{\varphi}_i}^{\Delta \underline{\varphi}_f} 
{\cal D}\Delta \underline{\varphi} } \nonumber \\
&   \times &  \exp \left\{ i \int_{t_i}^{t_f} {\rm d}^4x
\, \underline{\cal E}(x) \cdot \Delta \underline{\varphi}(x) - \frac{1}{2}
\int_{t_i}^{t_f} {\rm d}^4x\, {\rm d}^4y  
\sum_{a,b=1}^N \Delta \varphi_a (x)\, {\cal I}_{ab}(x,y) \, 
\Delta \varphi_b (y) \right\} \,\, . \label{reducedDM2}
\end{eqnarray}
The term quadratic in $\Delta \underline{\varphi}$
in the argument of the exponential can be rewritten introducing
auxiliary fields $\underline{\xi}= (\xi_1, \ldots, \xi_N)$,
\begin{equation}
\exp \left\{ - \frac{1}{2}
\int_{t_i}^{t_f} \!\! {\rm d}^4x\, {\rm d}^4y \sum_{a,b=1}^N 
\Delta \varphi_a (x)\,
{\cal I}_{ab}(x,y)\, \Delta \varphi_b (y)
\right\} \equiv  \int {\cal D} \underline{\xi} \, P[\underline{\xi}, 
\underline{\bar{\varphi}}] \, \exp \left\{ i \int_{t_i}^{t_f} \!\!
{\rm d}^4 x \, \underline{\xi}(x) \cdot \Delta \underline{\varphi} (x)
\right\}\,\,\, , \label{noise}
\end{equation}
where
\begin{equation} \label{probmeas}
P[\underline{\xi},\underline{\bar{\varphi}}] 
\equiv {\cal N}[\underline{\bar{\varphi}}] \, \exp \, 
\left\{ -\frac{1}{2} \int_{t_i}^{t_f} {\rm d}^4x\, {\rm d}^4y 
\sum_{a,b=1}^N \xi_a (x)\,
{\cal I}^{-1}_{ab}(x,y)\, \xi_b (y) \right\}
\end{equation}
is a normalized Gaussian measure. Note that
$P$ in general depends on $\underline{\bar{\varphi}}$ through ${\cal I}$. 
It is now possible to 
perform the functional integration over $\Delta \underline{\varphi}$ 
(except for the one at $t_i$) in expression (\ref{reducedDM2}),
with the result
\begin{equation}
\rho^{(\varphi)}(\underline{\bar{\varphi}}_f, \Delta \underline{\varphi}_f;t_f)
  \simeq  \int
{\cal D} \underline{\xi} \int {\rm D} \underline{\bar{\varphi}}_i \, 
{\rm D}\Delta \underline{\varphi}_i \,
\rho^{(\varphi)}(\underline{\bar{\varphi}}_i,\Delta \underline{\varphi}_i;t_i)
 \int_{\underline{\bar{\varphi}}_i}^{\underline{\bar{\varphi}}_f} 
{\cal D} \underline{\bar{\varphi}} \,\, \, P[\underline{\xi},
\underline{\bar{\varphi}}] \,\,\, \delta \left[ \underline{\cal E}(x) 
+ \underline{\xi} (x) \right]  \,\, .
\end{equation}
This result means that, for times $t_i < t < t_f$, the 
functional $\delta$ function forces the fields $\underline{\bar{\varphi}}$ 
to obey the {\em classical equations of motion\/}
\begin{equation} \label{cleom}
- \underline{\cal E}(x) \equiv - \frac{\delta 
S[\underline{\bar{\varphi}}]}{\delta \underline{\bar{\varphi}} (x)}  - 
\underline{\cal R}(x) = \underline{\xi} (x) \,\, .
\end{equation}
The condition $-\delta S[\underline{\bar{\varphi}}]/\delta 
\underline{\bar{\varphi}} (x) = 0$ is the usual classical
equation of motion. The new term $\underline{\cal R}$ characterizes 
the interactions of the soft, classical fields 
$\underline{\bar{\varphi}}$ with the hard, unobserved degrees of freedom.
As it will become clear in the following, a part of these interactions
is to be interpreted as dissipation. The associated
fluctuating noise field is represented by $\underline{\xi}$ 
on the right-hand side of the equations of motion (\ref{cleom}). 
These equations of motion are therefore Langevin--type equations.
Note that in general the noise is not white, since ${\cal I}$ need
not be proportional to $\delta (x_0 - y_0)$, and can be multiplicative,
due to the dependence of ${\cal I}$ on $\underline{\bar{\varphi}}$
\cite{cgbm,gleiser}. It should be mentioned at this point that
multiplicative noise terms are treated slightly differently in refs.\
\cite{cgbm,gleiser} than in the present work. While here there is only
one noise field with a $\underline{\bar{\varphi}}$--dependent
variance, in \cite{cgbm,gleiser} different noise fields are introduced with
$\underline{\bar{\varphi}}$--independent variances.

Another interesting aspect for systems with more than one field
degree of freedom is that, since ${\cal I}$ is in general not diagonal, 
${\cal I}_{ab}(x,y) \neq 0$ for $a \neq b$, there can be correlations 
between the noise terms in the equations of motion for two different field 
components $\bar{\varphi}_a$ and $\bar{\varphi}_b$.
In the following, this general formalism will be applied to the
$O(N)$ model, both in the symmetric case as well as with spontaneously
broken symmetry.

\section{The $O(N)$ model in the symmetric case}

The Lagrangian of the $O(N)$ model is:
\begin{equation} \label{ONlagrangian}
{\cal L} (\underline{\phi}) = \frac{1}{2} \left( \partial_\mu \,
\underline{\phi} \cdot \partial^\mu \, \underline{\phi} - m^2
\underline{\phi} \cdot \underline{\phi} \right) - \frac{\lambda}{N}
\left( \underline{\phi} \cdot \underline{\phi} \right)^2\,\, ,
\end{equation}
where $\underline{\phi} = (\phi_1,\, \phi_2, \ldots, \phi_N)$ is an
$N$--dimensional vector of real-valued scalar fields.

In the symmetric case, $m^2 >0$. The vacuum state of the model is at
the minimum of the potential 
\begin{equation} \label{U}
U(\underline{\phi}) \equiv \frac{m^2}{2}\, 
\underline{\phi} \cdot \underline{\phi} + \frac{\lambda}{N} \left(
\underline{\phi} \cdot \underline{\phi} \right)^2 \,\, ,
\end{equation}
which is $\underline{\phi}^{\rm vac} = 0$. Decomposing the field
$\underline{\phi}$ according to eqs.\
(\ref{soft},\ref{hard}), the action corresponding to interactions
between soft and hard fields reads:
\begin{eqnarray}
S_I[\underline{\varphi}, \underline{\Phi}] & = & 
- \frac{2\,\lambda}{N} \int_{t_i}^{t_f} {\rm d}^4x
\left[ 2\, \underline{\varphi}(x) \cdot \underline{\varphi}(x) \,\,
\underline{\varphi}(x) \cdot \underline{\Phi}(x) + 
\underline{\varphi}(x) \cdot \underline{\varphi}(x) \,\,
\underline{\Phi}(x) \cdot \underline{\Phi}(x) + 2
\left[\underline{\varphi}(x) \cdot \underline{\Phi}(x) \right]^2 
\right.\nonumber \\
&  & \hspace*{2.3cm}
 +\left. 2\, \underline{\varphi}(x) \cdot \underline{\Phi}(x)\,\,
\underline{\Phi}(x) \cdot \underline{\Phi}(x) \right]\,\,.
\end{eqnarray}
Consequently, 
\begin{mathletters} \label{48}
\begin{eqnarray}
\left\langle \frac{ \delta S_{I}[\bar{\underline{\varphi}},\underline{\Phi}] 
}{\delta \bar{\varphi}_a (x)} \right\rangle_{\bar{\underline{\varphi}}}  
& = & - \frac{4\,\lambda}{N} \left( 2 \, \bar{\varphi}_a(x)\,\,
\bar{\underline{\varphi}}(x) \cdot \left\langle \underline{\Phi} (x) 
\right\rangle_{\bar{\underline{\varphi}}} 
+ \bar{\varphi}_a(x) \,
\left\langle \underline{\Phi}(x) \cdot \underline{\Phi} (x) 
\right\rangle_{\bar{\underline{\varphi}}} + \bar{\underline{\varphi}}(x) \cdot
\bar{\underline{\varphi}}(x) \, \left\langle \Phi_a(x) 
\right\rangle_{\bar{\underline{\varphi}}} \right. \nonumber \\
&  & \hspace*{1cm} + \left. 2\, \bar{\underline{\varphi}}(x) \cdot 
\left\langle \underline{\Phi} (x) \, \Phi_a(x) 
\right\rangle_{\bar{\underline{\varphi}}} 
+ \left\langle \underline{\Phi}(x) \cdot \underline{\Phi}(x) \,\,
 \Phi_a(x) \right\rangle_{\bar{\underline{\varphi}}} \right)\,\, ,
\label{48a} \\
\left\langle \frac{ \delta^2 S_{I}[\bar{\underline{\varphi}},\underline{\Phi}] 
}{\delta \bar{\varphi}_a (x) \delta \bar{\varphi}_b (y)
} \right\rangle_{\bar{\underline{\varphi}}}  & = &
- \frac{4\,\lambda}{N} \left( 2 \,\delta_{ab}\,\,
\bar{\underline{\varphi}}(x) \cdot \left\langle \underline{\Phi} (x) 
\right\rangle_{\bar{\underline{\varphi}}} 
+ \delta_{ab} \, \left\langle \underline{\Phi}(x) \cdot \underline{\Phi} (x) 
\right\rangle_{\bar{\underline{\varphi}}} + 
2\, \bar{\varphi}_a(x) 
\left\langle \Phi_b(x) \right\rangle_{\bar{\underline{\varphi}}}
\right. \nonumber \\
&  & \hspace*{1cm} + \left. 2\, \bar{\varphi}_b(x) 
\left\langle \Phi_a(x) \right\rangle_{\bar{\underline{\varphi}}}
+ 2\, \left\langle \Phi_a (x) \, \Phi_b(x) 
\right\rangle_{\bar{\underline{\varphi}}} \right)\, \delta^{(4)}(x-y) \,\, .
\label{48b}
\end{eqnarray}
\end{mathletters}
In the following, the averages on the right-hand side 
will be computed perturbatively in the coupling constant $\lambda$.
The expansion in powers of $\lambda$ of the exponential
in the integrand of eq.\ (\ref{average}) reads:
\begin{eqnarray}
\lefteqn{\exp \left\{ i \left( S[\underline{\Phi}] + 
S_I[\underline{\bar{\varphi}},\underline{\Phi}]
-S[\underline{\Phi}'] - S_I[\underline{\bar{\varphi}},\underline{\Phi}']
\right) \right\} \simeq 
\exp \left\{ i \left( S_0[\underline{\Phi}] - S_0[\underline{\Phi}'] 
\right) \right\}} \nonumber \\
& \times & \left[ 1 - i\,\frac{\lambda}{N} \int_{t_i}^{t_f} {\rm d^4}y 
\left( 4\, \underline{\bar{\varphi}}(y) \cdot \underline{\bar{\varphi}}(y) \,\,
\underline{\bar{\varphi}}(y) \cdot \left[\underline{\Phi}(y) -
\underline{\Phi}'(y) \right]  + 2\, 
\underline{\bar{\varphi}}(y) \cdot \underline{\bar{\varphi}}(y) \, \left[
\underline{\Phi}(y) \cdot \underline{\Phi}(y) -
\underline{\Phi}'(y) \cdot \underline{\Phi}'(y) \right] \right. \right.
\nonumber \\
&  &  \hspace*{2cm}  + \, 4\,
\left[ \underline{\bar{\varphi}}(y) \cdot \underline{\Phi}(y) \right]^2 
- 4\, \left[ \underline{\bar{\varphi}}(y) \cdot \underline{\Phi}'(y) \right]^2 
+ 4\, \underline{\bar{\varphi}}(y) \cdot \left[
\underline{\Phi}(y)\,\, \underline{\Phi}(y) \cdot \underline{\Phi}(y) 
- \underline{\Phi}'(y)\,\, \underline{\Phi}'(y) \cdot \underline{\Phi}'(y)
\right] \nonumber \\
&   &\hspace*{2cm} +\left. \left. \left[ \underline{\Phi}(y)
\cdot \underline{\Phi}(y) \right]^2
 - \left[ \underline{\Phi}'(y) \cdot \underline{\Phi}'(y) \, \right]^2
\right) + O(\lambda^2) \right] , \label{expansion}
\end{eqnarray}
where $S_0[\underline{\Phi}]$ is the action for non-interacting hard fields.
The quartic self-interaction term of the hard fields has been
included in the perturbative treatment of $S_I$.
Let us define an average $\langle \,\, \cdot \,\, \rangle_0$ in analogy to 
(\ref{average}), where
$S[\underline{\Phi}] + S_I[\underline{\bar{\varphi}},\underline{\Phi}]- 
S[\underline{\Phi}'] - S_I[\underline{\bar{\varphi}},\underline{\Phi}']$ 
is replaced by $S_0[\underline{\Phi}]- S_0[\underline{\Phi}']$. If we
additionally assume the initial density matrix $\hat{\rho}^{(\Phi)}_i$ 
to be of the form $\hat{\rho}^{(\Phi)}_0 \equiv 
\exp \{ - \hat{H}_0/T\}/{\cal Z}$, this average is then 
the usual thermal average in a non-interacting system at temperature $T$.
Since this average involves a Gaussian measure in function space, 
the average (\ref{average}) of an arbitrary $n$--point function
$A(\underline{\Phi},\underline{\Phi}')$ vanishes for odd $n$ and
can be decomposed into a sum over products of 2--point functions for even 
$n$. These 2--point functions are \cite{LeBellac}:
\begin{mathletters}
\begin{eqnarray}
\langle \Phi_a(x) \, \Phi_b(y) \rangle_0 & \equiv & {\rm Tr}\, \left\{ 
\hat{\rho}^{(\Phi)}_0\, {\rm T} \left( \hat{\Phi}_a(x) \, \hat{\Phi}_b(y) 
\right) \right\} \equiv \delta_{ab}\, D_{++}(x-y) \nonumber \\
& \equiv & \delta_{ab} \left[ D_{>}(x-y)\, \Theta(x_0 - y_0)
+ D_{<}(x-y) \, \Theta(y_0-x_0) \right] \,\, , \label{G++} \\
\langle \Phi_a(x) \, \Phi_b'(y) \rangle_0 & \equiv & {\rm Tr}\, \left\{
\hat{\rho}^{(\Phi)}_0 \, \hat{\Phi}_a(x) \, \hat{\Phi}_b'(y) \right\}
\equiv  \delta_{ab}\, D_<(x-y)  \,\, ,\\
\langle \Phi_a'(x) \, \Phi_b(y) \rangle_0  & \equiv & {\rm Tr}\, \left\{
\hat{\rho}^{(\Phi)}_0 \, \hat{\Phi}_a'(x) \, \hat{\Phi}_b(y) \right\}
\equiv  \delta_{ab}\, D_>(x-y)  \,\, ,\\
\langle \Phi_a'(x) \, \Phi_b'(y) \rangle_0 & \equiv & {\rm Tr}\, \left\{
\hat{\rho}^{(\Phi)}_0 \, \tilde{\rm T} \left(
\hat{\Phi}_a'(x) \, \hat{\Phi}_b'(y) \right) \right\}
\equiv \delta_{ab}\, D_{--}(x-y) \nonumber \\
& \equiv & \delta_{ab} \left[ D_{<}(x-y)\, \Theta(x_0 - y_0)
+ D_{>}(x-y) \, \Theta(y_0-x_0) \right] \,\, ,
\end{eqnarray}
\end{mathletters}
where T stands for time ordering, $\tilde{\rm T}$ for anti-time ordering,
and where translational invariance in space--time has been assumed.
The functions $D_>$ and $D_<$ have the Fourier representation
\begin{equation} \label{Gi}
D_{i}(t, {\bf x}) = \int \frac{{\rm d}^3 {\bf k}}{(2\pi)^3} \, \Theta
(|{\bf k}|-k_c) \, e^{i \, {\bf k} \cdot {\bf x}} \, D_{i} 
(t, {\bf k}) \,\, ,  \,\,\, i \,= \,\, > \,\,{\rm or} \,\, < \,\,\, ,
\end{equation}
where
\begin{equation} \label{Glarger}
D_>(t, {\bf k}) \equiv \frac{1}{2E_{\bf k}} \left\{ \left[1 + n(E_{\bf k})
\right]\,e^{-i\, E_{\bf k}\, t} + n(E_{\bf k}) \,e^{i\, E_{\bf k}\, t}
\right\}\,\,\, , \,\,\,\,\, D_<(t, {\bf k}) \equiv D_>(-t, {\bf k}) \,\,.
\end{equation}
Here, $n(x) \equiv (e^{x/T}-1)^{-1}$ is the Bose--Einstein distribution
function, and $E_{\bf k} \equiv ({\bf k}^2 + m^2)^{1/2}$.

\subsection{Interactions between soft and hard fields 
to first order in $\lambda$}

\begin{figure}
\begin{center}
\epsfxsize=4cm
\epsfysize=3cm
\leavevmode
\hbox{ \epsffile{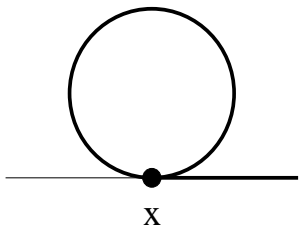}}
\end{center}
\caption{The order $\lambda$ contribution to the thermal mass.}
\label{Figmass1}
\end{figure}

To determine the interaction term $\underline{\cal R}$, eq.\ (\ref{R}), 
to order $\lambda$, 
one needs to compute the expectation values on the right-hand side of eq.\
(\ref{48a}) only to order $\lambda^0 = 1$, since there is already
an overall factor of $\lambda$ on the right-hand side. This means that
the averages $\langle \,\, \cdot \,\, \rangle_{\underline{\bar{\varphi}}}$ 
on the right-hand side can be replaced by $\langle \,\, \cdot \,\, \rangle_0$. 
This simplification can be traced back to the fact that the
functional derivatives of the influence functional in eq.\ (\ref{exp2ndorder})
were rewritten in terms of averages over functional derivatives of 
$S_I$, cf.\ eq.\ (\ref{22}), which itself is already of order $\lambda$.
Evaluation of the averages therefore yields
\begin{mathletters}
\begin{eqnarray}
\langle \Phi_a(x) \rangle_{\underline{\bar{\varphi}}}
& \rightarrow & \langle \Phi_a(x) \rangle_0 =0\,\, , \\
\langle \Phi_a(x)\, \Phi_b(x) \rangle_{\underline{\bar{\varphi}}}
& \rightarrow & \langle \Phi_a(x)\, \Phi_b(x) \rangle_0 = 
\delta_{ab} \, D_{++}(0) \,\, , \\
\langle \Phi_a(x)\, \Phi_b^2(x) \rangle_{\underline{\bar{\varphi}}}
& \rightarrow & \langle \Phi_a(x)\, \Phi_b^2(x) \rangle_0 = 0 \,\, .
\end{eqnarray}
\end{mathletters}
Thus,
\begin{equation}
{\cal R}_a(x) = - \frac{4\,  (N+2)\,  \lambda}{N}\, {\rm Re}\, D_{++}(0)\,
\bar{\varphi}_a(x) \,\, .
\end{equation}
With eqs.\ (\ref{G++}), (\ref{Gi}), and (\ref{Glarger})
one derives:
\begin{equation} \label{thermmass}
D_{++}(0) = \int \frac{{\rm d}^3 {\bf k}}{(2 \pi)^3}\, 
\Theta(|{\bf k}|-k_c) \, \frac{1}{2E_{\bf k}} \,
\left[ 1 + 2\, n(E_{\bf k}) \right]  \,\,.
\end{equation}
This expression diverges, due to the vacuum contribution 
corresponding to the 1 in brackets.
The divergence can be removed in the standard way,
for instance by introducing an appropriate counterterm in the Lagrangian.
In the following, it is implied that, wherever necessary, 
such divergences have been removed accordingly.
Note that for massless particles and in the limit $k_c \rightarrow 0$:
\begin{equation} \label{HTL}
\int \frac{{\rm d}^3 {\bf k}}{(2 \pi)^3}\, \frac{1}{E_{\bf k}} \,
n(E_{\bf k}) = \frac{T^2}{12}\,\, .
\end{equation}
Since the expression (\ref{thermmass}) is real-valued,
one obtains for the left-hand side of the classical
equation of motion for the field component $\bar{\varphi}_a$:
\begin{equation} \label{eomforphia}
{\cal E}_a(x) = - \left[ \Box + m^2(T) + \frac{4\,
\lambda}{N} \, \underline{\bar{\varphi}}(x) \cdot 
\underline{\bar{\varphi}}(x) \right]\,\bar{\varphi}_a (x) \,\,.
\end{equation}
Here,
\begin{equation}
m^2(T) = m^2  + \frac{4\,  (N+2)\,  \lambda}{N}\, D_{++}(0) 
\end{equation} 
is the usual thermal mass (squared) to first order in $\lambda$. The term
$[m^2(T) - m^2] \bar{\varphi}_a(x)$ can be graphically depicted as in 
Fig.\ \ref{Figmass1}. In this and
the following graphs, external legs correspond to classical fields
$\bar{\varphi}_a$, while internal lines correspond to 
propagators $D_{++}$ of hard modes.
The thin external leg on the left side of the vertex 
has no correspondence in the classical 
equation of motion, it corresponds to a factor
$\Delta \varphi_a (x)$ multiplying ${\cal E}_a(x)$ in the argument 
of the exponential function in eq.\ (\ref{reducedDM2}). It is included
here to show that the interaction in principle involves four particles.

The result (\ref{eomforphia}) 
means that to first order in $\lambda$, the only effect of the presence 
of the hard modes (the heat bath) is a modification of the mass term
in the classical equation of motion for the soft fields.
There is obviously no dissipation to this order in $\lambda$. 
Consequently, the fluctuating field $\xi_a$ has to vanish, too. This is 
verified by explicitly computing ${\cal I}_{ab}(x,y)$. One first notes that 
the last two terms in eq.\ (\ref{Iofphi}) are of order $\lambda^2$
and thus can be neglected to first order in $\lambda$. With
eq.\ (\ref{48b}), the remaining term yields:
\begin{equation}
{\cal I}_{ab}(x, y) = - \delta_{ab}\, \frac{4\,(N+2)\, \lambda }{2\,N}\,
{\rm Im} \, D_{++}(0) \, \delta^{(4)}(x-y) \,\, ,
\end{equation}
which obviously vanishes, since the expression (\ref{thermmass}) is
real-valued.
Therefore, to first order in $\lambda$, the introduction
of a fluctuating noise field $\underline{\xi}$ via (\ref{noise}) is obsolete.
This also implies that, to this lowest order in $\lambda$,
the classical field $\underline{\bar{\varphi}}(x)$ 
{\em has no means to equilibrate with the background of hard modes}.
One has to go to the next higher order in $\lambda$ to achieve this.

\subsection{Interactions between soft and hard fields
to second order in $\lambda$}

\begin{figure}
\begin{center}
\epsfxsize=8cm
\epsfysize=9cm
\leavevmode
\hbox{ \epsffile{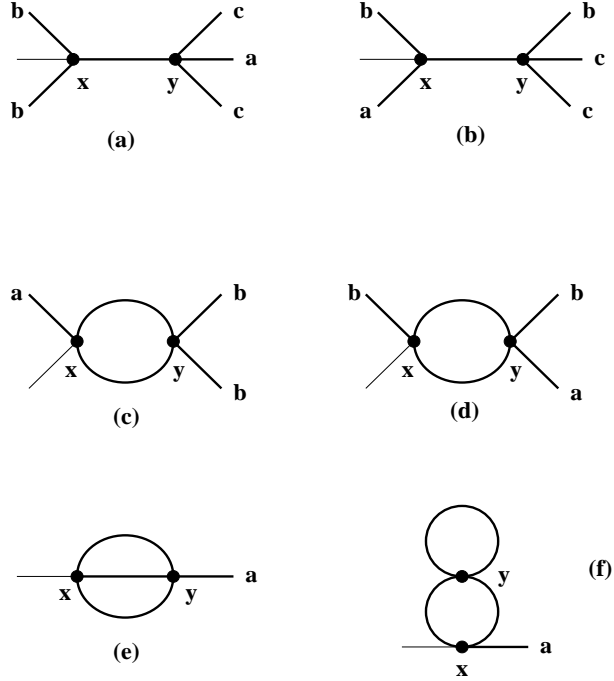}}
\end{center}
\caption{(a,b) The two terms in ${\cal T}_a^{(1)}$. A sum over 
$b,\, c$ is implied. (c,d) The two terms in ${\cal T}_a^{(2)}$. 
A sum over $b$ is implied. (e) The term ${\cal T}_a^{(3)}$. (f)
The order $\lambda^2$ contribution to the thermal mass.}
\label{Figmass2}
\end{figure}

To second order in $\lambda$, the expectation values on the
right-hand side of (\ref{48}) have to be evaluated to order $\lambda$.
The result (neglecting terms which vanish on account of momentum
conservation) is:
\begin{mathletters} \label{52}
\begin{eqnarray}
\langle \Phi_a(x) \rangle_{\underline{\bar{\varphi}}} 
& = & -i\, \frac{4\, \lambda}{N}
\int_{t_i}^{x_0} {\rm d}^4y \,\left[ D_>(x-y) -
D_<(x-y)\right]\, \underline{\bar{\varphi}}(y) \cdot
\underline{\bar{\varphi}}(y) \,\, \bar{\varphi}_a(y)\,\, , \\
\langle \Phi_a(x)\, \Phi_b(x) \rangle_{\underline{\bar{\varphi}}} 
& = &  \delta_{ab} \,D_{++}(0) 
 - i\, \frac{4\, \lambda}{N} \int_{t_i}^{x_0} {\rm d}^4y \, 
\left[ D_>^2(x-y) - D_<^2(x-y) \right]
\nonumber \\
&    &  \hspace*{2cm} \times
\left[\frac{}{} \delta_{ab}\,\, \underline{\bar{\varphi}}(y) \cdot
\underline{\bar{\varphi}}(y) + 2\, \bar{\varphi}_a(y)\,
\bar{\varphi}_b(y) + \delta_{ab} \, (N+2)\, D_{++}(0) \right]\,\, , \\
\langle \Phi_a(x)\, \Phi_b^2(x) \rangle_{\underline{\bar{\varphi}}} 
& = & -i\, \frac{8\,\lambda}{N}\, (1 + 2\, \delta_{ab})
\int_{t_i}^{x_0} {\rm d}^4y \, \left[ D_>^3(x-y) 
- D_<^3(x-y) \right] \, \bar{\varphi}_a(y)\,\,. \label{47c}
\end{eqnarray}
\end{mathletters}
In eq.\ (\ref{47c}) a term was omitted which vanishes on account of
momentum conservation in the expression $\int {\rm d}^4x\,
\underline{\cal E}(x) \cdot \Delta \underline{\varphi}(x)$.
The left-hand side of the equation of motion (\ref{cleom}) is then:
\begin{equation} \label{53}
{\cal E}_a(x) = - \left[  \Box + m^2(T) 
+ \frac{4\,\lambda}{N}\, \underline{\bar{\varphi}} (x) \cdot
\underline{\bar{\varphi}} (x) \right]
\, \bar{\varphi}_a(x) + \left(\frac{4\, \lambda}{N} \right)^2 \,
\sum_{i=1}^3  {\cal T}_a^{(i)}(x)\,\, ,
\end{equation}
where
\begin{mathletters} \label{dissipation}
\begin{eqnarray}
{\cal T}_a^{(1)} (x) & \equiv & i \int_{t_i}^{x_0} 
{\rm d}^4y  \left[ D_>(x-y) - D_<(x-y) \right]\, 
\left[ \underline{\bar{\varphi}}(x) \cdot \underline{\bar{\varphi}}(x)\,\, 
\bar{\varphi}_a(y) + 2\, \bar{\varphi}_a(x)\,\, 
\underline{\bar{\varphi}}(x) \cdot \underline{\bar{\varphi}}(y) \right]\, 
\underline{\bar{\varphi}}(y) \cdot \underline{\bar{\varphi}}(y)\,\, , 
\label{dissipation1} \\
{\cal T}_a^{(2)} (x) & \equiv & i \int_{t_i}^{x_0} 
{\rm d}^4y  \left[ D_>^2(x-y)  - D_<^2(x-y) \right]\,  
\left[ (N+4)\, \bar{\varphi}_a(x)\,\,
\underline{\bar{\varphi}}(y) \cdot \underline{\bar{\varphi}}(y) 
+ 4\, \underline{\bar{\varphi}}(x) \cdot \underline{\bar{\varphi}}(y)\,\,
\bar{\varphi}_a(y)\right] \,\, , \label{dissipation2} \\
{\cal T}_a^{(3)} (x) & \equiv & i \int_{t_i}^{x_0} 
{\rm d}^4y  \left[D_>^3(x-y)- D_<^3(x-y)\right] 
\,2 \, (N+2)  \, \bar{\varphi}_a (y)  \,\, , \label{dissipation3}
\end{eqnarray}
\end{mathletters}
and where the thermal mass $m(T)$ is given by:
\begin{eqnarray}
m^2(T) & = & m^2 + 
\frac{4\,(N+2)\, \lambda}{N}\, \left( 1 - 
 i\, \frac{4\,(N+2)\, \lambda}{N} 
\int_{t_i}^{x_0} {\rm d}^4y \, \left[D_>^2(x-y)  - D_<^2(x-y) \right]
 \,\right) D_{++}(0) \,\,. \label{54}
\end{eqnarray}
In eqs.\ (\ref{53}) -- (\ref{54}), use has been made of the fact that
$D_<^*(x-y) \equiv D_>(x-y)$, and thus $D_>^n(x-y) - D_<^n(x-y) \equiv
2i \,{\rm Im}\, D_>^n(x-y)$, such that all expressions in these
equations are real-valued. For $N=1$ and $\lambda \rightarrow g^2/4!$,
eqs.\ (\ref{52}) -- (\ref{54}) reduce to the corresponding
ones found in \cite{cgbm} in $\phi^4$ theory.
The terms ${\cal T}_a^{(i)}$ and the order $\lambda^2$ 
thermal mass correction are graphically depicted in Fig.\ \ref{Figmass2}. 
Since $-i [ D_>^n(x-y) - D_<^n(x-y) ] = 2\, {\rm Im}\, D_>^n(x-y)
\equiv 2\, {\rm Im}\, D_{++}^n(x-y)$ for $x_0 \geq y_0$, in order to make
this graphical correspondence it is implied
that one has to take the imaginary part of any combination of propagators 
linking the space-time points $x$ and $y$ in the diagrams
of Fig.\ \ref{Figmass2} (as well as in the following figures). 
Since all external legs 
correspond to real-valued classical fields, however, this is equivalent 
to taking the imaginary part of the {\em whole\/} diagram. Note here
the advantage of working with the functions $D$ instead of $G \equiv i\,D$ 
\cite{cgbm,gleiser}, which would require taking both real and imaginary
parts of the diagrams, depending on the number of internal lines.

The two graphs (a,b) correspond to the two terms in the expression 
(\ref{dissipation1}). A sum over the external legs $b$ and $c$ is implied. 
Graphs (c,d) correspond to the two terms in the integrand in 
(\ref{dissipation2}). 
A sum over the external legs $b$ is implied. Figure \ref{Figmass2} (e) 
corresponds to expression (\ref{dissipation3}). 
Finally, Fig.\ \ref{Figmass2} (f) 
is the second order contribution to the thermal mass 
(times $\bar{\varphi}_a(x)$).

In order to study the time evolution of the classical momentum mode 
functions $\bar{\varphi}_a(t,{\bf k})$, 
eq.\ (\ref{53}) is Fourier transformed with respect to ${\bf x}$
to obtain the classical equation of motion for these modes:
\begin{eqnarray} 
\lefteqn{ 
\left[ \partial_t^2 + (E_{\bf k}^*)^2 \right] \bar{\varphi}_a(t, {\bf k})
- \left( \frac{4\, \lambda}{N} \right)^2 \, \sum_{i=1}^3 {\cal T}_a^{(i)}
(t,{\bf k}) } \nonumber \\
& + & \frac{4\, \lambda}{N}\!
\int \frac{{\rm d}^3 {\bf p} {\rm d}^3 {\bf q}}{(2\pi)^6}
\Theta (k_c - |{\bf p}|) \Theta(k_c-|{\bf q}|)
\Theta (k_c - |{\bf k}-{\bf p}-{\bf q}|)\,
\underline{\bar{\varphi}}(t,{\bf p}) \cdot 
\underline{\bar{\varphi}}(t,{\bf q})\,
\bar{\varphi}_a(t,{\bf k}-{\bf p}-{\bf q}) = \xi_a (t,{\bf k}) 
 . \label{eqofmo}
\end{eqnarray}
Here $E_{\bf k}^* \equiv [{\bf k}^2 + m^2(T)]^{1/2}$. 
In order to see how dissipation enters in the equation of motion 
(\ref{eqofmo}),
one has to compute the three terms ${\cal T}_a^{(i)}(t,{\bf k}),\,i=1,2,3$. 
This calculation is rather similar to the one presented in \cite{cgbm} 
for $\phi^4$ theory. The details are
therefore referred to Appendix A and only the main results are outlined here.
From eq.\ (\ref{dissipation}) it is obvious that the 
time evolution of the classical fields prior to time $t$ influences
the value of the field at time $t$. The equation of motion
(\ref{eqofmo}) is therefore 
a so-called {\em delay-differential equation\/}. While 
such integro-differential equations can in principle be solved, it is
much simpler to make an Ansatz for the time evolution of the fields
$\underline{\bar{\varphi}}(y)$ in
the terms (\ref{dissipation}), which renders the equation of motion
local in time. Such an Ansatz is the so-called linear harmonic
approximation \cite{cgbm}, cf.\ Appendix A. In particular,
for the term ${\cal T}_a^{(3)}(t,{\bf k})$ one then obtains
\begin{equation}
{\cal T}_a^{(3)}(t,{\bf k}) = - 2\, (N+2)\, \Theta(k_c - |{\bf k}|) \left[
\,  {\rm P}\int \frac{{\rm d}\omega}{2 \pi}\,
\frac{{\cal M}_3 (\omega, {\bf k})}{E_{\bf k}-\omega} 
\, \bar{\varphi}_a(t, {\bf k}) +  
\frac{{\cal M}_3 (E_{\bf k}, {\bf k})}{2\, E_{\bf k}} \,
\partial_t\, {\bar{\varphi}}_a (t, {\bf k}) \right]\,\,, \label{massdamp}
\end{equation}
where ${\rm P}$ stands for the principal value and
${\cal M}_3(\omega,{\bf k})$ is the Fourier transform
of ${\cal M}_3(x) \equiv D_>^3(x) - D_<^3(x)$, cf.\ eq.\ (\ref{M3}).
The first term represents a thermal mass correction for the field 
$\bar{\varphi}_a(t,{\bf k})$, while the second yields
a damping term $ + \eta({\bf k}) \, 
\partial_t{\bar{\varphi}}_a(t,{\bf k})$
in the equation of motion (\ref{eqofmo}), where the damping coefficient is:
\begin{equation}
\eta ({\bf k}) \equiv \left( \frac{4\, \lambda}{N} \right)^2
2\,(N+2)\, \frac{{\cal M}_3(E_{\bf k}, {\bf k})}{2\, E_{\bf k}} \,\, .
\end{equation}
For $N=1$ and $\lambda \rightarrow g^2/4!$, 
this agrees with the result (52) of \cite{cgbm}, noting that
their function $i {\cal M}^{(c)} \equiv g^4 {\cal M}_3/6$
(cf.\ eq.\ (41) of \cite{cgbm}).
For ${\bf k}=0$ and $k_c \rightarrow 0$, 
the evaluation of ${\cal M}_3$ simplifies (cf.\ Appendix B) and yields
\begin{equation} \label{disscoeff}
\eta \equiv \eta ({\bf 0}) = \left( \frac{4\,\lambda}{N} \right)^2
\frac{3(N+2) \,T^2}{32\pi^3\, m}\, {\rm Li}_2 \left(
e^{-m/T} \right)\,\, ,
\end{equation}
where ${\rm Li}_2(x)$ is the dilogarithm (or Spence's integral,
cf.\ eq.\ (\ref{spence})). For $\phi^4$ theory, i.e., $N=1$ and
$\lambda \rightarrow g^2/4!$ the damping coefficient is 
\begin{equation}
\eta^{\phi^4} = \frac{g^4\, T^2}{128 \pi^3\, m} 
\, {\rm Li}_2 \left(e^{-m/T} \right)\,\, ,
\end{equation}
which is twice the on-shell damping rate at zero momentum,
computed to 2--loop order in $\phi^4$ theory \cite{jeong}. The fact
that the damping coefficient is twice the damping rate was explained
in detail in \cite{cgbm}.
For the $O(N)$ model with $N=4$, 
\begin{equation}
\eta^{O(4)} = \frac{9\, \lambda^2 \, T^2}{16\pi^3\, m}
\, {\rm Li}_2 \left(e^{-m/T} \right)\,\, ,
\end{equation}
in agreement with the result (4) of \cite{cgtb}.
Note that in the large--$N$ limit, the damping coefficient vanishes
$\sim 1/N$.

\section{The $O(N)$ model with spontaneously broken symmetry}

The $O(N)$ symmetry of the Lagrangian (\ref{ONlagrangian}) is 
spontaneously broken to $O(N-1)$ by taking $m^2 <0$.
Then, the potential (\ref{U}) assumes the well-known ``Mexican hat''
shape, with the chiral circle $|\underline{\phi}^{\rm vac}| \equiv f_\pi 
\equiv (-m^2 N/4\,\lambda)^{1/2}$ as global minimum. Adding a small explicitly
symmetry breaking term $H\,\phi_1$ to the Lagrangian (\ref{ONlagrangian})
``tilts the hat'' in $\phi_1$--direction, such that the vacuum state is
$\underline{\phi}^{\rm vac} = (f_\pi, {\bf 0})$, where now
$f_\pi \equiv (-m^2N/4\lambda)^{1/2} \times 2\, \cos [\alpha/3]/\sqrt{3}$,
$\cos \alpha = (HN/8\lambda)/(-m^2N/12\lambda)^{3/2}$. The
$O(N-1)$ symmetry is restored taking $H\rightarrow 0$ such that
$\cos [\alpha/3] \rightarrow \sqrt{3}/2$.

Let us introduce new fields $\sigma \equiv \phi_1 - f_\pi$ and 
$\mbox{\boldmath $\pi$} \equiv (\phi_2, \phi_3, \ldots, \phi_N)$,
and corresponding masses:
\begin{equation}
m^2_\sigma \equiv m^2 + \frac{12\, \lambda\, f_\pi^2}{N}
\,\,\, , \,\,\,\,
m^2_\pi \equiv m^2 + \frac{4\, \lambda\, f_\pi^2}{N}\,\, .
\end{equation}
Obviously, without explicit symmetry breaking ($H=0$), the pions
are true Goldstone bosons, $m_\pi = 0$. With these definitions
the Lagrangian (\ref{ONlagrangian}) becomes (constant terms are omitted):
\begin{equation} \label{linsig}
{\cal L}(\sigma, \mbox{\boldmath $\pi$})
= \frac{1}{2} \left( \partial_\mu \sigma \, \partial^\mu
\sigma - m^2_\sigma\, \sigma^2 \right) + 
\frac{1}{2} \left( \partial_\mu \mbox{\boldmath $\pi$} \cdot \partial^\mu
\mbox{\boldmath $\pi$} - m^2_\pi\,\mbox{\boldmath $\pi$} \cdot
\mbox{\boldmath $\pi$}  \right) - \frac{4\, \lambda\, f_\pi}{N}\,
\sigma (\sigma^2 + \mbox{\boldmath $\pi$} \cdot \mbox{\boldmath $\pi$})
- \frac{\lambda}{N} (\sigma^2 + \mbox{\boldmath $\pi$} \cdot 
\mbox{\boldmath $\pi$})^2\,\, .
\end{equation}
Taking $N=4$, this is the Lagrangian of the linear $\sigma$ model.
The parameters $m^2,\, \lambda$, and $H$ of the original
Lagrangian (\ref{ONlagrangian}) are related to the physical meson
masses $m_\sigma = 600$ MeV, $m_\pi= 139$ MeV, and the pion
decay constant $f_\pi = 93$ MeV via
$m^2 = -(m_\sigma^2 - 3\, m_\pi^2)/2$, $\lambda = N (m^2_\sigma - m^2_\pi)/
(8\, f_\pi^2)$, $H = m_\pi^2 \, f_\pi$.
There are two types of interaction vertices in the Lagrangian
(\ref{linsig}), a three--particle vertex proportional to 
$\lambda\, f_\pi$ and a four--particle vertex proportional to $\lambda$. 
Note that, since $f_\pi \sim \lambda^{-1/2}$, 
the three--particle vertex is formally of order $\lambda^{1/2}$.

The $\sigma$ and $\mbox{\boldmath $\pi$}$ meson fields have
the Fourier representation (cf.\ eq.\ (\ref{FR}))
\begin{equation}
\sigma(t,{\bf k}) = \int {\rm d}^3{\bf x}\, e^{-i\, {\bf k} \cdot {\bf x}}
\, \sigma(t,{\bf x})\,\,\,\, , \,\,\,\,\,\,
\mbox{\boldmath $\pi$}(t,{\bf k}) = \int {\rm d}^3{\bf x}\, 
e^{-i\, {\bf k} \cdot {\bf x}}\, \mbox{\boldmath $\pi$}(t,{\bf x})\,\,,
\end{equation}
and are decomposed into soft and hard modes as follows:
\begin{mathletters}
\begin{eqnarray}
\sigma (t,{\bf x}) =  \int \frac{{\rm d}^3 {\bf k}}{(2 \pi)^3}\,
e^{i\, {\bf k} \cdot {\bf x}}\, \sigma(t, {\bf k}) \, \Theta(k_c - 
|{\bf k}|)\,\,\,\,\, & , & \,\,\,\,\, 
\Sigma (t,{\bf x}) = \int \frac{{\rm d}^3 {\bf k}}{(2 \pi)^3}\,
e^{i\, {\bf k} \cdot {\bf x}}\, \sigma(t, {\bf k}) \, \Theta( 
|{\bf k}|-k_c)\,\, , \\
\mbox{\boldmath $\pi$} (t,{\bf x}) = \int 
\frac{{\rm d}^3 {\bf k}}{(2 \pi)^3}\,
e^{i\, {\bf k} \cdot {\bf x}}\, \mbox{\boldmath $\pi$}(t, {\bf k}) \, 
\Theta(k_c - |{\bf k}|)\,\,\,\,\, & , & \,\,\,\,\,
{\bf \Pi} (t,{\bf x}) = \int \frac{{\rm d}^3 {\bf k}}{(2 \pi)^3}\,
e^{i\, {\bf k} \cdot {\bf x}}\, \mbox{\boldmath $\pi$}(t, {\bf k}) \, 
\Theta(|{\bf k}|-k_c)\,\, .
\end{eqnarray}
\end{mathletters}
(The use of small letters is from now on reserved for the soft fields,
the hard fields are denoted by capital letters.)

The interaction between soft and hard fields is given by the action:
\begin{equation}
S_I [\sigma,\Sigma,\mbox{\boldmath $\pi$},{\bf \Pi}] \equiv
S_I^{(\lambda f_\pi)} [\sigma,\Sigma,\mbox{\boldmath $\pi$},{\bf \Pi}]
+ S_I^{(\lambda)} [\sigma,\Sigma,\mbox{\boldmath $\pi$},{\bf \Pi}]\,\, ,
\end{equation}
where
\begin{equation}
S_I^{(\lambda f_\pi)} [\sigma,\Sigma,\mbox{\boldmath $\pi$},{\bf \Pi}] =
-\frac{4\, \lambda\, f_\pi}{N} \int \left( \frac{}{} \!\!
\left[ 3\, \sigma^2 + \mbox{\boldmath $\pi$} \cdot \mbox{\boldmath $\pi$}
\right] \, \Sigma + 2 \left[ \sigma + \Sigma \right] 
\mbox{\boldmath $\pi$} \cdot {\bf \Pi} +  \sigma  \left[
3\, \Sigma^2 + {\bf \Pi} \cdot {\bf \Pi} \right]
\right)\,\, ,
\end{equation}
and
\begin{eqnarray}
S_I^{(\lambda)} [\sigma,\Sigma,\mbox{\boldmath $\pi$},{\bf \Pi}]
& = & 
-\frac{\lambda}{N} \int \left( \frac{}{} 4
\left[ \sigma^2 + \mbox{\boldmath $\pi$} \cdot \mbox{\boldmath $\pi$}
\right] \, \left[ \sigma\, \Sigma + \mbox{\boldmath $\pi$} \cdot
{\bf \Pi} \right] + 2 \left[ 3\, \sigma^2 +
\mbox{\boldmath $\pi$} \cdot \mbox{\boldmath $\pi$} \right]\, 
\Sigma^2  +  2 \left[ \sigma^2 + \mbox{\boldmath $\pi$} \cdot 
\mbox{\boldmath $\pi$} \right] {\bf \Pi} \cdot {\bf \Pi} \right. \nonumber \\
&   & \left. \hspace*{1cm} + \,\,\, 
4 \left[ 2\, \sigma\, \Sigma + \mbox{\boldmath $\pi$} \cdot
{\bf \Pi} \right]\, \mbox{\boldmath $\pi$} \cdot {\bf \Pi}
+ 4 \left[ \sigma\, \Sigma + 
\mbox{\boldmath $\pi$} \cdot {\bf \Pi} \right]\,
\left[ \Sigma^2 + {\bf \Pi} \cdot {\bf \Pi} \right]
 \frac{}{} \right)\,\, .
\end{eqnarray}
Therefore,
\begin{eqnarray}
\lefteqn{\left\langle \frac{ \delta S_{I}[\bar{\sigma},\Sigma, 
\bar{\mbox{\boldmath $\pi$}}, {\bf \Pi}] }{\delta \bar{\sigma} (x)} 
\right\rangle_{\bar{\sigma},\bar{\pi}}  } 
\nonumber \\  
& = & -\frac{4\, \lambda \, f_\pi}{N} \left( \frac{}{} 6\, \bar{\sigma}(x) \,
\langle \Sigma(x) \rangle_{\bar{\sigma},\bar{\pi}} 
+ 2\, \bar{\mbox{\boldmath $\pi$}}(x) \cdot \langle {\bf \Pi}(x) 
\rangle_{\bar{\sigma},\bar{\pi}} + 3\, 
\langle \Sigma^2(x) \rangle_{\bar{\sigma},\bar{\pi}} 
+ \langle {\bf \Pi}(x) \cdot {\bf \Pi}(x) 
\rangle_{\bar{\sigma},\bar{\pi}} \right) \nonumber \\
&   & - \, \frac{4\, \lambda}{N} \left( \frac{}{} \left[ 3\, \bar{\sigma}^2(x)
+ \bar{\mbox{\boldmath $\pi$}}(x) \cdot \bar{\mbox{\boldmath $\pi$}}(x)
\right] 
\langle \Sigma(x) \rangle_{\bar{\sigma},\bar{\pi}}
+ 2\, \bar{\sigma}(x)\,\, \bar{\mbox{\boldmath $\pi$}}(x) \cdot 
\langle {\bf \Pi}(x) \rangle_{\bar{\sigma},\bar{\pi}}
+ 2\, \bar{\mbox{\boldmath $\pi$}}(x) \cdot
\langle \Sigma(x)\, {\bf \Pi}(x)
\rangle_{\bar{\sigma},\bar{\pi}}\right. \nonumber \\
&   &  \hspace*{1cm} \left. + \,\,
\bar{\sigma}(x) \left[3\, \langle \Sigma^2(x) 
\rangle_{\bar{\sigma},\bar{\pi}}
+ \langle {\bf \Pi}(x) \cdot {\bf \Pi}(x)
\rangle_{\bar{\sigma},\bar{\pi}}
\right]
+ \langle \Sigma^3(x) \rangle_{\bar{\sigma},\bar{\pi}}
+ \langle \Sigma(x)\,\, {\bf \Pi}(x) \cdot {\bf \Pi}(x)
\rangle_{\bar{\sigma},\bar{\pi}} \frac{}{} \right)\,\, ,
\label{62}
\end{eqnarray}
while
\begin{eqnarray}
\lefteqn{\left\langle \frac{ \delta S_{I}[\bar{\sigma},\Sigma, 
\bar{\mbox{\boldmath $\pi$}}, {\bf \Pi}] }{\delta \bar{\pi}_a (x)} 
\right\rangle_{\bar{\sigma},\bar{\pi}}  
 =  - \frac{8\, \lambda\, f_\pi}{N} \left( \frac{}{} \bar{\pi}_a(x)\,
\langle \Sigma(x) \rangle_{\bar{\sigma},\bar{\pi}} 
+ \bar{\sigma}(x) \, 
\langle \Pi_a(x) \rangle_{\bar{\sigma},\bar{\pi}} 
+ \langle \Sigma(x)\, \Pi_a(x)
\rangle_{\bar{\sigma},\bar{\pi}} \right)} \nonumber \\
&   & - \frac{4\, \lambda}{N} \left( \frac{}{} 2\, \bar{\pi}_a(x)\, \left[ 
\bar{\sigma}(x)\, \langle \Sigma(x) 
\rangle_{\bar{\sigma},\bar{\pi}}  + 
\bar{\mbox{\boldmath $\pi$}}(x) \cdot \langle {\bf \Pi}(x) 
\rangle_{\bar{\sigma},\bar{\pi}}  \right]
+ \left[ \bar{\sigma}^2(x) + \bar{\mbox{\boldmath $\pi$}}(x) \cdot
\bar{\mbox{\boldmath $\pi$}}(x) \right]\, \langle \Pi_a(x)
\rangle_{\bar{\sigma},\bar{\pi}} \right. \nonumber \\
&   & \hspace*{1cm} + \,\, \bar{\pi}_a(x) \, \left[ \langle \Sigma^2(x)
\rangle_{\bar{\sigma},\bar{\pi}} 
+ \langle {\bf \Pi}(x) \cdot {\bf \Pi}(x)
\rangle_{\bar{\sigma},\bar{\pi}}  \right]
+ 2\, \left[ \bar{\sigma}(x)\, \langle \Sigma(x) \, \Pi_a(x)
\rangle_{\bar{\sigma},\bar{\pi}} 
+ \bar{\mbox{\boldmath $\pi$}}(x) \cdot \langle {\bf \Pi}(x)\, \Pi_a(x)
\rangle_{\bar{\sigma},\bar{\pi}}  \right] \nonumber \\
&   & \hspace*{1cm} +\left. \langle \Sigma^2(x)\, \Pi_a(x) 
\rangle_{\bar{\sigma},\bar{\pi}}  
+ \langle {\bf \Pi}(x) \cdot {\bf \Pi}(x) \,\, \Pi_a(x) 
\rangle_{\bar{\sigma},\bar{\pi}}  
\frac{}{} \right)\,\,. \label{63}
\end{eqnarray}

\subsection{Interactions between soft and hard fields 
to first order in $\lambda$}

In this subsection, the expectation values $\langle \, \cdot \,
\rangle_{\bar{\sigma},\bar{\pi}}$ on the right-hand side of eqs.\ 
(\ref{62}) and (\ref{63}) will be evaluated in perturbation theory.
Since $\lambda = N (m_\sigma^2 - m_\pi^2)/(8\, f_\pi^2) \simeq 20$ for
$N=4$ and realistic values of the
parameters $m_\sigma,\, m_\pi, \, f_\pi$, this is
certainly not a controlled approximation scheme. Therefore, the 
following results have to be viewed only as the first, but necessary, step 
to estimate the influence of an unobserved background of hard modes in the 
classical equations of motion for the $O(N)$ model in the phase
of spontaneously broken symmetry.

\begin{figure}
\begin{center}
\epsfxsize=11cm
\epsfysize=8cm
\leavevmode
\hbox{ \epsffile{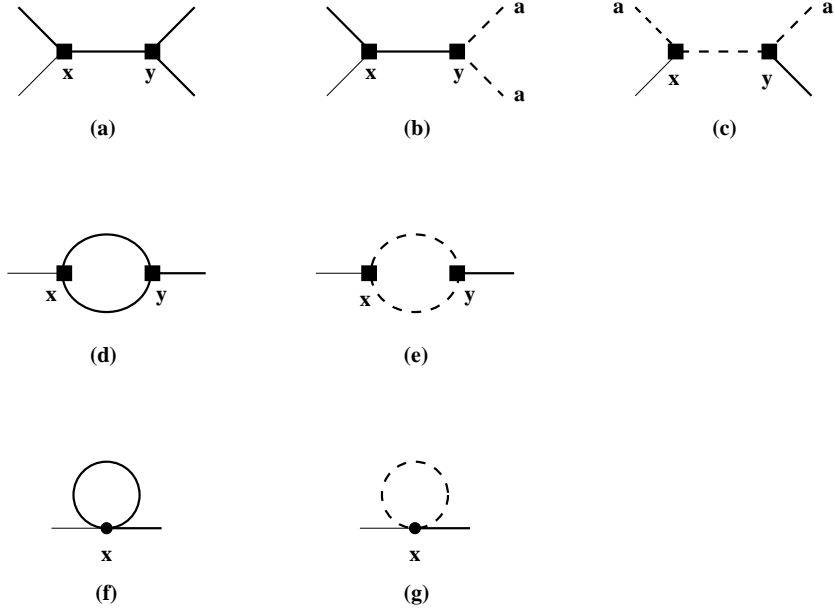}}
\end{center}
\caption{(a,b) The two terms in ${\cal T}_\sigma^{(1)}$. A sum over $a$
is implied. (c) The term ${\cal T}_\sigma^{(2)}$. A sum over $a$ is implied. 
(d,e) The two terms in ${\cal T}_\sigma^{(3)}$. (f,g)
Order $\lambda$ contributions to the thermal mass.}
\label{Figmass3}
\end{figure}

To determine the interaction terms ${\cal R}_\sigma$ and ${\cal R}_{\pi_a}$
in the classical equations of motion (\ref{cleom})
to first order in $\lambda$, due to the overall factors of $\lambda\, f_\pi
\sim \lambda^{1/2}$ and $\lambda$ in eqs.\ (\ref{62}) and (\ref{63})
the expectation values in the terms proportional to $\lambda \, f_\pi$ 
have to be computed only to order $\lambda \, f_\pi$, while those
in the terms proportional to $\lambda$ have to be computed only to order 1.
More explicitly, $\langle \Sigma(x) 
\rangle_{\bar{\sigma},\bar{\pi}} ,\, \langle
\Pi_a(x) \rangle_{\bar{\sigma},\bar{\pi}} ,\,
\langle \Sigma^2(x) \rangle_{\bar{\sigma},\bar{\pi}} ,\,
\langle \Pi_a^2(x) \rangle_{\bar{\sigma},\bar{\pi}} ,$
and $\langle \Sigma(x)\, \Pi_a(x)
\rangle_{\bar{\sigma},\bar{\pi}} $ have to be computed
up to order $\lambda\, f_\pi$, while
$\langle \Sigma^3(x) \rangle_{\bar{\sigma},\bar{\pi}} ,\,
\langle \Sigma^2(x)\, \Pi_a(x)
\rangle_{\bar{\sigma},\bar{\pi}} ,\,
\langle \Sigma(x)\, \Pi^2_a(x)
\rangle_{\bar{\sigma},\bar{\pi}} ,\,
\langle \Pi_a(x)\, \Pi_b(x)
\rangle_{\bar{\sigma},\bar{\pi}} ,$ and
$\langle \Pi_a(x)\, \Pi^2_b(x)
\rangle_{\bar{\sigma},\bar{\pi}} $ are required
to order 1. For the latter, this of course means that the 
average $\langle \,\, \cdot \,\, 
\rangle_{\bar{\sigma},\bar{\pi}} $ can be replaced
by $\langle \,\, \cdot \,\, \rangle_0$, and as a consequence, all 
expectation values vanish except for $\langle \Pi_a(x)\, \Pi_b(x)
\rangle_0 = \delta_{ab} \langle \Pi_a^2(x) \rangle_0$. For the non-vanishing
expectation values one obtains:
\begin{mathletters}
\begin{eqnarray}
\langle \Sigma(x) \rangle_{\bar{\sigma},\bar{\pi}} 
& = & -i\, \frac{4\, \lambda\, f_\pi}{N} \int_{t_i}^{x_0} {\rm d}^4y
\, \left[ D_>^{(\sigma)}(x-y) - D_<^{(\sigma)}(x-y) \right]\, 
\left[ 3\, \bar{\sigma}^2(y) +
\bar{\mbox{\boldmath $\pi$}}(y) \cdot \bar{\mbox{\boldmath $\pi$}}(y) \right]
\, ,\\
\langle \Pi_a(x) \rangle_{\bar{\sigma},\bar{\pi}} 
& = & -i\, \frac{4\, \lambda\, f_\pi}{N} \int_{t_i}^{x_0} {\rm d}^4y
\, \left[ D_>^{(\pi)}(x-y) - D_<^{(\pi)}(x-y) \right]
\,2\,\bar{\sigma}(y)\, \bar{\pi}_a(y)\, ,\\
\langle \Sigma^2(x) \rangle_{\bar{\sigma},\bar{\pi}} 
& = & D_{++}^{(\sigma)}(0) 
-i\, \frac{4\, \lambda\, f_\pi}{N} \int_{t_i}^{x_0} {\rm d}^4y
\, \left( \left[ D_>^{(\sigma)}(x-y) \right]^2
-  \left[ D_<^{(\sigma)}(x-y) \right]^2 \right) 6\, \bar{\sigma}(y)\, ,
\\
\langle \Pi^2_a(x) \rangle_{\bar{\sigma},\bar{\pi}} 
& = & D_{++}^{(\pi)}(0)  
-i\, \frac{4\, \lambda\, f_\pi}{N} \int_{t_i}^{x_0} {\rm d}^4y
\, \left( \left[ D_>^{(\pi)}(x-y) \right]^2
-  \left[ D_<^{(\pi)}(x-y) \right]^2 \right) 2\, \bar{\sigma}(y)\, ,
\end{eqnarray}
and
\begin{equation}
\langle \Sigma(x)\, \Pi_a(x) 
\rangle_{\bar{\sigma},\bar{\pi}}   = 
-i\, \frac{4\, \lambda\, f_\pi}{N} \int_{t_i}^{x_0} {\rm d}^4y \, 
\left[ D_>^{(\sigma)}(x-y) D_>^{(\pi)}(x-y) 
- D_<^{(\sigma)}(x-y) D_<^{(\pi)}(x-y) \right] \, 2\, \bar{\pi}_a(y) \,\,.
\end{equation}
\end{mathletters}
Again, as discussed above in the symmetric case, 
all expressions are real-valued.
For the left-hand side of the equation of motion for the classical
$\sigma$ field one therefore obtains:
\begin{eqnarray}
{\cal E}_{\sigma}(x) & = & - \left[  \Box + m^2_\sigma(T)
+ \frac{4\, \lambda}{N} \left(\bar{\sigma}^2(x)
+ \bar{\mbox{\boldmath $\pi$}}(x) \cdot \bar{\mbox{\boldmath $\pi$}}(x)
\right) \right] \bar{\sigma}(x)   \nonumber \\
& - &\frac{4\, \lambda\, f_\pi}{N} \left[ 3\, \bar{\sigma}^2(x) +
\bar{\mbox{\boldmath $\pi$}}(x) \cdot \bar{\mbox{\boldmath $\pi$}}(x)
+ 3\, D_{++}^{(\sigma)}(0) +(N-1)\,D_{++}^{(\pi)}(0) \right]
 + \left( \frac{4\,\lambda\,f_\pi}{N} \right)^2 \, 
\sum_{i=1}^3 {\cal T}^{(i)}_\sigma (x)\,\, , \label{67}
\end{eqnarray}
where
\begin{mathletters}
\begin{eqnarray}
{\cal T}^{(1)}_\sigma (x) & = & i \int_{t_i}^{x_0} {\rm d}^4y\,  
\left[ D_>^{(\sigma)}(x-y) - D_<^{(\sigma)} (x-y) \right] 
6\, \bar{\sigma}(x) \left[ 3\, \bar{\sigma}^2(y) +
\bar{\mbox{\boldmath $\pi$}}(y) \cdot \bar{\mbox{\boldmath $\pi$}}(y) \right]
\,\, , \label{Ts1} \\
{\cal T}^{(2)}_\sigma (x) & = & i \int_{t_i}^{x_0} {\rm d}^4y\, 
\left[D_>^{(\pi)}(x-y) -D_<^{(\pi)}(x-y) \right] 
4\,\bar{\mbox{\boldmath $\pi$}}(x) \cdot 
\bar{\mbox{\boldmath $\pi$}}(y)\, \bar{\sigma}(y) \,\, , \label{Ts2} \\
{\cal T}^{(3)}_\sigma (x) & = & i \int_{t_i}^{x_0} {\rm d}^4y\, \left\{
18 \left( \left[D_>^{(\sigma)}(x-y)\right]^2 - 
\left[D_<^{(\sigma)} (x-y) \right]^2 \right) \right. \nonumber \\
&   & \hspace*{0.4cm}  +\left. 2\, (N-1) \left(
\left[D_>^{(\pi)} (x-y)\right]^2 - 
\left[D_<^{(\pi)} (x-y) \right]^2 \right) \right\} \bar{\sigma}(y)
\,\, , \label{Ts3}
\end{eqnarray}
\end{mathletters}
and $m_\sigma^2(T)$ is given by:
\begin{equation} \label{msigmasquared}
m^2_\sigma(T) = m_\sigma^2 +
\frac{4 \, \lambda}{N}\, \left[
3\, D_{++}^{(\sigma)}(0) + (N-1) \, D_{++}^{(\pi)}(0) \right]\,\, .
\end{equation}
The terms ${\cal T}_\sigma^{(i)}$ and the thermal mass correction are
graphically displayed in Fig.\ \ref{Figmass3}. The conventions are as
before in Fig.\ \ref{Figmass2}. A dashed line
represents a pion and a solid line a sigma. Thin solid lines do not have
a correspondence in the classical equation of motion, they correspond to
a factor $\Delta \sigma(x)$ multiplying ${\cal E}_\sigma(x)$
in the phase of the reduced density matrix (\ref{reducedDM2}). 
A filled dotted vertex
corresponds to a four--particle vertex $\lambda$, a filled square vertex
to a three--particle vertex $\lambda\, f_\pi$. Graphs (a,b) correspond to
the two terms in eq.\ (\ref{Ts1}), graph (c) to (\ref{Ts2}), 
graphs (d,e) to the two terms in (\ref{Ts3}) and (f,g) to 
$[m_\sigma^2(T) - m_\sigma^2]\bar{\sigma}(x)$.

A Fourier transformation with respect to ${\bf x}$ yields the classical
equation of motion for the momentum mode functions $\bar{\sigma}(t,{\bf k})$
($|{\bf k}|\leq k_c$):
\begin{eqnarray}
\lefteqn{\left[ \partial_t^2 + 
\left({E_{\bf k}^{(\sigma)}}^*\right)^2 \right] \bar{\sigma}(t,{\bf k})
+ \frac{4\, \lambda \, f_\pi}{N} \left[ 3\, D_{++}^{(\sigma)}(0) +
(N-1)\, D_{++}^{(\pi)}(0) \right]\, (2 \pi)^3 \, \delta^{(3)}({\bf k})}
\nonumber \\
& + & \frac{4\, \lambda \, f_\pi}{N} \int \frac{{\rm d}^3 {\bf p}}{(2\pi)^3}
\,\Theta(k_c-|{\bf p}|)\, \Theta(k_c - |{\bf k} - {\bf p}|)\,
\left[ 3\, \bar{\sigma}(t,{\bf p})\,\bar{\sigma}(t,{\bf k} - {\bf p})
+ \bar{\mbox{\boldmath $\pi$}}(t,{\bf p}) \cdot \bar{\mbox{\boldmath $\pi$}}
(t,{\bf k} - {\bf p}) \right] \nonumber \\
& + & \frac{4\, \lambda}{N} \! \int 
\frac{{\rm d}^3 {\bf p} {\rm d}^3{\bf q}}{
(2\pi)^6} \Theta(k_c-|{\bf p}|)\Theta(k_c-|{\bf q}|) 
\Theta(k_c - |{\bf k} - {\bf p}-{\bf q}|)
\left[ \bar{\sigma}(t,{\bf p})\bar{\sigma}(t,{\bf q})
+ \bar{\mbox{\boldmath $\pi$}}(t,{\bf p}) \cdot \bar{\mbox{\boldmath $\pi$}}
(t,{\bf q}) \right] \bar{\sigma}(t,{\bf k} - {\bf p} -{\bf q})
\nonumber \\
& - & \left(\frac{4\, \lambda\, f_\pi}{N} \right)^2 \, \sum_{i=1}^3
{\cal T}_\sigma^{(i)} (t, {\bf k}) = \xi_{\sigma}(t,{\bf k})  \,\, .
\label{70a}
\end{eqnarray}
Here, ${E_{\bf k}^{(\sigma)}}^* \equiv [{\bf k}^2 +m_\sigma^2(T)]^{1/2}$.
An explicit calculation of the terms ${\cal T}_\sigma^{(i)} (t, {\bf k})$
in the linear harmonic approximation is referred to Appendix C.

For the classical pion fields one obtains:
\begin{equation} \label{Epi}
{\cal E}_{\pi_a}(x) =  - \left[  \Box + m^2_\pi(T)
+ \frac{4\, \lambda}{N} \left(\bar{\sigma}^2(x)
+ \bar{\mbox{\boldmath $\pi$}}(x) \cdot \bar{\mbox{\boldmath $\pi$}}(x)
+ 2\,f_\pi \, \bar{\sigma}(x) \right) \right] \bar{\pi}_a(x)
+ \left( \frac{4\,\lambda\,f_\pi}{N} \right)^2 \, \sum_{i=1}^3
{\cal T}_{\pi_a}^{(i)}(x)\,\, ,
\end{equation}
where
\begin{mathletters}
\begin{eqnarray}
{\cal T}_{\pi_a}^{(1)} (x) & = &  i \int_{t_i}^{x_0} {\rm d}^4y\, \left[
D_>^{(\sigma)}(x-y) - D_<^{(\sigma)}(x-y) \right] 
2\, \bar{\pi}_a(x) \left[ 3\, \bar{\sigma}^2(y) +
\bar{\mbox{\boldmath $\pi$}}(y) \cdot \bar{\mbox{\boldmath $\pi$}}(y) \right]
\,\, , \label{Tp1} \\
{\cal T}_{\pi_a}^{(2)} (x) & = & i \int_{t_i}^{x_0} {\rm d}^4y\, \left[
D_>^{(\pi)}(x-y) - D_<^{(\pi)}(x-y) \right] 
4\, \bar{\sigma}(x)\, \bar{\sigma}(y) \, \bar{\pi}_a(y)\,\, , \label{Tp2} \\
{\cal T}_{\pi_a}^{(3)} (x) & = & i \int_{t_i}^{x_0} {\rm d}^4y\, \left[ 
D_>^{(\sigma)}(x-y)\, D_>^{(\pi)}(x-y) -
D_<^{(\sigma)} (x-y) \, D_<^{(\pi)}(x-y) \right] 4 \, \bar{\pi}_a(y)
\,\, , \label{Tp3} 
\end{eqnarray}
\end{mathletters}
and
\begin{equation} \label{70}
m^2_\pi(T) = m_\pi^2  + 
\frac{4 \, \lambda}{N}\, \left[
D_{++}^{(\sigma)}(0) + (N+1)D_{++}^{(\pi)}(0) \right]\,\,.
\end{equation}
At first glance, $m_\pi(T)$ seems to be the thermal mass 
of soft (classical) pionic excitations. This, however, cannot be true,
since in the chiral limit $m_\pi \rightarrow 0$, the pions are 
massless Goldstone bosons, even at finite temperature \cite{kapusta}, whereas
$m_\pi(T)$ as given by eq.\ (\ref{70}) is finite (cf.\ eq.\ (\ref{HTL})).
This apparent violation of Goldstone's theorem will be resolved below.

\begin{figure}
\begin{center}
\epsfxsize=14cm
\epsfysize=6cm
\leavevmode
\hbox{ \epsffile{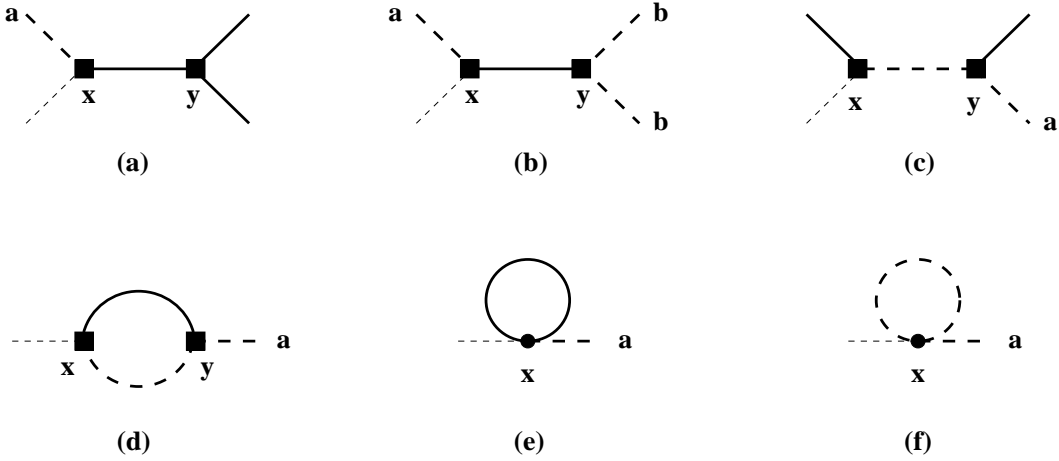}}
\end{center}
\caption{(a,b) The two terms in ${\cal T}_{\pi_a}^{(1)}$. A sum over $b$
is implied. (c) The term ${\cal T}_{\pi_a}^{(2)}$. 
(d) The term ${\cal T}_{\pi_a}^{(3)}$. (e,f)
Order $\lambda$ contributions to the thermal mass.}
\label{Figmass4}
\end{figure}

The terms ${\cal T}_{\pi_a}^{(i)}$ and the thermal mass correction are
graphically displayed in Fig.\ \ref{Figmass4}. The notation is the same
as in Fig.\ \ref{Figmass3}. Graphs (a,b) correspond to
the two terms in eq.\ (\ref{Tp1}), graph (c) to (\ref{Tp2}), 
graph (d) to (\ref{Tp3}) and (e,f) to 
$[m_\pi^2(T) - m_\pi^2]\bar{\pi}_a(x)$.

Fourier transforming eq.\ (\ref{Epi}) with respect to ${\bf x}$ yields
the classical equation of motion for the mode functions $\bar{\pi}_a
(t,{\bf k})$:
\begin{eqnarray}
\lefteqn{\left[ \partial_t^2 + 
\left({E_{\bf k}^{(\pi)}}^*\right)^2 \right] \bar{\pi}_a(t,{\bf k})
+ \frac{8\, \lambda \, f_\pi}{N} \int \frac{{\rm d}^3 {\bf p}}{(2\pi)^3}
\,\Theta(k_c-|{\bf p}|)\, \Theta(k_c - |{\bf k} - {\bf p}|)\,
\bar{\sigma}(t,{\bf p})\,\bar{\pi}_a(t,{\bf k} - {\bf p})}
\nonumber \\
& + & \frac{4\, \lambda}{N} \! \int 
\frac{{\rm d}^3 {\bf p} {\rm d}^3{\bf q}}{
(2\pi)^6} \Theta(k_c-|{\bf p}|)\Theta(k_c-|{\bf q}|) 
\Theta(k_c - |{\bf k} - {\bf p}-{\bf q}|)
\left[ \bar{\sigma}(t,{\bf p})\bar{\sigma}(t,{\bf q})
+ \bar{\mbox{\boldmath $\pi$}}(t,{\bf p}) \cdot \bar{\mbox{\boldmath $\pi$}}
(t,{\bf q}) \right] \bar{\pi}_a(t,{\bf k} - {\bf p} -{\bf q})
\nonumber \\
& - & \left(\frac{4\, \lambda\, f_\pi}{N} \right)^2 \, \sum_{i=1}^3
{\cal T}_{\pi_a}^{(i)} (t, {\bf k}) = \xi_{\pi_a}(t,{\bf k})  \,\, .
\label{74}
\end{eqnarray}
Here, ${E_{\bf k}^{(\pi)}}^* \equiv [{\bf k}^2 +m_\pi^2(T)]^{1/2}$.
An explicit calculation of the terms ${\cal T}_{\pi_a}^{(i)} (t, {\bf k})$
in the linear harmonic approximation is referred to Appendix D.

For the variances of the noise terms one obtains 
with eq.\ (\ref{Iofphi}) up to order $\lambda$:
\begin{mathletters} \label{variance}
\begin{eqnarray}
{\cal I}_{\sigma \sigma}(x,y) & = & \left( \frac{4\, \lambda\, f_\pi}{N}
\right)^2 \left\{ 18\, \bar{\sigma}(x)\,  \bar{\sigma}(y) \, \left[
D_>^{(\sigma)}(x-y) + D_<^{(\sigma)}(x-y) \right] \right. \nonumber \\
&   & \hspace*{1.51cm} + \,\,
2\, \bar{\mbox{\boldmath $\pi$}}(x) \cdot \bar{\mbox{\boldmath $\pi$}}(y) \, 
\left[D_>^{(\pi)}(x-y) + D_<^{(\pi)}(x-y) \right] \nonumber \\
+ & 9 & \left.\!\!\! \left(
\left[D_>^{(\sigma)}(x-y)\right]^2 + \left[D_<^{(\sigma)}(x-y) \right]^2
\right) + (N-1) \left(
\left[D_>^{(\pi)}(x-y)\right]^2 + \left[D_<^{(\pi)}(x-y) \right]^2
\right) \right\} , \\
{\cal I}_{\sigma \pi_a}(x,y) & = & \left( \frac{4\, \lambda\, f_\pi}{N}
\right)^2 \left\{ 6\, \bar{\sigma}(x)\, \bar{\pi}_a(y) \, \left[
D_>^{(\sigma)}(x-y) + D_<^{(\sigma)}(x-y) \right]  \right. \nonumber \\
&  & \hspace*{1.5cm} \left. +\,\,
2\, \bar{\pi}_a(x) \,\bar{\sigma}(y) \, \left[
D_>^{(\pi)}(x-y) + D_<^{(\pi)}(x-y) \right] \right\} \,\, , \\
{\cal I}_{\pi_a \pi_b}(x,y) & = & \left( \frac{4\, \lambda\, f_\pi}{N}
\right)^2 \left\{ 2\, \bar{\pi}_a(x)\, \bar{\pi}_b(y)\,  \left[
D_>^{(\sigma)}(x-y) + D_<^{(\sigma)}(x-y) \right] \right. \nonumber \\
&   & \hspace*{1.51cm} + \,\,
2\, \delta_{ab}\, \bar{\sigma}(x)\, \bar{\sigma}(y) \, \left[
D_>^{(\pi)}(x-y) + D_<^{(\pi)}(x-y) \right] \nonumber \\
&  & \hspace*{1.51cm} + \,  \left. 2 \, \delta_{ab} \, \left[
D_>^{(\sigma)}(x-y)\, D_>^{(\pi)}(x-y) + D_<^{(\sigma)}(x-y)\, 
D_<^{(\pi)}(x-y) \right] \right\} \,\, .
\end{eqnarray}
\end{mathletters}
This confirms the existence of correlations between the noise
fields $\xi_\sigma$ and $\xi_{\pi_a}$ as well as between
$\xi_{\pi_a}$ and $\xi_{\pi_b}$, $a \neq b$, as mentioned at the
end of section III. There also exists an obvious graphical representation
for the variances. The difference to the graphs of Fig.\ \ref{Figmass3}
and \ref{Figmass4} is that one thick line on the vertex at
space--time point $y$ is replaced by a thin line, 
corresponding to the second factor 
$\Delta \sigma(y)$ or $\Delta \pi_b(y)$ in the phase 
in eq.\ (\ref{reducedDM2}). Also, due to 
$D_>^n(x) + D_<^n(x) = 2\, {\rm Re}\, D_{++}^n(x)$
one has to take the {\em real\/} instead of the imaginary part of
the respective diagrams.

\subsection{Spatially homogeneous solutions of the classical equations of
motion}

\begin{figure} \hspace*{3cm} 
\psfig{figure=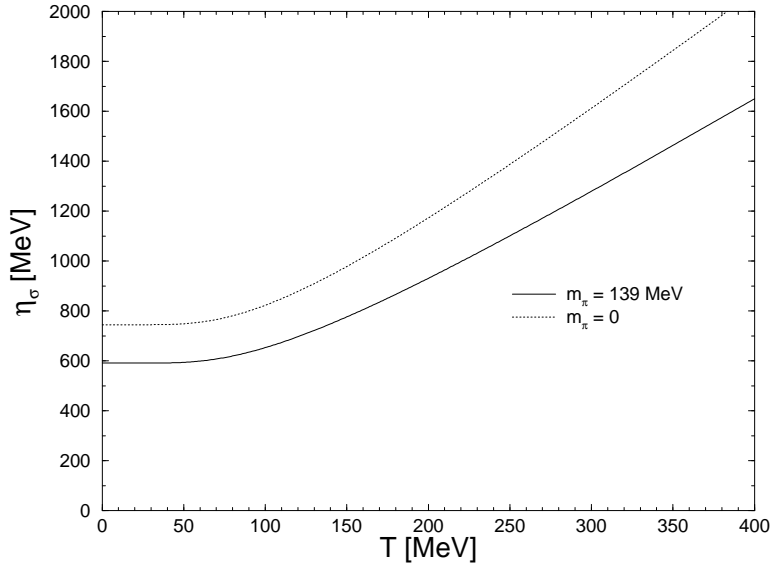,width=2.5in,height=3.3in,angle=270}
\vspace*{-1cm}
\caption{The temperature dependence of the
dissipation coefficient $\eta_\sigma$ in the case $m_\pi=139$ MeV 
(solid) and in the chiral limit $m_\pi =0$ (dotted).}
\label{Figetasigma}
\end{figure}

In the following, let us focus on the time evolution of the zero--momentum
mode functions $\bar{\sigma}(t,{\bf 0})$ and $\bar{\pi}_a(t,{\bf 0})$. 
For the sake of simplicity, let us also
take the limit $k_c \rightarrow 0$, i.e., only spatially homogeneous field
configurations are considered to be classical, and let us assume
$t_i \rightarrow - \infty,\, t_f \rightarrow +\infty$, to facilitate
Fourier transformations. Then, the explicit form of the functions 
${\cal T}_{\sigma,\pi_a}^{(i)} (t, {\bf k})$
given in Appendices C, D 
shows that ${\cal T}_{\sigma,\pi_a}^{(1)} (t, {\bf 0}) = 
{\cal T}_{\sigma,\pi_a}^{(2)} (t, {\bf 0}) = 0$, while
\begin{eqnarray}
{\cal T}^{(3)}_{\sigma} (t,{\bf 0}) & = &
- 2\, {\rm P}\int \frac{{\rm d}\omega}{2 \pi}\,
\frac{9\,{\cal M}_2^{(\sigma\sigma)}(\omega, {\bf 0})
+(N-1)\,{\cal M}_2^{(\pi\pi)}(\omega, {\bf 0}) }{m_{\sigma}-\omega }
\,\,  \bar{\sigma}(t, {\bf 0}) \nonumber \\
&   & 
-\,2\, \frac{9\, {\cal M}^{(\sigma\sigma)}_2 (m_\sigma, {\bf 0})
+(N-1)\,{\cal M}^{(\pi\pi)}_2 (m_\sigma, {\bf 0}) }{2\, m_\sigma} \,\,
\partial_t\, \bar{\sigma} (t, {\bf 0})\,\,, 
\end{eqnarray}
and 
\begin{equation}
{\cal T}^{(3)}_{\pi_a} (t,{\bf 0})= 
- 4\, {\rm P}\int \frac{{\rm d}\omega}{2 \pi}\,
\frac{{\cal M}_2^{(\sigma\pi)}(\omega, {\bf 0})}{m_\pi - \omega }
\, \bar{\pi}_a(t, {\bf 0}) -  
4\, \frac{ {\cal M}^{(\sigma\pi)}_2 (m_\pi, {\bf 0})}{2\, m_\pi} \,\,
\partial_t\, \bar{\pi}_a (t, {\bf 0})\,\,, 
\end{equation}
where ${\cal M}_2^{(ij)}(\omega, {\bf k})$ is the Fourier
transform of ${\cal M}_2^{(ij)}(x) \equiv D_>^{(i)}(x)\,D_>^{(j)}(x)
-  D_<^{(i)}(x)\,D_<^{(j)}(x)$, $i,j = \sigma$ or $\pi$, cf.\ eqs.\
(\ref{M2sigmapi}) and (\ref{M2ij}). Using $(2\pi)^3\, 
\delta^{(3)}({\bf q}) \equiv V \delta^{(3)}_{{\bf q},{\bf 0}}$, where
$V$ is the total 3--volume of the system, as well as 
$V\int {\rm d}^3 {\bf p}/(2\pi)^3 \equiv \sum_{\bf k}$ for the 
momentum integrals, and defining $\bar{\sigma}(t) \equiv \bar{\sigma}
(t,{\bf 0})/V,\, \bar{\mbox{\boldmath $\pi$}}(t) \equiv \bar{\mbox{\boldmath 
$\pi$}}(t,{\bf 0})/V,\, \xi_\sigma(t) \equiv \xi_\sigma(t,{\bf 0})/V,\, 
\xi_{\pi_a}(t) \equiv \xi_{\pi_a}(t,{\bf 0})/V$, 
one derives from (\ref{70a}) the classical equation of motion for 
$\bar{\sigma}(t)$:
\begin{eqnarray}
\left[ \partial_t^2 + \tilde{m}_\sigma^2(T) \right]\, \bar{\sigma}(t)
& + & \frac{4\, \lambda\, f_\pi}{N} \left[3\, D_{++}^{(\sigma)}(0) 
+ (N-1)\, D_{++}^{(\pi)}(0) \right]+ 
\frac{4\, \lambda\, f_\pi}{N} \left[3\, \bar{\sigma}^2(t) 
+ \bar{\mbox{\boldmath $\pi$}}(t) \cdot \bar{\mbox{\boldmath $\pi$}}(t) \right]
  \nonumber \\
& + & \frac{4\, \lambda}{N} \left[\bar{\sigma}^2(t) + 
\bar{\mbox{\boldmath $\pi$}}(t) \cdot \bar{\mbox{\boldmath $\pi$}}(t) 
\right] \bar{\sigma}(t)
+ \eta_\sigma \, \partial_t \, \bar{\sigma}(t) = \xi_\sigma(t)\,\, ,
\label{eqofmosig}
\end{eqnarray} 
with
\begin{eqnarray}
\tilde{m}^2_\sigma(T) & = & m_\sigma^2 (T)
+ 2\, \left( \frac{4\, \lambda\, f_\pi}{N} \right)^2 \, {\rm P}
\int \frac{{\rm d} \omega}{2 \pi} \, \frac{9\,{\cal M}_2^{(\sigma\sigma)}
(\omega,{\bf 0}) + (N-1)\,{\cal M}_2^{(\pi\pi)}(\omega,{\bf 0}) }{
m_\sigma-\omega} \nonumber \\
& = & m_\sigma^2 + \frac{4 \, \lambda}{N}\, \left[
3\, D_{++}^{(\sigma)}(0) + (N-1) \, D_{++}^{(\pi)}(0) \right] \nonumber \\
& +  & \frac{2}{\pi^2}\, \left( \frac{4\, \lambda\, f_\pi}{N} \right)^2 \, 
\left[9 \int_{m_\sigma}^\infty {\rm d}E\, \frac{\sqrt{E^2-m_\sigma^2}}{
m_\sigma^2 -4 E^2}\, \frac{1}{e^{E/T}-1} + 
(N-1) \int_{m_\pi}^\infty {\rm d}E\, \frac{\sqrt{E^2-m_\pi^2}}{m_\sigma^2
-4 E^2}\, \frac{1}{e^{E/T}-1} \right]\,\, ,
\end{eqnarray}
where use has been made of eqs.\ (\ref{msigmasquared}) and 
(\ref{M2ij}), and divergent terms have been removed by renormalization. 
The apparent singularity in the last
integral poses no problem, since it is integrable. 
The dissipation coefficient is (cf.\ Appendix E):
\begin{eqnarray} 
\eta_\sigma & = & \left(\frac{4\,\lambda\,f_\pi}{N} \right)^2
2\, \frac{ 9\, {\cal M}_2^{(\sigma\sigma)}(m_\sigma, {\bf 0}) 
+ (N-1) \, {\cal M}_2^{(\pi\pi)}(m_\sigma , {\bf 0})}{ 2\, m_\sigma} \nonumber
\\
& = & 
\left(\frac{4\, \lambda\, f_\pi}{N} \right)^2 
\frac{N-1}{8\pi\, m_\sigma}\, \sqrt{1 - \frac{4\, m_\pi^2}{m_\sigma^2}}
\,\, {\rm coth} \,\frac{m_\sigma}{4\, T}\,\, .\label{etasigma}
\end{eqnarray}
The dissipation coefficient
$\eta_\sigma$ corresponds to the imaginary part of the diagram
Fig.\ \ref{Figmass3} (e), where the incoming (outgoing) $\sigma$
particle is on-shell and at rest. 
Therefore, the dissipation occurs physically due to the decay of the 
$\sigma$ into two $\pi$'s \cite{weldon}.
The temperature dependence of $\eta_\sigma$ is shown in Fig.\ 
\ref{Figetasigma}.
Note that even at $T=0$, the dissipation coefficient does not vanish.
This is physically plausible, since even then
a $\sigma$ can always decay into two $\pi$'s. This means, however,
that dissipation (and associated fluctuations) persist
{\em even in the absence of a heat bath}. In that case, the
fluctuations have to interpreted as {\em quantum\/} rather than
thermal fluctuations.

For $T=0$, and for the parameters of the linear sigma model,
$\eta_\sigma = 591.45$ MeV, which is on the order $m_\sigma$, and
thus quite large. In the chiral limit, $m_\pi \rightarrow 0$ and at
$T=0$, the dissipation coefficient
$\eta_\sigma \rightarrow 3\, m_\sigma^3/(32\pi\, f_\pi^2) = 745.26$
MeV, which is even larger. $\eta_\sigma$ increases with $T$ because of
Bose--Einstein enhancement of the final two--pion state at finite $T$.

\vspace*{0.7cm}

For the variance of the noise field $\xi_\sigma(t)$ the so-called
``white-noise'' approximation is employed (cf.\ Appendix G),
which is consistent with the linear harmonic approximation (\ref{linharm})
that made the equation of motion for the $\sigma$ field local
in time and led to the term $\eta_\sigma\, \partial_t \,\bar{\sigma}(t)$ 
in eq.\ (\ref{eqofmosig}). 
The variance of the noise field $\xi_\sigma(t)$ becomes
\begin{equation} \label{varsig}
\langle \xi_\sigma(t) \, \xi_\sigma(t') \rangle_{\xi} = \frac{1}{V}\,
\delta(t-t')\,\eta_\sigma\, m_\sigma\,\coth \left[\frac{m_\sigma}{2T}\right]
\,\, ,
\end{equation}
where $\langle \, \cdot \, \rangle_\xi$ denotes the average with respect
to the Gaussian measure (\ref{probmeas}).
In the high-temperature limit, $\coth [m_\sigma/2T] \rightarrow 2T/m_\sigma$,
and the variance coincides with what is known from
the classical fluctuation--dissipation relation, which is employed in most 
treatments of the subject \cite{cgtb}.
However, due to $m_\sigma \gg T$ for the range of temperatures of interest, 
this limit is not really applicable. Moreover, it would predict that the
fluctuations vanish at $T=0$, while the dissipation (\ref{etasigma}) 
persists. This certainly contradicts the fluctuation--dissipation theorem. 
The more general expression (\ref{varsig}) resolves this apparent 
contradiction, since $\coth[m_\sigma/2T] \rightarrow 1$ for $T \rightarrow 0$. 
Physically, the fluctuations at $T=0$ are quantum fluctuations originating
from the decay of the $\sigma$ into two $\pi$'s.

\begin{figure} \hspace*{3cm} 
\psfig{figure=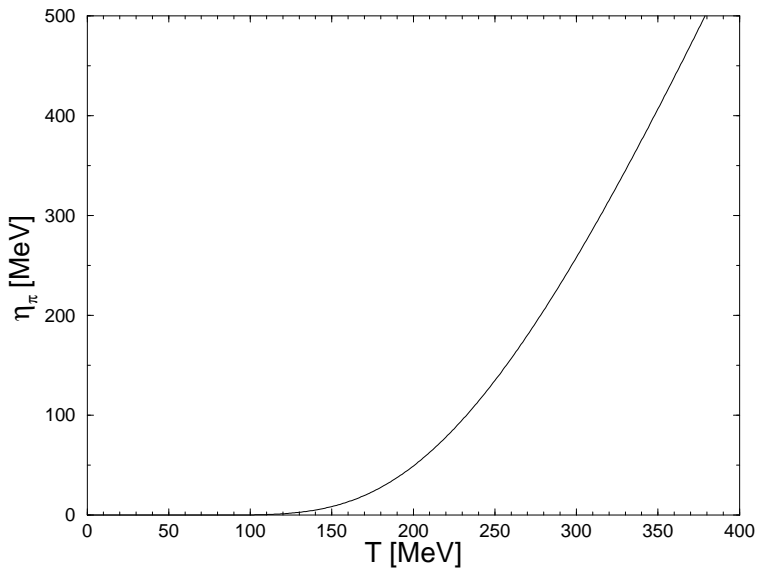,width=2.5in,height=3.3in,angle=-90}
\vspace*{-1cm}
\caption{The temperature dependence of the
dissipation coefficient $\eta_\pi$.}
\label{Figetapi}
\end{figure}

For the equation of motion of the pionic mode
$\bar{\pi}_a(t) = \bar{\pi}_a(t,{\bf 0})/V$ one obtains from eq.\ (\ref{74}):
\begin{equation}
\left[ \partial_t^2 + \tilde{m}_\pi^2(T) \right]\, \bar{\pi}_a(t)
+  \frac{8\, \lambda\, f_\pi}{N}\, \bar{\sigma}(t)\, \bar{\pi}_a(t)   
+  \frac{4\, \lambda}{N} \left[\bar{\sigma}^2(t) + 
\bar{\mbox{\boldmath $\pi$}}(t) \cdot \bar{\mbox{\boldmath $\pi$}}(t) 
\right] \bar{\pi}_a(t)
+ \eta_\pi\, \partial_t \,\bar{\pi}_a(t) = \xi_{\pi_a}(t)\,\, ,
\label{eqofmopi}
\end{equation} 
where
\begin{eqnarray}
\lefteqn{\tilde{m}^2_\pi(T) = m_\pi^2 (T)
+ 4\, \left( \frac{4\, \lambda\, f_\pi}{N} \right)^2 \, {\rm P}
\int \frac{{\rm d} \omega}{2 \pi} \, \frac{{\cal M}_2^{(\sigma\pi)}
(\omega,{\bf 0})}{m_\pi-\omega}  =  m_\pi^2 + \frac{4 \, \lambda}{N}\, \left[
D_{++}^{(\sigma)}(0) + (N+1)D_{++}^{(\pi)}(0) \right] }\nonumber \\
& + & \frac{2}{\pi^2}\, \left( \frac{4\, \lambda\, f_\pi}{N} \right)^2 
\left[ \int_{m_\sigma}^\infty {\rm d}E\, \frac{\sqrt{E^2-m_\sigma^2}}{
 m_\sigma^4-4 E^2 m_\pi^2 } \, \frac{m_\sigma^2}{e^{E/T}-1} -
\int_{m_\pi}^\infty {\rm d}E\, \frac{\sqrt{E^2-m_\pi^2}}{
(m_\sigma^2-2m_\pi^2)^2-4 E^2 m_\pi^2} \, 
\frac{m_\sigma^2 - 2m_\pi^2}{e^{E/T}-1}
\right] \,\,, \label{pimass}
\end{eqnarray}
and (cf.\ Appendix F)
\begin{eqnarray}
\eta_\pi & = & \left(\frac{4\,\lambda\,f_\pi}{N} \right)^2
4\, \frac{ {\cal M}_2^{(\sigma\pi)}(m_\pi, {\bf 0}) }{ 2\, m_\pi} \nonumber \\
& = & \left(\frac{4\, \lambda\, f_\pi}{N} \right)^2 
\frac{m_\sigma^2}{4\pi\, m_\pi^3}\, \sqrt{1 - \frac{4\, m_\pi^2}{m_\sigma^2}}
\,\, \frac{1-\exp[-m_\pi/T]}{1-\exp[-m_\sigma^2/2m_\pi T]} \, 
\frac{1}{\exp[(m_\sigma^2-2m_\pi^2)/2 m_\pi T] -1} \,\, .
\end{eqnarray}
Damping of pions arises due to the processes $\pi\,\pi \rightarrow \sigma$
and $\pi\, \sigma \rightarrow \pi$, where one $\pi$ and the $\sigma$
in the incoming channel come from the heat bath of hard modes.
The temperature dependence of $\eta_\pi$ is shown in Fig.\ \ref{Figetapi}.
Note that $\eta_\pi$ is small as compared to $\eta_\sigma$
in the temperature range of interest. The reason is the large $\sigma$ mass
which strongly suppresses the phase space for the processes $\pi\, \pi
\rightarrow \sigma$ and $\pi \, \sigma \rightarrow \pi$ (remember that
one pion is at rest).
Obviously, $\eta_\pi$ vanishes at $T=0$, because then there is no
background of hard $\pi$'s or $\sigma$'s. It also vanishes in the chiral limit
$m_\pi \rightarrow 0$ for {\em all\/} temperatures: truly massless
Goldstone bosons are not damped (at least to first order in $\lambda$ and
for ${\bf k}=0$). 
Correspondingly, the associated noise $\xi_{\pi_a}$ has to vanish as well.
This can be explicitly seen from the expression for the variance
of $\xi_{\pi_a}(t)$ in white-noise approximation (cf.\ Appendix G):
\begin{equation}
\langle \xi_{\pi_a}(t) \, \xi_{\pi_b}(t') \rangle_\xi
= \frac{1}{V}\, \delta_{ab}\, \delta(t-t')\, \eta_\pi\, m_\pi\, \coth
\left[ \frac{m_\pi}{2T} \right]\,\,.
\end{equation}
Finally, it should be mentioned that for the spatially homogeneous solutions
under consideration, cross correlations between different noise terms
vanish (cf.\ Appendix G).

\vspace*{0.7cm}

\subsection{Static, spatially 
homogeneous solutions of the classical equations of motion}

The equations of motion (\ref{eqofmosig}) and (\ref{eqofmopi})
have the following interesting consequence. Let us
consider a small perturbation $(\delta \bar{\sigma}, \delta 
\bar{\mbox{\boldmath $\pi$}})$ of the vacuum ground state $(\bar{\sigma},
\bar{\mbox{\boldmath $\pi$}})^{\rm vac} = 0$ and let us consider
the equations of motion only to lowest order in 
$\lambda\, f_\pi \sim \lambda^{1/2}$. All terms of order
$\lambda$ as well as higher order terms in $\delta \bar{\sigma}$
and $\delta \bar{\mbox{\boldmath $\pi$}}$ will be neglected.
To this lowest order in $\lambda$, the variances of the noise terms vanish, 
cf.\ eqs.\ (\ref{variance}), 
thus also $\xi_\sigma = \xi_{\pi_a}\equiv 0$, 
and the equations of motion have the solution
\begin{equation} \label{delsig}
\delta \bar{\sigma} = - \frac{4\, \lambda\, f_\pi}{N\, m_\sigma^2}\,
\left[ 3\, D_{++}^{(\sigma)}(0) + (N-1)\, D_{++}^{(\pi)}(0) \right]\,\,,
\,\,\,\,  \delta \bar{\mbox{\boldmath $\pi$}} = 0\,\, .
\end{equation}
This {\em static\/} solution
corresponds to a {\em constant, temperature-dependent
shift\/} of the ground state $\bar{\sigma}^{\rm vac}=0 
\rightarrow \bar{\sigma}^{\rm vac} = \delta \bar{\sigma}(T) $. 
This shift is identical to the well-known change of the
vacuum ground state at finite temperatures \cite{tytpis}. To see this,
let us consider the chiral limit $m_\pi \rightarrow 0$, where (for
$N=4$) $\lambda = m_\sigma^2/(2\, f_\pi^2)$, and let us assume that
$T \ll m_\sigma$, such that $D_{++}^{(\sigma)}(0)$ may be
neglected in comparison to $D_{++}^{(\pi)}(0)$ in eq.\ (\ref{delsig}).
Then one obtains with eq.\ (\ref{HTL}):
\begin{equation}
\phi_0^{\rm vac} \simeq 
f_\pi \left( 1 - \frac{T^2}{8\, f_\pi^2} \right) \,\, ,
\end{equation}
in agreement with eq.\ (30) of \cite{tytpis}.

Let us now resolve the apparent contradiction of Goldstone's theorem
mentioned earlier.
It has to be shown that in the chiral limit $m_\pi \rightarrow 0$,
the mass parameter for the static and homogeneous solution of eq.\
(\ref{eqofmopi}) vanishes. In the present perturbative
treatment, which is accurate to order $\lambda$, one may employ the static
solution (\ref{delsig}) for the $\sigma$ and $\pi$ fields
in the equation of motion (\ref{eqofmopi}) in all terms which are
proportional to at least one power of $\lambda\, f_\pi \sim \lambda^{1/2}$. 
As shown above, the noise
and fluctuation terms for the $\pi$ field vanish in the chiral limit.
Thus, to order $\lambda$, one is left with the equation of motion:
\begin{equation} \label{eqofmopi2}
\hat{m}^2_\pi(T) \, \bar{\pi}_a = 0 \,\, ,
\end{equation}
where
\begin{eqnarray}
\hat{m}^2_\pi(T) & \equiv & \lim _{m_\pi \rightarrow 0} \left\{
m^2_\pi + \frac{4 \, \lambda}{N}\, \left[ D_{++}^{(\sigma)}(0) 
+ (N+1)\, D_{++}^{(\pi)}(0) - \frac{8 \, \lambda \, f_\pi^2}{N\, m_\sigma^2}
\left( 3\, D_{++}^{(\sigma)}(0) + (N-1)\, D_{++}^{(\pi)}(0) \right) 
\right. \right. \nonumber \\
&   & \left. \left. \hspace*{3cm} + \, \frac{16\, \lambda\, f_\pi^2}{N} \,\,
{\rm P} \int \frac{{\rm d}\omega}{2 \pi}\,
\frac{{\cal M}_2^{(\sigma\pi)}(\omega, {\bf 0})}{m_\pi-\omega}
\right] \right\}\,\,.
\end{eqnarray}
In the limit $m_\pi \rightarrow 0$, the last integral
can be easily calculated (cf.\ eq.\ (\ref{pimass})) to yield
\begin{eqnarray}
\hat{m}_\pi^2(T) & = & \frac{4 \, \lambda}{N}\, \left[ 
D_{++}^{(\sigma)}(0) + (N+1)\,D_{++}^{(\pi)}(0) - 
\frac{8 \, \lambda \, f_\pi^2}{N\, m_\sigma^2}
\left( 3\, D_{++}^{(\sigma)}(0) + (N-1)\, D_{++}^{(\pi)}(0) \right)
 \right.  \nonumber \\
&  & \left. \hspace*{4.9cm} +\,
\frac{16\, \lambda\, f_\pi^2}{N\,m_\sigma^2}
\left(D_{++}^{(\sigma)}(0) -D_{++}^{(\pi)}(0) \right) \right]\,\, .
\end{eqnarray}
In the chiral limit,  the coupling
constant $\lambda \rightarrow m_\sigma^2/ (2\, f_\pi^2)$, and all
contributions to the thermal pion mass cancel, 
$\hat{m}^2_\pi(T) \rightarrow 0$, which completes the proof that
Goldstone's theorem remains valid. 
In other words, the vanishing of $\hat{m}_\pi(T)$ implies 
that there exist non-trivial massless, static, homogeneous solutions 
to the equation of motion (\ref{eqofmopi2}), which are, of course,
nothing but the $N-1$ Goldstone bosons. Note that the above cancellation of 
terms is equivalent to the arguments presented in \cite{kapusta}.

\section{Numerical solutions}

In this section, numerical solutions of the classical equations of motion
(\ref{eqofmosig}) and (\ref{eqofmopi}) are presented, to assess whether
DCC's can form in the presence of dissipation and fluctuation.
First, note that the sign of the quantity
\begin{equation} \label{meffpi}
\left[ m_\pi^{\rm eff}(t,T)\right]^2 = 
\tilde{m}_\pi^2(T) +  \frac{8\, \lambda\, f_\pi}{N}\, \bar{\sigma}(t)   
+  \frac{4\, \lambda}{N} \left[\bar{\sigma}^2(t) + 
\bar{\mbox{\boldmath $\pi$}}(t) \cdot \bar{\mbox{\boldmath $\pi$}}(t) \right] 
\end{equation}
determines the time evolution of the pion fields in the absence of
fluctuations or dissipation. For $\left[m_\pi^{\rm eff}(t,T)\right]^2>0$,
the pion fields simply perform oscillations with a constant amplitude, 
while for $\left[m_\pi^{\rm eff}(t,T)\right]^2<0$ their amplitude grows 
exponentially.
This exponential growth leads to large amplitude oscillations and,
in turn, to a large number of pions in a given charge state (say $\pi_3
\equiv \pi_0$). The characteristic probability 
to find the $\mbox{\boldmath $\pi$}$ field aligned
in a certain direction in isospin space leads to a probability
for the ratio $R$ of neutral to all pions of $P(R) \sim 1/\sqrt{R}$. 
This characteristic behavior was suggested as experimental
signature for the formation of DCC's \cite{DCC}.
Note that $\left[m_\pi^{\rm eff}(t,T)\right]^2$ can be negative
only if the second term in (\ref{meffpi}), i.e., the $\bar{\sigma}$ field, 
is large and negative.
An explicit calculation confirms that the mass parameter 
$\tilde{m}^2_\pi(T)$, eq.\ (\ref{pimass}), is an increasing function of $T$. 
Therefore, exponential growth of the pion fields (and thus formation of DCC's) 
is most likely (and fastest) at $T=0$. The following considerations will 
therefore be restricted to the case of vanishing temperature.
In that case, however, there is no heat bath. Dissipation and fluctuation
arise solely from the decay $\sigma \rightarrow \pi \pi$.

The equations of motion (\ref{eqofmosig}), (\ref{eqofmopi})
are solved with a standard fourth-order Runge--Kutta method. The time step
width was chosen to be $\Delta t = 0.002 \times 2 \pi/m_\sigma$. The
value of the fluctuating field $\xi_\sigma$ is a Gaussian random number 
with variance $\eta_\sigma \, m_\sigma/(V\, \Delta t)$ \cite{reinhard}. 
It is chosen at the beginning of each time step and kept fixed during 
the Runge--Kutta step. 

\vspace*{-1cm}
\begin{figure} \hspace*{3cm} 
\psfig{figure=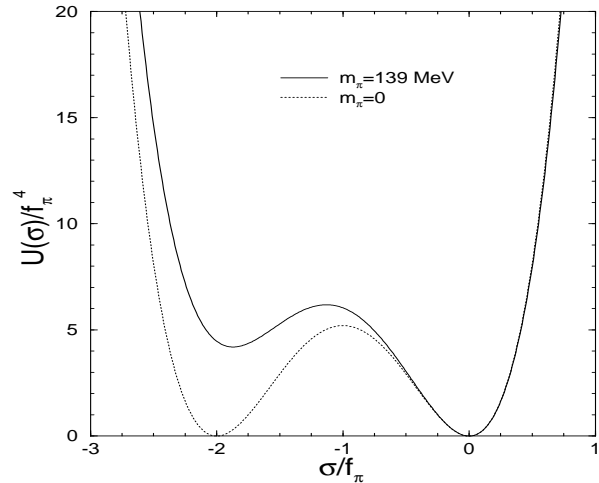,width=3.5in,height=3in,angle=-90}
\vspace*{0.5cm}
\caption{The potential $U(\sigma)$ (in units of $f_\pi^4$)
as function of $\sigma$ (in units of $f_\pi$). The solid line is
for $m_\pi=139$ MeV and the dotted line for the chiral limit $m_\pi =0$.}
\label{Figpotential}
\end{figure}

In Fig.\ \ref{Figpotential} the potential 
\begin{equation} \label{potential}
U(\sigma) = \frac{m_\sigma^2}{2}\, \sigma^2
+ \frac{4\, \lambda\, f_\pi}{N}\, \sigma^3 +
 \frac{\lambda}{N}\, \sigma^4
\end{equation}
is shown for $N=4$, $m_\sigma = 600$ MeV, and $\lambda = (m_\sigma^2 - m_\pi^2)
/(2\, f_\pi^2)$, where $m_\pi = 139$ MeV, $f_\pi = 93$ MeV
(solid line). This function represents a cut through
the potential energy surface of the linear sigma model at 
$\mbox{\boldmath $\pi$} = 0$, cf.\ eq.\ (\ref{linsig}).
The absolute minimum, corresponding to the ground state, 
is at $\sigma^{(1)} = 0$. There is another local minimum at 
$\sigma^{(2)} \simeq -1.87\, f_\pi$, and a local maximum
at $\sigma^{(3)} \simeq -1.13\, f_\pi$. In the chiral limit, 
$m_\pi \rightarrow 0$, $\sigma^{(2)} \rightarrow -2\, f_\pi$, 
while $\sigma^{(3)} \rightarrow 0$ (cf.\ dotted
line). Trajectories which are likely candidates for DCC formation
obviously start at $\sigma < \sigma^{(3)}$, and small non-zero
values of $\mbox{\boldmath $\pi$}$ (this is necessary because otherwise
$\mbox{\boldmath $\pi$} = 0$ remains a solution throughout the system's
evolution). 

\begin{figure} \hspace*{1cm} 
\psfig{figure=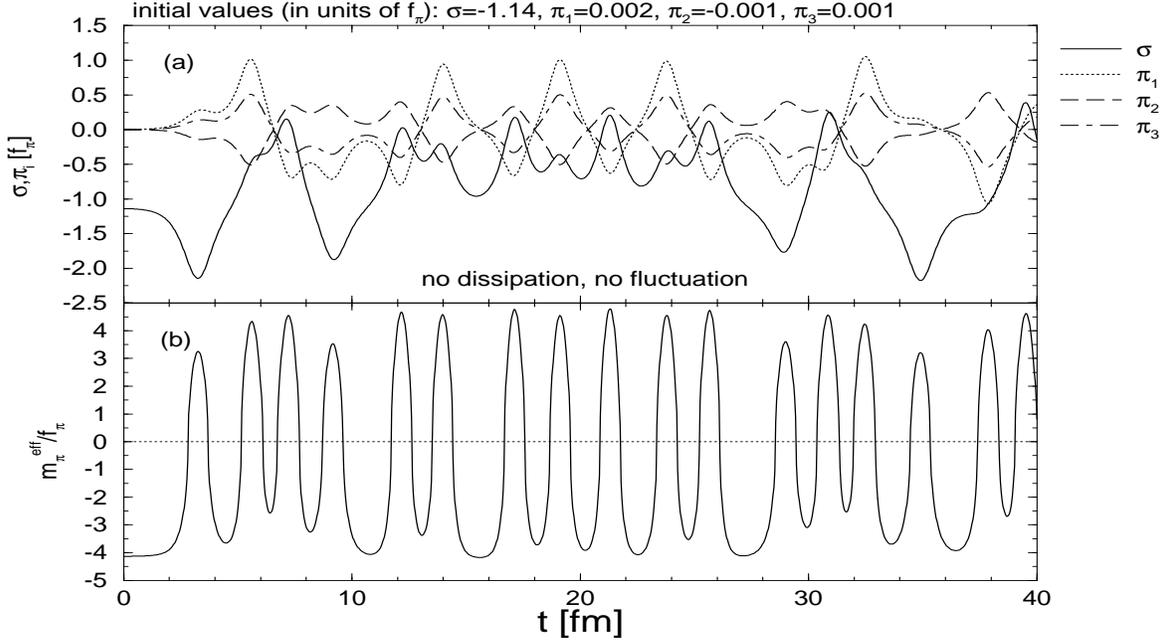,width=3.5in,height=4in,angle=-90}
\vspace*{-2cm}
\caption{(a) Classical trajectories without dissipation and
fluctuations for $\bar{\sigma}$ (solid), $\bar{\pi}_1$ (dotted),
$\bar{\pi}_2$ (dashed), and $\bar{\pi}_3$ (dash-dotted), in units of $f_\pi$.
Initial values are $\bar{\sigma} = -1.14\, f_\pi,\, \bar{\pi}_1 = 0.002\, 
f_\pi, \, \bar{\pi}_2 = - 0.001\, f_\pi,\, \bar{\pi}_3 = 0.001\,
f_\pi$.
(b) The function ${\rm sgn}\left\{[m_\pi^{\rm eff}]^2\right\}\, 
|m_\pi^{\rm eff}|$ (in units of $f_\pi$).}
\label{Figclasstraj}
\end{figure}

A representative candidate is shown in Fig.\ \ref{Figclasstraj}.
In part (a), the time evolution of $\bar{\sigma}$ and 
$\bar{\mbox{\boldmath $\pi$}}$ fields are shown for the initial conditions
$\bar{\sigma} = -1.14\, f_\pi,\, \bar{\pi}_1 = 0.002\, 
f_\pi, \, \bar{\pi}_2 = - 0.001\, f_\pi.\, \bar{\pi}_3 = 0.001\, f_\pi$.
The derivatives of the fields are taken to be zero initially. 
Dissipation and fluctuation terms have also been set to zero for the
solution shown in Fig.\ \ref{Figclasstraj}.
One observes that the $\bar{\sigma}$ field first ``rolls'' towards the minimum 
$\sigma^{(2)}$ of the potential (\ref{potential}). Since this minimum
is unstable in the direction of the $\bar{\mbox{\boldmath $\pi$}}$ fields,
the $\bar{\sigma}$ field ``rolls'' on towards the absolute minimum
$\sigma^{(1)}=0$. During this process the $\bar{\mbox{\boldmath $\pi$}}$ 
fields grow. This growth is characterized by negative values of
$\left[ m_\pi^{\rm eff}(T) \right]^2$ or of ${\rm sgn} 
\left\{ [m_\pi^{\rm eff}(T) ]^2\right\}\, | m_\pi^{\rm eff}(T) |$, as
shown in part (b) of Fig.\ \ref{Figclasstraj}. At this level, the evolution
is conservative, and the system continues to oscillate around the ground 
state, since there is no way to dissipate the ``potential energy''
associated with the chosen initial state. This potential energy is simply
converted into kinetic energy and vice versa in the course of the
evolution. This also explains why the system periodically reaches the vicinity
of the unstable minimum $\sigma^{(2)}$. This changes once dissipation
and fluctuation is taken into account (it would also change if we solved
the equations of motion in expanding geometries \cite{cgtb,randrup}).
The large-amplitude fluctuations of the pion fields make the observation
of DCC formation experimentally possible.

\vspace*{0.7cm}

\vspace*{0.7cm}
\begin{figure} \hspace*{1cm} 
\psfig{figure=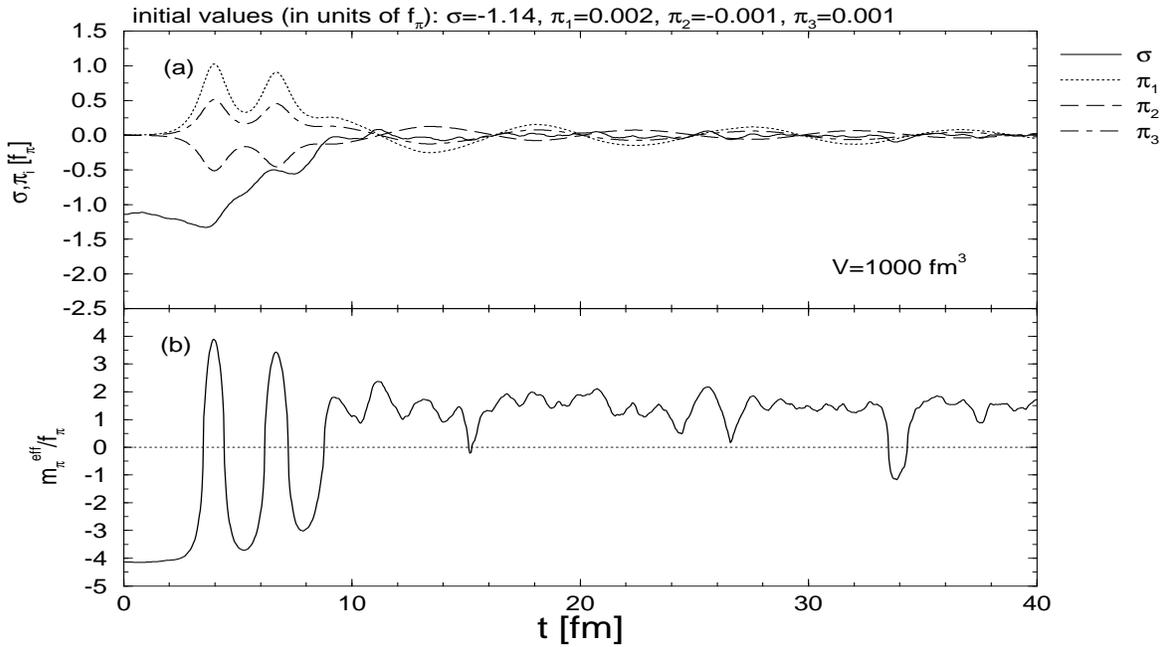,width=3.5in,height=4in,angle=-90}
\vspace*{-2cm}
\caption{(a) Trajectories with dissipation and
fluctuations for $\bar{\sigma}$ (solid), $\bar{\pi}_1$ (dotted),
$\bar{\pi}_2$ (dashed), and $\bar{\pi}_3$ (dash-dotted), in units of $f_\pi$.
Initial values are the same as in Fig.\ 8, the volume
of the system is $V=1000\, {\rm fm}^3$.
(b) The function ${\rm sgn}\left\{[m_\pi^{\rm eff}]^2\right\}\, 
|m_\pi^{\rm eff}|$ (in units of $f_\pi$).}
\label{Figdissfluc1000}
\end{figure}

Fig.\ \ref{Figdissfluc1000} (a) shows a sample trajectory including
dissipation and fluctuation terms in a comparatively large volume
$V=1000\, {\rm fm}^3$ with the same initial conditions as in Fig.\
\ref{Figclasstraj}. Dissipation and fluctuations damp the
oscillations of the fields, and they settle into the ground state.
The time scale for this to happen is, for the sample trajectory shown here,
about 10 fm. During this time, the pion fields oscillate strongly. 
Therefore, DCC formation would still be observable, if the 
system were to decouple after the first 10 fm.

This conclusion holds, however, only for the {\em one particular\/} trajectory
shown in Fig.\ \ref{Figdissfluc1000}, i.e., for one particular choice
for the time evolution of the randomly fluctuating forces $\xi_\sigma$. 
For a {\em different\/} random sequence, the time scale for damping could 
be larger or smaller. In a strict sense, one would have to average over an 
ensemble of time evolutions for $\xi_\sigma$ for a given set of initial 
conditions for the fields \cite{reinhard}. 
The result of such an averaging is, however, predictable: on the average the
fluctuations tend to cancel, and the ensemble averages of $\bar{\sigma}$ and
$\bar{\mbox{\boldmath $\pi$}}$ fields are constant in time. 
The assumed values depend, however, on the strength of the fluctuations, i.e.,
since $\xi_\sigma \sim 1/\sqrt{V}$, on the volume of the system.
In large volumes, the fluctuations are not strong enough to ``kick'' the 
$\bar{\sigma}$ field too far out of the ground state.  On the other
hand, in small volumes the fluctuations may have enough strength to 
let the $\bar{\sigma}$ field reach the unstable minimum $\sigma^{(2)}$. 
Therefore, the ensemble average of the $\bar{\sigma}$ field decreases
towards $- f_\pi$ as the volume of the system decreases (on the average,
the $\bar{\sigma}$ field tends to be anywhere between the two minima 
$\sigma^{(1)}$ and 
$\sigma^{(2)}$, i.e., on the average, close to $-f_\pi$). The
ensemble averages of the pion fields, however, are always zero
(since the potential $U(\sigma,\mbox{\boldmath $\pi$})$ is symmetric with 
respect to $\mbox{\boldmath $\pi$} \rightarrow -\mbox{\boldmath $\pi$}$). 

\begin{figure} \hspace*{1cm} 
\psfig{figure=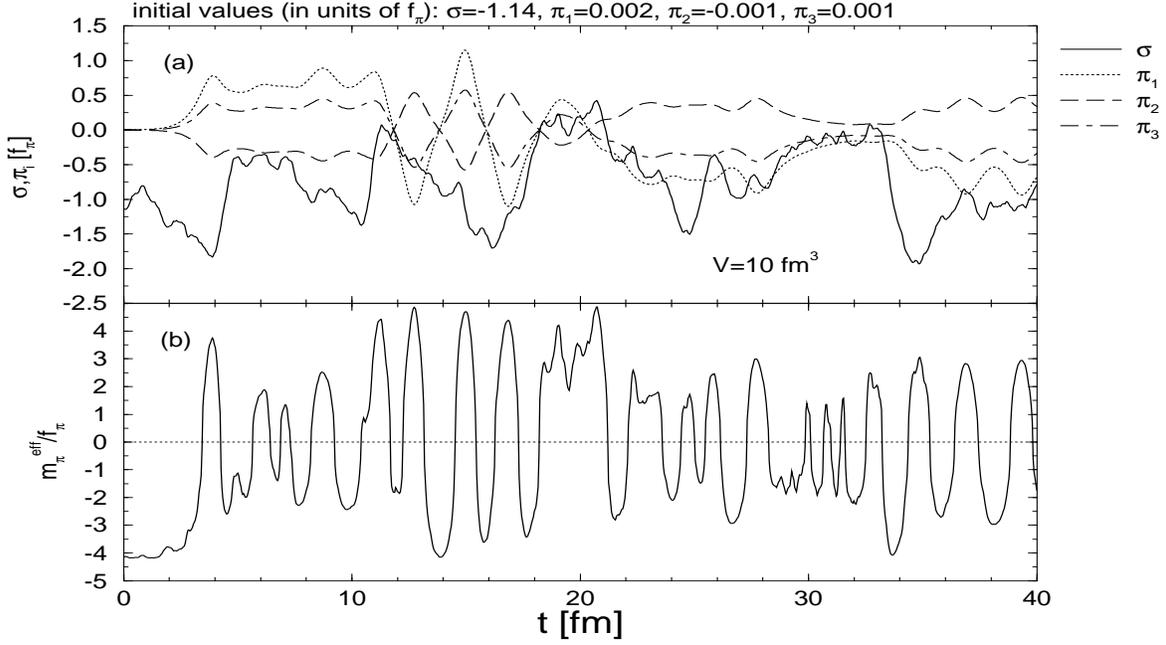,width=3.5in,height=4in,angle=-90}
\vspace*{-2cm}
\caption{(a) Trajectories with dissipation and
fluctuations for $\bar{\sigma}$ (solid), $\bar{\pi}_1$ (dotted),
$\bar{\pi}_2$ (dashed), and $\bar{\pi}_3$ (dash-dotted), in units of $f_\pi$.
Initial values are the same as in Fig.\ 8, the volume
of the system is $V=10\, {\rm fm}^3$.
(b) The function ${\rm sgn}\left\{[m_\pi^{\rm eff}]^2\right\}\, 
|m_\pi^{\rm eff}|$ (in units of $f_\pi$).}
\label{Figdissfluc10}
\end{figure}

In Fig.\ \ref{Figdissfluc10} the time evolution of the fields is shown
for a small volume $V=10\,{\rm fm}^3$, for the same initial
conditions (and the same random sequence for $\xi_\sigma$) 
as in the previous two figures. As discussed above, in this case 
fluctuations are large enough to drive the system out of the ground state 
and induce disorientation of the $\bar{\mbox{\boldmath $\pi$}}$ fields.
One could speculate that the formation of DCC's is facilitated in 
{\em smaller\/} volumes, i.e., it should be more likely to observe them
in collisions of {\em lighter\/} ions, or even $pp$--collisions. 
This could also provide an explanation for the CENTAURO events 
\cite{centauro}, where heavy ions are not likely to play any role as 
collision partner. To confirm this, however, a more detailed
investigation in an expanding geometry, with a realistic
evolution for the temperature in the collision, and including
modes with finite ${\bf k}$ (to study domain formation) is necessary
\cite{krdhr}.

\begin{figure} \hspace*{1cm} 
\psfig{figure=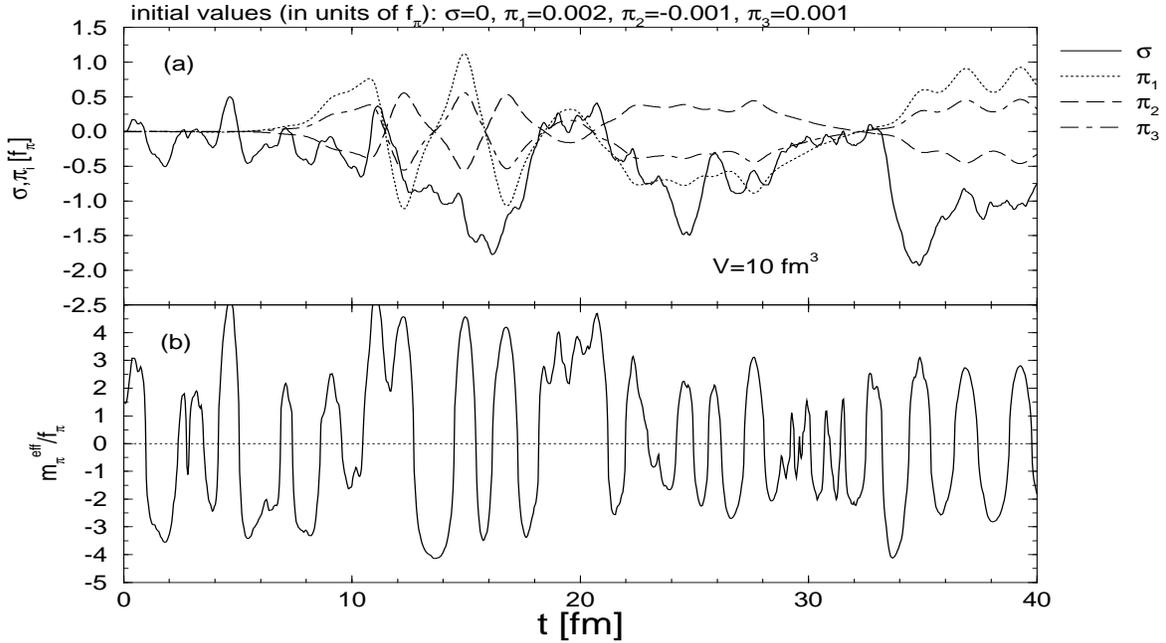,width=3.5in,height=4in,angle=-90}
\vspace*{-2cm}
\caption{(a) Trajectories with dissipation and
fluctuations for $\bar{\sigma}$ (solid), $\bar{\pi}_1$ (dotted),
$\bar{\pi}_2$ (dashed), and $\bar{\pi}_3$ (dash-dotted), in units of $f_\pi$.
Initial values are $\bar{\sigma} = 0$, $\bar{\pi}_1 = 0.002\, f_\pi$,
$\bar{\pi}_2 = -0.001\, f_\pi$, $\bar{\pi}_3 = 0.001\, f_\pi$,
the volume of the system is $V=10\, {\rm fm}^3$.
(b) The function ${\rm sgn}\left\{[m_\pi^{\rm eff}]^2\right\}\, 
|m_\pi^{\rm eff}|$ (in units of $f_\pi$).}
\label{Figdissfluc10a}
\end{figure}

The problem with large-scale fluctuations in small volumes is, however,
that one does not necessarily need to first restore chiral symmetry to
observe them. In Fig.\ \ref{Figdissfluc10a} a time evolution is shown
in a volume $V=10\,{\rm fm}^3$ for the initial condition
$\bar{\sigma} = 0,\, \bar{\pi}_1 = 0.002\, f_\pi,\,
\bar{\pi}_2 = -0.001\, f_\pi,\, \bar{\pi}_3 = 0.001\, f_\pi$, i.e., the
initial values of the $\bar{\mbox{\boldmath $\pi$}}$ fields are the same
as before, but the $\bar{\sigma}$ field is taken to be zero. The strong
fluctuations $\xi_\sigma$ drive the $\bar{\sigma}$ field out of the
ground state, and the small initial perturbations of the pion fields are
strongly enhanced over a time scale of $\simeq 10$ fm
to produce large amplitude oscillations (and thus DCC formation).
Note that the $\bar{\sigma}$ field is always subject to these volume-dependent 
fluctuations, but if one starts with the true ground state, $\bar{\sigma} = 
\bar{\mbox{\boldmath $\pi$}}=0$, as initial condition,
the pion fields are not affected and remain zero throughout the evolution
of the system. The conclusion would be that, in the presence of strong
fluctuations, DCC formation is not necessarily a signal for restoration
of chiral symmetry, a small perturbation of the ground state in the
$\mbox{\boldmath $\pi$}$ direction seems to suffice. However,
it is likely that this at first glance rather interesting phenomenon
is physically identical with (and thus indistiguishable from)
ordinary fluctuations in finite volumes.

Note that there is a minimum volume $V_{\rm min}$ below
which the dissipation coefficient $\eta_\sigma$ becomes zero. In order
to have $\eta_\sigma >0$, the $\sigma$ at rest has to be able to
decay into two pions with finite, but opposite momenta. The lowest
non-zero momentum state for a particle in box volume $V=L^3$ is
${\bf k} = (\pi/L,0,0)$. Energy conservation in the decay process
requires $2\,E_{\bf k}^{(\pi)} \equiv m_\sigma$, or $L_{\rm min} = 2\pi/
\sqrt{m_\sigma^2 - 4 m_\pi^2}$, or $V_{\rm min} \simeq 12.68\,{\rm fm}^3$,
i.e., the case $V=10\, {\rm fm}^3$ considered above is just on the order of
the physically possible minimum volume.

\section*{Acknowledgments}

The author thanks C.\ Greiner, M.\ Gyulassy, 
U.\ Heinz, E.\ Iancu, S.\ Jeon, J.\ Kapusta, T.D.\ Lee, S.\ Leupold, 
L.\ McLerran, R.\ Pisarski, K.\ Rajagopal, and D.\ Son for
valuable discussions, C.\ Greiner, B.\ M\"uller, R.\ Pisarski, and 
K.\ Rajagopal for
a critical reading of the manuscript and suggestions, and Columbia 
University's Nuclear Theory Group for access to their computing facilities.

\newpage
\vspace*{0.7cm}

\vspace*{0.7cm}

\appendix

\section{Evaluation of the interaction terms in the $O(N)$ symmetric case}

In this Appendix, the evaluation of the interaction terms ${\cal T}_a^{(i)}(t,
{\bf k})$ in linear harmonic approximation is presented 
for the $O(N)$ model in the symmetric case. Let us first define the functions
\begin{mathletters} \label{A1}
\begin{eqnarray}
{\cal M}_1 (x) & = & D_>(x) - D_<(x) \,\, ,\\
{\cal M}_2 (x) & = & D_>^2(x) - D_<^2(x) \,\, ,\\
{\cal M}_3 (x) & = & D_>^3(x) - D_<^3(x) \,\, 
\end{eqnarray}
\end{mathletters}
(called ``memory kernels'' in \cite{cgbm}),
such that with eqs.\ (\ref{Gi}), (\ref{Glarger}): 
\begin{mathletters} \label{A2}
\begin{eqnarray}
\lefteqn{ {\cal M}_1 (\omega,{\bf k}) = \Theta(|{\bf k}|-k_c)\, 
\frac{2 \pi}{2\,E_{\bf k}} \left[ \delta(\omega- E_{\bf k}) - 
\delta(\omega+E_{\bf k}) \right] \,\, ,\label{M1}}\,\,\,\,  \\
\lefteqn{{\cal M}_2 (\omega,{\bf k})  =  2 \pi \int 
\frac{{\rm d}^3{\bf p}}{(2 \pi)^3}\, \Theta(|{\bf p}|-k_c)\, 
\Theta (|{\bf k} - {\bf p}|-k_c) \, 
\frac{1}{4\, E_{\bf p}\, E_{{\bf k}- {\bf p}}} }\,\, , \nonumber \\
& \times & \left\{ \,\,\, \left( \frac{}{} \!\!
\left[1+n(E_{\bf p}) \right] \left[1+n(E_{{\bf k}- {\bf p}}) \right] 
- n(E_{\bf p}) n(E_{{\bf k}- {\bf p}}) \, \right) 
\left[ \frac{}{} \delta(\omega- E_{\bf p} - E_{{\bf k}- {\bf p}} ) - 
    \delta(\omega+ E_{\bf p} + E_{{\bf k}- {\bf p}}) \, \right] \right. 
\nonumber \\
&   &  \left.
+ \left(\frac{}{} \!\! \left[1+n(E_{\bf p}) \right] n(E_{{\bf k}- {\bf p}}) 
- n(E_{\bf p}) \left[1+ n(E_{{\bf k}- {\bf p}}) \right]\,  \right) 
\left[\frac{}{} \delta(\omega-E_{\bf p} + E_{{\bf k}- {\bf p}}) - 
 \delta(\omega+E_{\bf p} - E_{{\bf k}- {\bf p}}) \, \right]
\right\}\, , \label{M2} \\
\lefteqn{ {\cal M}_3 (\omega,{\bf k}) = 2 \pi \int \frac{{\rm d}^3{\bf p}
\,{\rm d}^3{\bf q}}{(2 \pi)^6}
\, \Theta(|{\bf p}|-k_c)\, \Theta(|{\bf q}|-k_c) \, 
\Theta (|{\bf k} - {\bf p}-{\bf q}|-k_c) \, \frac{1}{
8\, E_{\bf p}\, E_{\bf q}\, E_{{\bf k}- {\bf p}-{\bf q}}} } \,\,\,\, 
\nonumber \\
& \times & \left\{ \,\,\, \left( \frac{}{} \!\!
\left[1+n(E_{\bf p}) \right] \left[1+n(E_{\bf q}) \right] 
\left[1+n(E_{{\bf k}- {\bf p}-{\bf q}}) \right] 
- n(E_{\bf p})\, n(E_{\bf q})\, n(E_{{\bf k}- {\bf p}-{\bf q}})\, \right) 
\right. \nonumber \\
&   & \hspace*{0.5cm} \times \left[\frac{}{} 
\delta(\omega-E_{\bf p}-E_{\bf q} - E_{{\bf k}- {\bf p}-{\bf q}}) - 
\delta(\omega+E_{\bf p}+E_{\bf q} + E_{{\bf k}- {\bf p}-{\bf q}})\, \right] 
\nonumber \\
&   & + \, 3\, \left( \frac{}{} \!\!
\left[1+n(E_{\bf p}) \right] \left[1+n(E_{\bf q}) \right] 
\,n(E_{{\bf k}- {\bf p}-{\bf q}}) 
- n(E_{\bf p})\, n(E_{\bf q})\, \left[1+n(E_{{\bf k}- {\bf p}-{\bf q}})
\right]\, \right) 
\nonumber \\
&   & \left. \hspace*{0.5cm} \times \left[\frac{}{} 
\delta(\omega-E_{\bf p}-E_{\bf q} + E_{{\bf k}- {\bf p}-{\bf q}}) - 
\delta(\omega+E_{\bf p}+E_{\bf q} - E_{{\bf k}- {\bf p}-{\bf q}})\, \right] 
\right\}\,\, . \label{M3}
\end{eqnarray}
\end{mathletters}
Note that all functions fulfill the symmetry relation ${\cal M}_i
(\omega, {\bf k}) = - {\cal M}_i(-\omega,{\bf k})$.
Then, eqs.\ (\ref{dissipation}) can be written as
\begin{mathletters}
\begin{eqnarray}
{\cal T}^{(1)}_a(x) & \equiv &\sum_b \left[ \underline{\bar{\varphi}}(x)
\cdot \underline{\bar{\varphi}}(x)\, \delta_{ab} + 2\, \bar{\varphi}_a(x)\,
\bar{\varphi}_b(x) \right] \int \frac{{\rm d}^3 {\bf k}}{(2 \pi)^3}\,
e^{i\, {\bf k} \cdot {\bf x}} \, {\cal Y}_b^{(1)} (x_0,{\bf k})\,\, , \\
{\cal T}^{(2)}_a(x) & \equiv & \sum_{b,c} \left[ (N+4) \bar{\varphi}_a(x) \,
\delta_{bc} + 4\, \bar{\varphi}_b(x)\, \delta_{ac} \right] 
\int \frac{{\rm d}^3 {\bf k}}{(2 \pi)^3}\,
e^{i\, {\bf k} \cdot {\bf x}} \, {\cal Y}_{bc}^{(2)} (x_0,{\bf k})\,\, , \\
{\cal T}^{(3)}_a(x) & \equiv & 2\, (N+2)
\int \frac{{\rm d}^3 {\bf k}}{(2 \pi)^3}\,
e^{i\, {\bf k} \cdot {\bf x}} \, {\cal Y}_{a}^{(3)} (x_0,{\bf k})\,\, ,
\end{eqnarray}
\end{mathletters}
where (for $t_i \rightarrow - \infty$):
\begin{mathletters}
\begin{eqnarray}
{\cal Y}_b^{(1)} (x_0,{\bf k}) & \equiv &
\int \frac{{\rm d} \omega}{2 \pi}\, {\cal M}_1(\omega,{\bf k})
\int \frac{{\rm d}^3{\bf p}\, {\rm d}^3 {\bf q}}{(2 \pi)^6}\,
\Theta(k_c - | {\bf p}|)\, \Theta(k_c-|{\bf q}|)\,
\Theta(k_c -|{\bf k}- {\bf p} - {\bf q}|) \nonumber \\ 
& \times & 
i \int_0^{\infty} {\rm d} \tau\, e^{-i \omega  \tau}\, 
\bar{\varphi}_b(x_0-\tau,{\bf k} - {\bf p}-{\bf q})\,\, 
\underline{\bar{\varphi}}(x_0 - \tau,{\bf p}) \cdot
\underline{\bar{\varphi}}(x_0 - \tau,{\bf q})\,\, , \\
{\cal Y}_{bc}^{(2)} (x_0,{\bf k}) & \equiv &
\int \frac{{\rm d} \omega}{2 \pi}\, {\cal M}_2(\omega,{\bf k})
\int \frac{{\rm d}^3{\bf p}}{(2 \pi)^3}\,
\Theta(k_c - | {\bf p}|)\, \Theta(k_c -|{\bf k}- {\bf p}|) \nonumber \\ 
& \times & 
i \int_0^{\infty} {\rm d} \tau\, e^{-i \omega  \tau}\, 
\bar{\varphi}_b(x_0-\tau,{\bf p})\,
\bar{\varphi}_c(x_0 - \tau,{\bf k} -{\bf p}) \,\, , \\
{\cal Y}_{a}^{(3)} (x_0,{\bf k}) & \equiv & \int \frac{{\rm d}\omega}{2 \pi}
\, {\cal M}_3(\omega , {\bf k}) \, \Theta(k_c-|{\bf k}|) \,
i \int_0^\infty {\rm d} \tau \, e^{-i \omega \tau}\, \bar{\varphi}_a
(x_0 - \tau, {\bf k})\,\, .
\end{eqnarray}
\end{mathletters}
Further evaluation is possible by employing the so-called
linear harmonic approximation \cite{cgbm} for the time evolution of the
classical fields:
\begin{equation} \label{linharm}
\underline{\bar{\varphi}}(t-\tau,{\bf k}) \simeq
\cos(E_{\bf k} \tau) \, \underline{\bar{\varphi}}(t,{\bf k})
- \frac{\sin(E_{\bf k} \tau)}{E_{\bf k}}
\,\partial_t\, \underline{\bar{\varphi}}(t, {\bf k}) \,\,.
\end{equation}
This eliminates the $\tau$ integrals over the history of the classical
fields, rendering the final expressions local in time.
Using the symmetry of ${\cal M}_i$ under $\omega
\rightarrow - \omega$ and the relation
\begin{equation} \label{A7}
\int_0^\infty {\rm d}\tau\, e^{i\, (x-\omega)\, \tau}
= i\, {\rm P} \frac{1}{x-\omega} + \pi\, \delta(x-\omega)\,\,,
\end{equation}
where P denotes the principal value, one obtains:
\begin{eqnarray}
\lefteqn{{\cal Y}_b^{(1)}(t,{\bf k})  \simeq  \int
\frac{{\rm d}^3{\bf p}\, {\rm d}^3 {\bf q}}{(2 \pi)^6}\,
\Theta(k_c - | {\bf p}|)\, \Theta(k_c-|{\bf q}|)\,
\Theta(k_c -|{\bf k}- {\bf p} - {\bf q}|)} \nonumber \\ 
& \times & \left\{
- \frac{1}{4} \,{\rm P}\, \int \frac{{\rm d}\omega}{2\pi}\,
{\cal M}_1(\omega,{\bf k})\, \left[ \frac{1}{E_1+E_2+E_3-\omega} + 
\frac{2}{E_1+E_2-E_3-\omega} + \frac{1}{E_1-E_2-E_3-\omega} \right]\, 
\right. \nonumber \\
&   & \hspace*{7cm} \times \, 
\bar{\varphi}_b(t,{\bf k} - {\bf p}-{\bf q})\,\, 
\underline{\bar{\varphi}}(t,{\bf p}) \cdot
\underline{\bar{\varphi}}(t,{\bf q})  \nonumber \\
&    & +
\frac{1}{4\, E_2\, E_3} \,{\rm P}\, \int \frac{{\rm d}\omega}{2\pi}\,
{\cal M}_1(\omega,{\bf k})\, \left[ \frac{1}{E_1+E_2 + E_3-\omega} - 
\frac{2}{E_1+E_2-E_3-\omega} + \frac{1}{E_1-E_2-E_3-\omega} \right]
\nonumber \\
&    & \hspace*{7cm} \times \,
\bar{\varphi}_b(t,{\bf k} - {\bf p}-{\bf q})\,\, 
\partial_t \, \underline{\bar{\varphi}}(t,{\bf p}) \cdot
\partial_t\, \underline{\bar{\varphi}}(t,{\bf q}) \nonumber \\
&    & + \frac{1}{2\, E_1\, E_3} 
\,{\rm P}\,\int \frac{{\rm d}\omega}{2\pi}\,
{\cal M}_1(\omega,{\bf k})\,  \left[ \frac{1}{E_1+E_2+E_3-\omega} - 
\frac{1}{E_1+E_2 - E_3-\omega} + \frac{1}{E_1-E_2+E_3-\omega} \right.
\nonumber \\
&    &\hspace*{4.6cm} - \left. \frac{1}{E_1-E_2- E_3-\omega} \right] 
\, \partial_t\, \bar{\varphi}_b(t,{\bf k} - {\bf p}-{\bf q})\,\, 
\underline{\bar{\varphi}}(t,{\bf p}) \cdot
\partial_t\, \underline{\bar{\varphi}}(t,{\bf q}) \nonumber \\
&    & - \frac{1}{4\,E_3} \left[ 
{\cal M}_1(E_1 +E_2+ E_3,{\bf k}) - {\cal M}_1(E_1 +E_2 - E_3,{\bf k})
+ {\cal M}_1(E_1 -E_2 + E_3,{\bf k}) -{\cal M}_1(E_1 -E_2 - E_3,{\bf k}) 
\right] \nonumber \\
&    & \hspace*{7cm} \times \,
\bar{\varphi}_b(t,{\bf k} - {\bf p}-{\bf q})\,\, 
\underline{\bar{\varphi}}(t,{\bf p}) \cdot
\partial_t\, \underline{\bar{\varphi}}(t,{\bf q})  \nonumber \\
&   & - \frac{1}{8\,E_1} \left[ 
{\cal M}_1(E_1 +E_2 + E_3,{\bf k}) +2\,{\cal M}_1(E_1 +E_2 - E_3,{\bf k})
+ {\cal M}_1(E_1 -E_2 - E_3,{\bf k})
\right] \nonumber \\
&    & \hspace*{7cm} \times \,
\partial_t\, \bar{\varphi}_b(t,{\bf k} - {\bf p}-{\bf q})\,\, 
\underline{\bar{\varphi}}(t,{\bf p}) \cdot
\underline{\bar{\varphi}}(t,{\bf q})  \nonumber \\
&    & + \frac{1}{8\,E_1\, E_2\, E_3} 
\left[ {\cal M}_1(E_1 +E_2 + E_3,{\bf k}) -2\,{\cal M}_1(E_1+E_2-E_3,{\bf k})
+ {\cal M}_1(E_1 -E_2 - E_3,{\bf k})
\right] \nonumber \\
&    & \hspace*{7cm} \times \, 
\left. \partial_t\, \bar{\varphi}_b(t,{\bf k} - {\bf p}-{\bf q})\,\, 
\partial_t\, \underline{\bar{\varphi}}(t,{\bf p}) \cdot
\partial_t\, \underline{\bar{\varphi}}(t,{\bf q}) \frac{}{}\right\} \,\, ,
\end{eqnarray}
where $E_1 \equiv E_{{\bf k}-{\bf p}-{\bf q}},\,
E_2 \equiv E_{\bf p},\, E_3 \equiv E_{\bf q}$. For $N=1$, this
expression reduces (up to a constant prefactor due to the difference 
between ${\cal M}^{(e)}$ and ${\cal M}_1$)
to eq.\ (50) of \cite{cgbm}. Furthermore,
\begin{eqnarray}
{\cal Y}_{bc}^{(2)} (t,{\bf k}) & \simeq &
\int \frac{{\rm d}^3{\bf p}}{(2 \pi)^3}\,
\Theta(k_c - | {\bf p}|)\, \Theta(k_c -|{\bf k}- {\bf p}|) \nonumber \\ 
& \times & 
\left\{- \frac{1}{2} \, {\rm P} 
\int \frac{{\rm d} \omega}{2 \pi}\, {\cal M}_2(\omega,{\bf k})
\left[ \frac{1}{E_1+E_2-\omega} 
+ \frac{1}{E_1 -E_2-\omega} \right]\, \bar{\varphi}_b(t,{\bf p})\,
\bar{\varphi}_c(t, {\bf k} - {\bf p}) \right. \nonumber \\
&    & +
\frac{1}{2\, E_1\, E_2}\, {\rm P} \int \frac{{\rm d} \omega}{2 \pi}\, 
{\cal M}_2(\omega,{\bf k}) \left[ \frac{1}{E_1+E_2-\omega} 
- \frac{1}{E_1 -E_2-\omega} \right]\, 
\partial_t\, \bar{\varphi}_b(t,{\bf p})\,
\partial_t\, \bar{\varphi}_c(t, {\bf k} - {\bf p}) \nonumber \\
&    & -\, \frac{1}{4\, E_1} \, \left[ {\cal M}_2(E_1 +E_2,{\bf k})
+ {\cal M}_2(E_1 - E_2,{\bf k}) \right]\, \bar{\varphi}_b(t,{\bf p})\,
\partial_t\, \bar{\varphi}_c(t, {\bf k} - {\bf p}) \nonumber \\
&    & -\left.  \frac{1}{4\, E_2} \, \left[ {\cal M}_2 (E_1 +E_2,{\bf k})
- {\cal M}_2(E_1 - E_2,{\bf k}) \right]\, 
\partial_t\, \bar{\varphi}_b(t,{\bf p})\,
\bar{\varphi}_c(t, {\bf k} - {\bf p}) 
\frac{}{} \right\} \,\, ,
\end{eqnarray}
where $E_1 \equiv E_{{\bf k}- {\bf p}},\, E_2 \equiv E_{\bf p}$. For $N=1$,
this is equivalent to eq.\ (49) of \cite{cgbm}. (There is a factor of
2 missing in front of the last term of that equation.)
Finally,
\begin{equation}
{\cal Y}_{a}^{(3)} (t,{\bf k}) \simeq - \Theta(k_c - |{\bf k}|) \left[
\,  {\rm P}\int \frac{{\rm d}\omega}{2 \pi}\,
\frac{{\cal M}_3(\omega, {\bf k})}{E_{\bf k}-\omega} 
\, \bar{\varphi}_a(t, {\bf k}) +  
\frac{{\cal M}_3(E_{\bf k}, {\bf k})}{2\, E_{\bf k}} \,
\partial_t\,\bar{\varphi}_a (t, {\bf k}) \right]\,\,,
\end{equation}
which is equivalent to eq.\ (48) of \cite{cgbm}.
Then, the final expressions for the functions ${\cal T}_a^{(i)}(t,{\bf k})$
read:
\begin{mathletters}
\begin{eqnarray}
{\cal T}_a^{(1)}(t,{\bf k}) & \simeq & \int
\frac{{\rm d}^3{\bf p}\, {\rm d}^3 {\bf q}}{(2 \pi)^6}\,
\Theta(k_c - | {\bf p}|)\, \Theta(k_c-|{\bf q}|)\, \sum_b
\left[ \underline{\bar{\varphi}} (t,{\bf p}) \cdot
 \underline{\bar{\varphi}} (t,{\bf q})\, \delta_{ab}
+ 2\, \bar{\varphi}_a(t,{\bf p})\, \bar{\varphi}_b(t,{\bf q}) \right] 
\nonumber \\
&    & \hspace*{7cm} \times \, 
{\cal Y}_b^{(1)}(t,{\bf k}-{\bf p}-{\bf q}) \,\, , \\
{\cal T}_a^{(2)}(t,{\bf k}) & \simeq & \int
\frac{{\rm d}^3{\bf p}}{(2 \pi)^3}\,\Theta(k_c - | {\bf p}|)\,
\sum_{b,c} \left[ (N+4)\,\bar{\varphi}_a(t,{\bf p})\, \delta_{bc}
+ 4\, \bar{\varphi}_b(t,{\bf p}) \, \delta_{ac} \right]\,
{\cal Y}_{bc}^{(2)} (t,{\bf k} -{\bf p})\,\, ,\\
{\cal T}_a^{(3)}(t,{\bf k}) & \simeq & 
2\, (N+2)\, {\cal Y}_a^{(3)}(t,{\bf k}) \,\,.
\end{eqnarray}
\end{mathletters}

\section{The dissipation coefficient in the $O(N)$ symmetric case}

In order to derive the dissipation coefficient 
(\ref{disscoeff}) one has to compute the function
${\cal M}_3(m,{\bf 0})$.
In the limit $k_c \rightarrow 0$ and for $\omega = m>0$, ${\bf k} =0$, 
one first observes that the $\delta$ functions corresponding to the 
decay of one particle into three and the reverse reaction 
(the second and third line in eq.\ (\ref{M3}))
have no support in this kinematic range. The only remaining contribution
comes from the scattering of the particle at rest with a particle
from the heat bath of hard modes (the
last two lines in eq.\ (\ref{M3})). Since the particles from the
heat bath are supposed to be in thermal equilibrium, $n(x) =
[e^{x/T}-1]^{-1}$, the reaction rates observe the detailed
balance criterion, or in other words, with the energy-conserving
$\delta$ functions one can rewrite the contribution
from the reverse reaction in terms of a factor $e^{-\omega/T}$ times
the original reaction rate. Then,
\begin{eqnarray}
{\cal M}_3(m,{\bf 0}) & = & \frac{3\pi}{4}\,
\left( 1 - \exp[-m/T] \right)\, \int 
\frac{{\rm d}^3 {\bf p} \, {\rm d}^3 {\bf q}}{(2 \pi)^6} \,
\frac{1}{E_{\bf p} \, E_{\bf q} \,E_{{\bf p}+{\bf q}}} \nonumber \\
& \times & \left[ 1 + n(E_{\bf p})\right]\,
\left[ 1 + n(E_{\bf q}) \right]\,  n(E_{{\bf p}+{\bf q}})\,
\delta \left(E_{{\bf p}+{\bf q}} + m -
E_{\bf p} - E_{\bf q} \right) \,\,.
\end{eqnarray}
(The last $\delta$ function in the last line of eq.\ 
(\ref{M3}) has also no support for $\omega = m >0$.)
This integral is most easily evaluated as follows. Let us first define
$p \equiv |{\bf p}|\, ,\,\, q \equiv |{\bf q}|\, ,\,\,
E_x \equiv \sqrt{x^2 +m^2}$. The angular
integration (involving the angle between ${\bf p}$ and ${\bf q}$) is
substituted by an integration over $E_{{\bf p}+{\bf q}}$,
with the Jacobian ${\rm d}E_{{\bf p}+{\bf q}}/{\rm d} \cos
({\bf p}, {\bf q}) = p\,q/E_{{\bf p}+{\bf q}}$. This
allows for a simple evaluation of this integral with the help of
the $\delta$ function:
\begin{eqnarray}
\lefteqn{{\cal M}_3(m,{\bf 0}) = 
\frac{3}{32\, \pi^3} \, \left( 1 - \exp[-m/T] \right)\, \int_0^\infty
\frac{{\rm d}p\, p}{E_p} \, 
\left[ 1 + n(E_p)\right]\,
\int_0^\infty \frac{{\rm d}q\, q}{E_q} \, 
\left[ 1 + n(E_q) \right] } \nonumber \\
 & \times &
n(E_p + E_q - m) \, \Theta \left(E_{p+q} +  m-
E_p - E_q \right) \, \Theta \left(E_p + E_q - m - E_{p-q} \right)\,\, .
\end{eqnarray}
The $\Theta$ functions are equivalent to the constraints
\begin{equation}
p\, q \geq \left(E_p - m \right)\,
\left(E_q - m \right) \geq - \, p \, q\,\,.
\end{equation}
The right inequality is trivially fulfilled ($E_p \geq m$), while the
left is also true for all values of $p$ and $q$, since
$E_p^2 \leq (p + m)^2$. Therefore, the $\Theta$ functions do not
impose additional constraints on the $p$ and $q$ integrations and can
be simply omitted. Introducing the dimensionless variables
$a \equiv m/T$, $x \equiv E_p/T$, $y\equiv
E_q/T$, abbreviating $N(x) \equiv (e^x-1)^{-1}$,
and noting that 
\begin{equation}
(1+N(y))\,N(x+y-a) = N(x-a) \, \frac{{\rm d}}{{\rm d} y}\,
\ln \frac{1-e^{-y}}{e^{a-x-y}-1}\,\, ,
\end{equation}
one arrives at
\begin{equation}
{\cal M}_3(m,{\bf 0}) =  
\frac{3\, T^2}{32\, \pi^3} \, \left( 1 - e^{-a} \right)\, \int_a^\infty
{\rm d}x \, [1+N(x)]\, N(x-a)\, \ln \frac{1-e^{-x}}{1-e^{-a}}\,\,.
\end{equation}
Introducing $t \equiv e^{-a}$, substituting $u=e^{-x}$, and then 
$z = (u-t)/(1-u)$, one obtains the final result
\begin{equation}
{\cal M}_3(m,{\bf 0}) = \frac{3\, T^2}{32\, \pi^3} \, {\rm Li}_2(e^{-m/T})
\,\, ,
\end{equation}
where
\begin{equation} \label{spence}
{\rm Li}_2(t) \equiv - \int_0^1 \frac{{\rm d}z}{z}\, \ln(1-zt)
\end{equation}
is the dilogarithm, or Spence's integral.

\section{The interaction terms in the equation of motion for the $\sigma$
field}

In analogy to eqs.\ (\ref{A1}) let us define
\begin{mathletters} \label{M1M2}
\begin{eqnarray}
{\cal M}_1^{(i)} (x) & = & D_>^{(i)}(x) - D_<^{(i)}(x) \,\, ,\\
{\cal M}_2^{(ij)} (x) & = & D_>^{(i)}(x)\, D_>^{(j)}(x) - 
D_<^{(i)}(x) \, D_<^{(j)}(x) \,\, ,
\,\,\, i,j = \sigma\,\,\, {\rm or} \,\,\, \pi\,\, .
\label{M2sigmapi}
\end{eqnarray}
\end{mathletters}
The Fourier transforms are rather similar to those in eqs.\ (\ref{A2}):
\begin{mathletters} 
\begin{eqnarray}
{\cal M}_1^{(i)} (\omega,{\bf k}) & = & \Theta(|{\bf k}|-k_c)\, 
\frac{2 \pi}{2\,E_{\bf k}^{(i)}} \left[ \delta(\omega- E_{\bf k}^{(i)}) - 
\delta(\omega+E_{\bf k}^{(i)}) \right] \,\, ,\label{M1i}\,\,\,\,  \\
{\cal M}_2^{(ij)} (\omega,{\bf k}) & = & 2 \pi \int 
\frac{{\rm d}^3{\bf p}}{(2 \pi)^3}\, \Theta(|{\bf p}|-k_c)\, 
\Theta (|{\bf k} - {\bf p}|-k_c) \, 
\frac{1}{4\, E_{\bf p}^{(i)}\, E_{{\bf k}- {\bf p}}^{(j)}} \,\, , 
\nonumber \\
& \times & \left\{  \left( \left[1+n \left(E_{\bf p}^{(i)}\right) \right] 
\left[1+n \left(E_{{\bf k}- {\bf p}}^{(j)}\right) \right] 
- n\left(E_{\bf p}^{(i)}\right) n\left(E_{{\bf k}- {\bf p}}^{(j)}\right) 
\, \right) \right. \nonumber \\
&   & \hspace*{0.5cm} \times
\left[ \delta \left( \omega- E_{\bf p}^{(i)} - E_{{\bf k}- {\bf p}}^{(j)} 
\right)  -  \delta \left( \omega+ E_{\bf p}^{(i)} + 
E_{{\bf k}- {\bf p}}^{(j)} \right) \, \right]  \nonumber \\
&   &  
+ \left( \left[1+n \left(E_{\bf p}^{(i)} \right) \right] 
n \left(E_{{\bf k}- {\bf p}}^{(j)} \right) 
- n \left(E_{\bf p}^{(i)} \right) \left[1+ n \left(
E_{{\bf k}- {\bf p}}^{(j)} \right) \right]\,  \right) \nonumber \\
&   & \left.\hspace*{0.5cm} \times\,
\left[\delta \left(\omega-E_{\bf p}^{(i)} + E_{{\bf k}- {\bf p}}^{(j)} \right) 
-  \delta \left(\omega+E_{\bf p}^{(i)} - E_{{\bf k}- {\bf p}}^{(j)} \right) 
\, \right] \right\}\,\,. \label{M2ij}
\end{eqnarray}
\end{mathletters}
Then, the terms ${\cal T}_\sigma^{(i)}(x)$
in the classical equation of motion for the $\sigma$ field read:
\begin{mathletters}
\begin{eqnarray}
{\cal T}^{(1)}_\sigma(x) & \equiv &6\, \bar{\sigma}(x)
\int \frac{{\rm d}^3 {\bf k}}{(2 \pi)^3}\,
e^{i\, {\bf k} \cdot {\bf x}} \, {\cal S}^{(1)} (x_0,{\bf k})\,\, , \\
{\cal T}^{(2)}_\sigma(x) & \equiv & 4 \sum_{a} \bar{\pi}_a(x)
\int \frac{{\rm d}^3 {\bf k}}{(2 \pi)^3}\,
e^{i\, {\bf k} \cdot {\bf x}} \, {\cal S}_a^{(2)} (x_0,{\bf k})\,\, , \\
{\cal T}^{(3)}_\sigma(x) & \equiv & 
\int \frac{{\rm d}^3 {\bf k}}{(2 \pi)^3}\,
e^{i\, {\bf k} \cdot {\bf x}} \, {\cal S}^{(3)} (x_0,{\bf k})\,\, ,
\end{eqnarray}
\end{mathletters}
where (for $t_i \rightarrow - \infty$):
\begin{mathletters}
\begin{eqnarray}
{\cal S}^{(1)} (x_0,{\bf k}) & \equiv &
\int \frac{{\rm d} \omega}{2 \pi}\, {\cal M}_1^{(\sigma)}(\omega,{\bf k})
\int \frac{{\rm d}^3{\bf p}}{(2 \pi)^3}\,
\Theta(k_c - | {\bf p}|)\, \Theta(k_c -|{\bf k}- {\bf p}|) \nonumber \\ 
& \times & 
i \int_0^{\infty} {\rm d} \tau\, e^{-i \omega  \tau}\, 
\left[ 3\, \bar{\sigma}(x_0-\tau,{\bf k} - {\bf p})\,
\bar{\sigma}(x_0-\tau,{\bf p}) +
\bar{\mbox{\boldmath $\pi$}}(x_0 - \tau,{\bf k}-{\bf p}) \cdot
\bar{\mbox{\boldmath $\pi$}}(x_0 - \tau,{\bf p}) \right]\,\, ,  \label{S1}\\
{\cal S}_{a}^{(2)} (x_0,{\bf k}) & \equiv &
\int \frac{{\rm d} \omega}{2 \pi}\, {\cal M}_1^{(\pi)}(\omega,{\bf k})
\int \frac{{\rm d}^3{\bf p}}{(2 \pi)^3}\,
\Theta(k_c - | {\bf p}|)\, \Theta(k_c -|{\bf k}- {\bf p}|) \nonumber \\ 
& \times & 
i \int_0^{\infty} {\rm d} \tau\, e^{-i \omega  \tau}\, 
\bar{\pi}_a(x_0-\tau,{\bf p})\, \bar{\sigma}(x_0 - \tau,{\bf k} -{\bf p}) 
\,\, , \label{S2}\\
{\cal S}^{(3)} (x_0,{\bf k}) & \equiv & 2 \int \frac{{\rm d}\omega}{2 \pi}
 \left[ 9\, {\cal M}_2^{(\sigma\sigma)}(\omega , {\bf k}) 
+ (N-1) \, {\cal M}_2^{(\pi\pi)}(\omega , {\bf k}) \right] 
\Theta(k_c-|{\bf k}|) \,
i \int_0^\infty {\rm d} \tau \, e^{-i \omega \tau}\, \bar{\sigma}
(x_0 - \tau, {\bf k})\, .
\end{eqnarray}
\end{mathletters}
In the linear harmonic approximation (\ref{linharm}) these terms become:
\begin{mathletters} \label{S}
\begin{eqnarray}
{\cal S}^{(1)} (t,{\bf k}) & \simeq &
\int \frac{{\rm d}^3{\bf p}}{(2 \pi)^3}\,
\Theta(k_c - | {\bf p}|)\, \Theta(k_c -|{\bf k}- {\bf p}|) \nonumber \\ 
& \times & 
\left\{ - \frac{3}{2}\, {\rm P} \int \frac{{\rm d} \omega}{2 \pi}\, 
{\cal M}_1^{(\sigma)}(\omega,{\bf k}) \left[ \frac{1}{E_1^{(\sigma)}
+E_2^{(\sigma)}-\omega} + \frac{1}{E_1^{(\sigma)} - E_2^{(\sigma)}-\omega}
\right] \, \bar{\sigma}(t,{\bf p})\, \bar{\sigma}(t,{\bf k}- {\bf p}) \right.
\nonumber \\
& +  &  \frac{3}{2\, E_1^{(\sigma)}\, E_2^{(\sigma)}} \, {\rm P} 
\int \frac{{\rm d} \omega}{2 \pi}\, {\cal M}_1^{(\sigma)}(\omega,{\bf k})
\left[ \frac{1}{E_1^{(\sigma)} + E_2^{(\sigma)}-\omega} - 
\frac{1}{E_1^{(\sigma)} - E_2^{(\sigma)}-\omega} \right] \,
\partial_t\, \bar{\sigma}(t,{\bf p}) \, \partial_t\, \bar{\sigma}(t,{\bf k}
-{\bf p}) \nonumber \\
& -  &  \frac{1}{2} \, {\rm P} \int \frac{{\rm d} \omega}{2 \pi}\, 
{\cal M}_1^{(\sigma)}(\omega,{\bf k})
\left[ \frac{1}{E_1^{(\pi)} +E_2^{(\pi)}-\omega} + 
\frac{1}{E_1^{(\pi)} - E_2^{(\pi)}-\omega} \right] \, 
\bar{\mbox{\boldmath $\pi$}}(t,{\bf p})\cdot 
\bar{\mbox{\boldmath $\pi$}}(t,{\bf k}- {\bf p}) \nonumber \\
& +  &  \frac{1}{2\, E_1^{(\pi)}\, E_2^{(\pi)}} \, {\rm P} 
\int \frac{{\rm d} \omega}{2 \pi}\, {\cal M}_1^{(\sigma)}(\omega,{\bf k})
\left[ \frac{1}{E_1^{(\pi)} + E_2^{(\pi)}-\omega} - 
\frac{1}{E_1^{(\pi)} - E_2^{(\pi)}-\omega} \right] \,
\partial_t\, \bar{\mbox{\boldmath $\pi$}}(t,{\bf p}) \cdot \partial_t\, 
\bar{\mbox{\boldmath $\pi$}}(t,{\bf k} -{\bf p}) \nonumber \\
& -  &  \frac{3}{2\, E_2^{(\sigma)}} 
\left[{\cal M}_1^{(\sigma)}\left(E_1^{(\sigma)} +E_2^{(\sigma)},{\bf k}\right) 
-{\cal M}_1^{(\sigma)}\left(E_1^{(\sigma)} - E_2^{(\sigma)},{\bf k} \right)
\right] \, \partial_t \, \bar{\sigma}(t,{\bf p})\, 
\bar{\sigma}(t,{\bf k}-{\bf p}) \nonumber \\
& -  & \left. \frac{1}{2\,E_2^{(\pi)}} 
\left[ {\cal M}_1^{(\sigma)}\left(E_1^{(\pi)} +E_2^{(\pi)},{\bf k}\right) 
- {\cal M}_1^{(\sigma)}\left(E_1^{(\pi)} - E_2^{(\pi)},{\bf k}\right) 
\right] \, \partial_t\, \bar{\mbox{\boldmath $\pi$}}(t,{\bf p})\cdot 
 \bar{\mbox{\boldmath $\pi$}}(t,{\bf k}-{\bf p})\,\right\}\,\,, \\
{\cal S}_{a}^{(2)} (t,{\bf k}) & \simeq &
\int \frac{{\rm d}^3{\bf p}}{(2 \pi)^3}\,
\Theta(k_c - | {\bf p}|)\, \Theta(k_c -|{\bf k}- {\bf p}|) \nonumber \\ 
& \times & 
\left\{- \frac{1}{2}\, {\rm P} \int \frac{{\rm d} \omega}{2 \pi}\, 
{\cal M}_1^{(\pi)}(\omega,{\bf k})\left[ \frac{1}{E_1^{(\sigma)}
+E_2^{(\pi)}-\omega} + \frac{1}{E_1^{(\sigma)} - E_2^{(\pi)}-\omega}
\right] \, \bar{\pi}_a(t,{\bf p})\, \bar{\sigma}(t,{\bf k}- {\bf p}) \right.
\nonumber \\
& +  &  \frac{1}{2\, E_1^{(\sigma)}\, E_2^{(\pi)}} \, {\rm P} 
\int \frac{{\rm d} \omega}{2 \pi}\, {\cal M}_1^{(\pi)}(\omega,{\bf k})
\left[ \frac{1}{E_1^{(\sigma)} + E_2^{(\pi)}-\omega} - 
\frac{1}{E_1^{(\sigma)} - E_2^{(\pi)}-\omega} \right] \,
\partial_t\, \bar{\pi}_a(t,{\bf p}) \, \partial_t\, \bar{\sigma}(t,{\bf k}
-{\bf p}) \nonumber \\
& -  &  \frac{1}{4\,E_1^{(\sigma)}} 
\left[ {\cal M}_1^{(\pi)}\left(E_1^{(\sigma)} +E_2^{(\pi)},{\bf k}\right) 
+ {\cal M}_1^{(\pi)}\left(E_1^{(\sigma)} - E_2^{(\pi)},{\bf k} \right) 
\right] \, \bar{\pi}_a(t,{\bf p})\, 
\partial_t\, \bar{\sigma}(t,{\bf k}-{\bf p}) \nonumber \\
& -  & \left. \frac{1}{4\, E_2^{(\pi)}} 
\left[ {\cal M}_1^{(\pi)}\left(E_1^{(\sigma)} +E_2^{(\pi)},{\bf k}\right) -
{\cal M}_1^{(\pi)}\left(E_1^{(\sigma)} - E_2^{(\pi)},{\bf k}\right) \right] \, 
\partial_t\, \bar{\pi}_a(t,{\bf p})\,\bar{\sigma}(t,{\bf k}-{\bf p})
\,\right\}\,\,, \\
{\cal S}^{(3)} (t,{\bf k}) & \simeq & - 2\, \Theta(k_c-|{\bf k}|) 
\left[ {\rm P} \int \frac{{\rm d}\omega}{2 \pi}\,
 \frac{ 9\, {\cal M}_2^{(\sigma\sigma)}(\omega , {\bf k}) 
+ (N-1) \, {\cal M}_2^{(\pi\pi)}(\omega , {\bf k})}{
E_{\bf k}^{(\sigma)}-\omega}\,\, \bar{\sigma}(t,{\bf k}) \right. \nonumber \\
&    &  \hspace*{2cm}+ \left.
\, \frac{ 9\, {\cal M}_2^{(\sigma\sigma)}(E_{\bf k}^{(\sigma)}, {\bf k}) 
+ (N-1) \, {\cal M}_2^{(\pi\pi)}(E_{\bf k}^{(\sigma)} , {\bf k})}{
2\, E_{\bf k}^{(\sigma)}}  \, \partial_t\, \bar{\sigma}(t,{\bf k})\right]\,\,,
\label{S3}
\end{eqnarray}
\end{mathletters}
where $E_1^{(i)} = E_{{\bf k}-{\bf p}}^{(i)}$ and 
$E_2^{(i)} = E_{\bf p}^{(i)}$.
The final expressions for ${\cal T}_\sigma^{(i)}(t,{\bf k})$ are:
\begin{mathletters} \label{Tsigma}
\begin{eqnarray}
{\cal T}_\sigma^{(1)}(t,{\bf k}) & \simeq & 6\,
\int \frac{{\rm d}^3{\bf p}}{(2 \pi)^3}\, \Theta(k_c - | {\bf p}|)\,
\bar{\sigma}(t,{\bf p})\, {\cal S}^{(1)}(t,{\bf k} - {\bf p}) \,\, , \\
{\cal T}_\sigma^{(2)}(t,{\bf k}) & \simeq & 4
\int \frac{{\rm d}^3{\bf p}}{(2 \pi)^3}\, \Theta(k_c - | {\bf p}|)\,
\sum_a \bar{\pi}_a(t,{\bf p})\, {\cal S}_a^{(2)}(t,{\bf k} - {\bf p}) \,\, , \\
{\cal T}_\sigma^{(3)}(t,{\bf k}) & \simeq & 
{\cal S}^{(3)}(t,{\bf k}) \,\, . 
\end{eqnarray}
\end{mathletters}
Since ${\cal M}_1^{(i)}(\omega,{\bf 0}) \equiv 0$, cf.\ eq.\ (\ref{M1i}),
it follows from relations (\ref{S}) and (\ref{Tsigma})
that for $k_c \rightarrow 0$, ${\cal T}_\sigma^{(1)}(t,{\bf 0})=
{\cal T}_\sigma^{(2)}(t,{\bf 0}) \equiv 0$. 

\section{The interaction terms in the equation of motion for the $\pi$ field}

The interaction terms in the equation of motion for the $\bar{\pi}$ field
are:
\begin{mathletters}
\begin{eqnarray}
{\cal T}^{(1)}_{\pi_a}(x) & \equiv &2\, \bar{\pi}_a(x)
\int \frac{{\rm d}^3 {\bf k}}{(2 \pi)^3}\,
e^{i\, {\bf k} \cdot {\bf x}} \, {\cal P}^{(1)} (x_0,{\bf k})\,\, , \\
{\cal T}^{(2)}_{\pi_a}(x) & \equiv & 4\, \bar{\sigma}(x)
\int \frac{{\rm d}^3 {\bf k}}{(2 \pi)^3}\,
e^{i\, {\bf k} \cdot {\bf x}} \, {\cal P}_a^{(2)} (x_0,{\bf k})\,\, , \\
{\cal T}^{(3)}_{\pi_a}(x) & \equiv & 
\int \frac{{\rm d}^3 {\bf k}}{(2 \pi)^3}\,
e^{i\, {\bf k} \cdot {\bf x}} \, {\cal P}^{(3)}_a (x_0,{\bf k})\,\, ,
\end{eqnarray}
\end{mathletters}
where ${\cal P}^{(1)}(x_0,{\bf k}) \equiv {\cal S}^{(1)}(x_0,{\bf k})$,
and ${\cal P}_a^{(2)}(x_0,{\bf k}) \equiv {\cal S}_a^{(2)}(x_0,{\bf k})$,
cf.\ eqs.\ (\ref{S1}) and (\ref{S2}), while
\begin{equation}
{\cal P}^{(3)}_a(x_0,{\bf k}) \equiv 4 \int \frac{{\rm d}\omega}{2 \pi}
 {\cal M}_2^{(\sigma\pi)}(\omega , {\bf k})\, \Theta(k_c-|{\bf k}|) \,
i \int_0^\infty {\rm d} \tau \, e^{-i \omega \tau}\, \bar{\pi}_a
(x_0 - \tau, {\bf k})\, .
\end{equation}
In linear harmonic approximation we therefore obtain:
\begin{mathletters} \label{Tpi}
\begin{eqnarray}
{\cal T}_{\pi_a}^{(1)}(t,{\bf k}) & \simeq & 2
\int \frac{{\rm d}^3{\bf p}}{(2 \pi)^3}\, \Theta(k_c - | {\bf p}|)\,
\bar{\pi}_a(t,{\bf p})\, {\cal S}^{(1)}(t,{\bf k} - {\bf p}) \,\, , \\
{\cal T}_{\pi_a}^{(2)}(t,{\bf k}) & \simeq & 4
\int \frac{{\rm d}^3{\bf p}}{(2 \pi)^3}\, \Theta(k_c - | {\bf p}|)\,
\bar{\sigma}(t,{\bf p})\, {\cal S}_a^{(2)}(t,{\bf k} - {\bf p}) \,\, , \\
{\cal T}_{\pi_a}^{(3)}(t,{\bf k}) & \simeq & -
4\, \Theta(k_c-|{\bf k}|) 
\left[ {\rm P} \int \frac{{\rm d}\omega}{2 \pi}\,
 \frac{ {\cal M}_2^{(\sigma\pi)}(\omega , {\bf k}) }{
E_{\bf k}^{(\pi)}-\omega}\,\, \bar{\pi}_a(t,{\bf k}) 
+  \frac{ {\cal M}_2^{(\sigma\pi)}(E_{\bf k}^{(\pi)}, {\bf k}) }{
2\, E_{\bf k}^{(\pi)}}  \, \partial_t\, \bar{\pi}_a(t,{\bf k})\right]\,\,.
\label{Tpi3} 
\end{eqnarray}
\end{mathletters}
Again, since ${\cal M}_1^{(i)}(\omega,{\bf 0}) \equiv 0$, cf.\ eq.\ (\ref{M1}),
it follows for $k_c \rightarrow 0$ that ${\cal T}_{\pi_a}^{(1)}(t,{\bf 0})=
{\cal T}_{\pi_a}^{(2)}(t,{\bf 0}) \equiv 0$. 

\section{The dissipation coefficient for the $\sigma$ field}

The dissipation coefficient $\eta_\sigma$ in the classical 
equation of motion for $\bar{\sigma}(t)$, eq.\ (\ref{eqofmosig})
is defined by
\begin{equation} \label{etasigma2}
\eta_\sigma \equiv \left(\frac{4\,\lambda\,f_\pi}{N} \right)^2
2\, \frac{ 9\, {\cal M}_2^{(\sigma\sigma)}(m_\sigma, {\bf 0}) 
+ (N-1) \, {\cal M}_2^{(\pi\pi)}(m_\sigma , {\bf 0})}{ 2\, m_\sigma} \,\, ,
\end{equation}
where ${\cal M}_2^{(ij)}(\omega,{\bf k})$ is given by eq.\ (\ref{M2ij}). 
${\cal M}_2^{(\sigma\sigma)}(m_\sigma,{\bf 0})$ describes 
the decay of a $\sigma$ at rest into two $\sigma$'s (and the
reverse reaction, corresponding to 
the second and third line of eq.\ (\ref{M2ij})), and the absorption of
a $\sigma$ by a $\sigma$ at rest, producing a $\sigma$ 
(corresponding to the fourth and fifth line of eq.\ (\ref{M2ij})).

However, a $\sigma$ at rest cannot decay into two $\sigma$'s by energy
conservation (similarly, two $\sigma$'s cannot annihilate to form one), 
and also the absorption of a $\sigma$ by a $\sigma$ at rest, producing a
$\sigma$, is also impossible. Therefore, ${\cal M}_2^{(\sigma \sigma)}
(m_\sigma,{\bf 0}) \equiv 0$. Mathematically, this is immediately obvious 
inspecting the arguments of the $\delta$ functions in eq.\ (\ref{M2ij})
with $\omega \equiv m_\sigma,\, {\bf k}=0$.

Similar arguments prevent the absorption of a $\pi$ by a $\sigma$ at rest,
producing a $\pi$,
but the decay of a heavy $\sigma$ at rest into two light $\pi$'s (and the
corresponding reverse reaction of $\pi \pi \rightarrow \sigma$)
is always possible (provided $m_\sigma > 2 m_\pi$).
Therefore (utilizing detailed balance, and taking the limit 
$k_c \rightarrow 0$),
\begin{eqnarray}
{\cal M}_2^{(\pi \pi)} (m_\sigma,{\bf 0}) & = & 
2\pi\, \left( 1 - \exp[-m_\sigma/T] \right)\,
\int \frac{{\rm d^3}{\bf k}}{(2 \pi)^3}\, 
\left[2\, E_{\bf k}^{(\pi)}\right]^{-2} \, 
\left[ 1 + n \left( E_{\bf k}^{(\pi)} \right) \right]^2\, 
\delta \left(m_\sigma - 2\, E_{\bf k}^{(\pi)} \right) \nonumber \\
& = & \frac{1}{8\pi}\, \sqrt{1-\frac{4 m_\pi^2}{m_\sigma^2}}\,\,
\, \frac{1+\exp[-m_\sigma/2T]}{1-\exp[-m_\sigma/2T]}\,\, .
\end{eqnarray}
Note that this expression is non-zero even at vanishing temperature.
The dissipation coefficient becomes:
\begin{equation}
\eta_\sigma = \left(\frac{4\, \lambda\, f_\pi}{N} \right)^2 
\frac{N-1}{8\pi\, m_\sigma}\, \sqrt{1 - \frac{4\, m_\pi^2}{m_\sigma^2}}
\,\, {\rm coth} \,\frac{m_\sigma}{4\, T}\,\, .
\end{equation}

\section{The dissipation coefficient for the $\pi$ field}

The dissipation coefficient $\eta_\pi$ in the classical 
equation of motion for $\bar{\pi}_a(t)$, eq.\ (\ref{eqofmopi}),
is defined by
\begin{equation} \label{etapi2}
\eta_\pi \equiv \left(\frac{4\,\lambda\,f_\pi}{N} \right)^2
4\, \frac{ {\cal M}_2^{(\sigma\pi)}(m_\pi, {\bf 0}) }{ 2\, m_\pi} \,\, ,
\end{equation}
where ${\cal M}_2^{(ij)}(\omega,{\bf k})$ is the Fourier transform
of ${\cal M}_2^{(ij)}(x)$, eq.\ (\ref{M2ij}). 
${\cal M}_2^{(\sigma\pi)}(m_\pi,{\bf 0})$ describes 
the decay of a $\pi$ at rest into a $\pi$ and a $\sigma$ (and the
reverse reaction, corresponding to 
the second and third line of eq.\ (\ref{M2ij})), and the absorption of
a $\pi$ or $\sigma$ by a $\pi$ at rest, 
producing a $\sigma$ or a $\pi$, respectively 
(corresponding to the fourth and fifth line of eq.\ (\ref{M2ij})).

The decay of a $\pi$ into $\sigma$ and $\pi$ and the reverse process
is kinematically forbidden. The remaining contribution 
reads with detailed balance (and taking the limit 
$k_c \rightarrow 0$):
\begin{eqnarray}
{\cal M}_2^{(\sigma \pi)} (m_\pi,{\bf 0}) & = & 2\pi\, (1-\exp[-m_\pi/T])
\int \frac{{\rm d}^3 {\bf k}}{(2 \pi)^3} \, 
\frac{1+n \left(E_{\bf k}^{(\sigma)}\right)}{2\, E_{\bf k}^{(\sigma)}}
\, \frac{ n \left(E_{\bf k}^{(\pi)} \right)}{2\, E_{\bf k}^{(\pi)}}
\, \delta \left(E_{\bf k}^{(\pi)} + m_\pi - E_{\bf k}^{(\sigma)}
\right) \nonumber \\
& = & \frac{1}{8 \pi} \, \frac{m_\sigma^2}{m_\pi^2} \,
\sqrt{ 1-\frac{4 m_\pi^2}{m_\sigma^2} }\, \frac{1-\exp[-m_\pi/T]}{
1-\exp[-m_\sigma^2/2m_\pi T]} \, \frac{1}{\exp[(m_\sigma^2-2m_\pi^2)/2
m_\pi T] -1} \,\, .
\end{eqnarray}
The dissipation coefficient therefore becomes
\begin{equation}
\eta_\pi = \left(\frac{4\, \lambda\, f_\pi}{N} \right)^2 
\frac{m_\sigma^2}{4\pi\, m_\pi^3}\, \sqrt{1 - \frac{4 m_\pi^2}{m_\sigma^2}}
\,\, \frac{1-\exp[-m_\pi/T]}{1-\exp[-m_\sigma^2/2m_\pi T]} \, 
\frac{1}{\exp[(m_\sigma^2-2m_\pi^2)/2 m_\pi T] -1} \,\, .
\end{equation}
This expression vanishes at $T=0$ and in the limit $m_\pi \rightarrow 0$.

\section{The variances of the noise fields}

The variances of the noise fields are given by eqs.\ (\ref{variance}).
In order to evaluate them, let us define in analogy to eqs.\ (\ref{M1M2})
the ``noise kernels''
\begin{mathletters}
\begin{eqnarray}
{\cal N}_1^{(i)} (x) & \equiv & D_>^{(i)}(x) + D_<^{(i)}(x) \,\, , \\
{\cal N}_2^{(ij)} (x) & \equiv & D_>^{(i)}(x) \, D_>^{(i)}(x) 
+ D_<^{(i)}(x)\, D_<^{(i)}(x) \,\, , \,\,\,\,i,j= \sigma \,\,\,\,
{\rm or}\,\,\,\, \pi\,\, .
\end{eqnarray}
\end{mathletters}
Their Fourier transforms are quite similar to those of 
${\cal M}_1^{(i)}$ and ${\cal M}_2^{(ij)}$: 
\begin{mathletters} 
\begin{eqnarray}
{\cal N}_1^{(i)} (\omega,{\bf k}) & = & \Theta(|{\bf k}|-k_c)\, 
\frac{2 \pi}{2\,E_{\bf k}^{(i)}} 
\left[1+2\,n\left(E_{\bf k}^{(i)} \right) \right]\,
\left[ \delta(\omega- E_{\bf k}^{(i)}) +
\delta(\omega+E_{\bf k}^{(i)}) \right] \,\, ,\label{N1i}\,\,\,\,  \\
{\cal N}_2^{(ij)} (\omega,{\bf k}) & = & 2 \pi \int 
\frac{{\rm d}^3{\bf p}}{(2 \pi)^3}\, \Theta(|{\bf p}|-k_c)\, 
\Theta (|{\bf k} - {\bf p}|-k_c) \, 
\frac{1}{4\, E_{\bf p}^{(i)}\, E_{{\bf k}- {\bf p}}^{(j)}} \,\, , 
\nonumber \\
& \times & \left\{  \left( \left[1+n \left(E_{\bf p}^{(i)}\right) \right] 
\left[1+n \left(E_{{\bf k}- {\bf p}}^{(j)}\right) \right] 
+ n\left(E_{\bf p}^{(i)}\right) n\left(E_{{\bf k}- {\bf p}}^{(j)}\right) 
\, \right) \right. \nonumber \\
&   & \hspace*{0.5cm} \times
\left[ \delta \left( \omega- E_{\bf p}^{(i)} - E_{{\bf k}- {\bf p}}^{(j)} 
\right)  +  \delta \left( \omega+ E_{\bf p}^{(i)} + 
E_{{\bf k}- {\bf p}}^{(j)} \right) \, \right]  \nonumber \\
&   &  
+ \left( \left[1+n \left(E_{\bf p}^{(i)} \right) \right] 
n \left(E_{{\bf k}- {\bf p}}^{(j)} \right) 
+ n \left(E_{\bf p}^{(i)} \right) \left[1+ n \left(
E_{{\bf k}- {\bf p}}^{(j)} \right) \right]\,  \right) \nonumber \\
&   & \left.\hspace*{0.5cm} \times\,
\left[\delta \left(\omega-E_{\bf p}^{(i)} + E_{{\bf k}- {\bf p}}^{(j)} \right) 
+  \delta \left(\omega+E_{\bf p}^{(i)} - E_{{\bf k}- {\bf p}}^{(j)} \right) 
\, \right] \right\}\,\,. 
\end{eqnarray}
\end{mathletters}
Using the fact that
the hard modes are distributed according to the Bose--Einstein 
distribution function $n(E) = (e^{E/T}-1)^{-1}$, one can show that
(cf.\ eq.\ (66) of \cite{cgbm})
\begin{mathletters} \label{coth}
\begin{eqnarray}
{\cal N}_1^{(i)}(\omega,{\bf k}) & \equiv & {\cal M}_1^{(i)}(\omega,{\bf k})\,
\coth \left[ \frac{\omega}{2T} \right] \,\,, \\
{\cal N}_2^{(ij)}(\omega,{\bf k}) & \equiv & {\cal M}_2^{(ij)}
(\omega,{\bf k})\, \coth \left[ \frac{\omega}{2T} \right] \,\,.
\end{eqnarray}
\end{mathletters}
For the Fourier transforms of ${\cal I}_{ab}(x,y)$ one therefore
derives ($t_i \rightarrow - \infty,\, t_f \rightarrow +\infty$):
\begin{mathletters}
\begin{eqnarray}
{\cal I}_{\sigma\sigma}(k_0,{\bf k};q_0,{\bf q}) & \equiv  &
\int_{t_i}^{t_f} {\rm d}^4x\, {\rm d}^4y \, e^{i\, (k_0 x_0 - {\bf k} \cdot
{\bf x} + q_0 y_0 - {\bf q} \cdot {\bf y})}\, {\cal I}_{\sigma\sigma}(x,y) 
\nonumber \\
& = & \left( \frac{4\,\lambda\,f_\pi}{N} \right)^2 \left\{
\int \frac{{\rm d}p_0 \, {\rm d}^3{\bf p}}{(2 \pi)^4}\, 
\Theta(k_c-|{\bf p}|)\, \Theta (k_c-|{\bf k}+{\bf q} - {\bf p}|) \right.
\nonumber \\
&   & \times 
\left[ \frac{}{}
18\, \bar{\sigma}(p_0,{\bf p})\,{\cal N}_1^{(\sigma)}(k_0-p_0,
{\bf k} - {\bf p})\,\bar{\sigma}(k_0+q_0-p_0, {\bf k} +{\bf q} - {\bf p})
\right. \nonumber \\
&    & \left. +\, 2 
\sum_a \bar{\pi}_a (p_0,{\bf p})\,{\cal N}_1^{(\pi)}(k_0-p_0,
{\bf k} - {\bf p})\,\bar{\pi}_a(k_0+q_0-p_0, {\bf k} +{\bf q} - {\bf p})
\right] \nonumber \\
&    & + \left. (2 \pi)^4 \,\delta(k_0+q_0)\,\delta^{(3)}({\bf k}+{\bf q})
\left[ 9\, {\cal N}_2^{(\sigma\sigma)}(k_0,{\bf k})
+ (N-1)\, {\cal N}_2^{(\pi\pi)}(k_0,{\bf k}) \right] \frac{}{} \right\}
\,\, , \\
{\cal I}_{\sigma\pi_a}(k_0,{\bf k};q_0,{\bf q}) & = & 
\left( \frac{4\,\lambda\,f_\pi}{N} \right)^2 
\int \frac{{\rm d}p_0 \, {\rm d}^3{\bf p}}{(2 \pi)^4}\, 
\Theta(k_c-|{\bf p}|)\, \Theta (k_c-|{\bf k}+{\bf q} - {\bf p}|) 
\nonumber \\
&   & \times 
\left[ \frac{}{}
6\, \bar{\sigma}(p_0,{\bf p})\,{\cal N}_1^{(\sigma)}(k_0-p_0,
{\bf k} - {\bf p})\,\bar{\pi}_a(k_0+q_0-p_0, {\bf k} +{\bf q} - {\bf p})
\right. \nonumber \\
&    & \left. +\, 2 \,
\bar{\pi}_a (p_0,{\bf p})\,{\cal N}_1^{(\pi)}(k_0-p_0,
{\bf k} - {\bf p})\,\bar{\sigma}(k_0+q_0-p_0, {\bf k} +{\bf q} - {\bf p})
\right]\,\, , \\
{\cal I}_{\pi_a\pi_b}(k_0,{\bf k};q_0,{\bf q}) & = & 
\left( \frac{4\,\lambda\,f_\pi}{N} \right)^2 \left\{
\int \frac{{\rm d}p_0 \, {\rm d}^3{\bf p}}{(2 \pi)^4}\, 
\Theta(k_c-|{\bf p}|)\, \Theta (k_c-|{\bf k}+{\bf q} - {\bf p}|) \right.
\nonumber \\
&   & \times 
\left[ \frac{}{}
2\, \bar{\pi}_a(p_0,{\bf p})\,{\cal N}_1^{(\sigma)}(k_0-p_0,
{\bf k} - {\bf p})\,\bar{\pi}_b(k_0+q_0-p_0, {\bf k} +{\bf q} - {\bf p})
\right. \nonumber \\
&    & \left. +\, 2 \,\delta_{ab}\,
\bar{\sigma} (p_0,{\bf p})\,{\cal N}_1^{(\pi)}(k_0-p_0,
{\bf k} - {\bf p})\,\bar{\sigma}(k_0+q_0-p_0, {\bf k} +{\bf q} - {\bf p})
\right] \nonumber \\
&    & + \left. 2\, \delta_{ab}\,
(2 \pi)^4 \,\delta(k_0+q_0)\,\delta^{(3)}({\bf k}+{\bf q})
\, {\cal N}_2^{(\sigma\pi)}(k_0,{\bf k}) \frac{}{} \right\} \,\, .
\end{eqnarray}
\end{mathletters}
The focus of interest are the spatially homogeneous noise terms,
${\bf k} = {\bf q} = 0$. For this case, all integrals 
vanish in these expressions due to the fact that ${\cal N}_1^{(i)}(\omega,
{\bf k})$ is proportional to $\Theta(|{\bf k}|-k_c)$, cf.\ eq.\
(\ref{N1i}). This has the further consequence that all
cross correlations between the noise fields vanish,
${\cal I}_{ab} \sim \delta_{ab}$. Using $(2\pi)^3\delta^{(3)}({\bf
k} + {\bf q}) \equiv V \delta^{(3)}_{{\bf k}+{\bf q},0}$ one arrives at:
\begin{mathletters}
\begin{eqnarray}
{\cal I}_{\sigma\sigma}(k_0,{\bf 0};q_0,{\bf 0}) & = &
V\, 2\pi \, \delta(k_0+q_0) \, \left( \frac{4\,\lambda\,f_\pi}{N} \right)^2
\left[ 9\, {\cal N}_2^{(\sigma\sigma)}(k_0,{\bf 0}) + (N-1)\,
{\cal N}_2^{(\pi\pi)}(k_0,{\bf 0}) \right]\,\, , \\
{\cal I}_{\pi_a\pi_b}(k_0,{\bf 0};q_0,{\bf 0}) & = & \delta_{ab}\,
V\, 2\pi \, \delta(k_0+q_0) \, \left( \frac{4\,\lambda\,f_\pi}{N} \right)^2
2\, {\cal N}_2^{(\sigma\pi)}(k_0,{\bf 0}) \,\, .
\end{eqnarray}
\end{mathletters}
For the variance of the noise fields $\xi_\sigma(t)$ and $\xi_{\pi_a}(t)$
one therefore obtains:
\begin{mathletters}
\begin{eqnarray}
\langle \xi_\sigma(t)\,\xi_\sigma(t') \rangle_\xi & \equiv &
\left\langle \frac{\xi_\sigma(t,{\bf 0})}{V} \,
\frac{\xi_\sigma(t',{\bf 0})}{V} \right\rangle_\xi
= \int \frac{{\rm d}k_0\,{\rm d}q_0}{(2 \pi)^2}\,
e^{i(k_0t+q_0t')} \left\langle \frac{\xi_\sigma(-k_0,{\bf 0})}{V} \,
\frac{\xi_\sigma(-q_0,{\bf 0})}{V} \right\rangle_\xi \nonumber \\
& = & \int \frac{{\rm d}k_0\,{\rm d}q_0}{(2 \pi)^2}\,
e^{i(k_0t+q_0t')} \,\frac{1}{V^2}\,
{\cal I}_{\sigma\sigma}(k_0,{\bf 0};q_0,{\bf 0}) \nonumber \\
&  = & \int \frac{{\rm d}k_0}{2 \pi}\,
e^{ik_0(t-t')}\, \frac{1}{V} \left( \frac{4\,\lambda\,f_\pi}{N} \right)^2
\left[ 9\, {\cal N}_2^{(\sigma\sigma)}(k_0,{\bf 0}) + (N-1)\,
{\cal N}_2^{(\pi\pi)}(k_0,{\bf 0}) \right]\,\, , \\
\langle \xi_{\pi_a}(t)\,\xi_{\pi_b}(t') \rangle_\xi & = & \delta_{ab}
\int \frac{{\rm d}k_0}{2 \pi}\,
e^{ik_0(t-t')}\, \frac{1}{V} \left( \frac{4\,\lambda\,f_\pi}{N} \right)^2
2\, {\cal N}_2^{(\sigma\pi)}(k_0,{\bf 0})\,\, .
\end{eqnarray}
\end{mathletters}
Here, $\langle \, \cdot \, \rangle_\xi$ is the average with respect to
the Gaussian measure (\ref{probmeas}).
Further evaluation is simplified by approximating 
${\cal N}_2^{(\sigma\sigma)}(k_0,{\bf 0}) \simeq {\cal N}_2^{(\sigma\sigma)}
(m_\sigma,{\bf 0})$, 
${\cal N}_2^{(\pi\pi)}(k_0,{\bf 0}) \simeq {\cal N}_2^{(\pi\pi)}
(m_\sigma,{\bf 0})$, 
${\cal N}_2^{(\sigma\pi)}(k_0,{\bf 0}) \simeq {\cal N}_2^{(\sigma\pi)}
(m_\pi,{\bf 0})$, i.e., taking the energy $k_0$ to be the on-shell energy 
(i.e., since ${\bf k}=0$ for both $\sigma$ and $\pi$ fields, the mass) of
the respective particle. This approximation is consistent with
the linear harmonic approximation which puts the energy
in the ${\cal M}_2$ functions on-shell. The consequence is that
the $k_0$--integral can be performed, yielding with eqs.\ (\ref{coth}),
(\ref{etasigma2}), and (\ref{etapi2}):
\begin{mathletters} \label{variance2}
\begin{eqnarray}
\langle \xi_\sigma(t)\,\xi_\sigma(t') \rangle_\xi & \simeq & \delta(t-t') \,
\frac{1}{V}  \, m_\sigma \, \eta_\sigma \, 
\coth\left[ \frac{m_\sigma}{2T} \right] \,\, , \\
\langle \xi_{\pi_a}(t)\,\xi_{\pi_b}(t') \rangle_\xi & \simeq & \delta_{ab}\,
\delta(t-t') \, \frac{1}{V} \, m_\pi \, \eta_\pi \, 
\coth\left[ \frac{m_\pi}{2T} \right] \,\, , 
\end{eqnarray}
\end{mathletters}
The $\delta$ function corresponds to white noise. 
Therefore, this approximation will be called ``white-noise'' approximation.

\end{document}